\UseRawInputEncoding
\documentclass[10pt]{article}
\usepackage{authblk}
\usepackage{amsmath}
\usepackage{amsfonts,amssymb}
\usepackage{epsfig}

\newcommand{\ui}{\mathrm{i}}
\newcommand{\ud}{\mathrm{d}}
\begin{document}

\title{Geometric descriptions for the polarization of nonparaxial light: a tutorial}
\author{Miguel A. Alonso}
\affil{Aix Marseille Univ., CNRS, Centrale Marseille, Institut Fresnel, UMR 7249, 13397 Marseille Cedex 20, France\\
The Institute of Optics and Laboratory for Laser Energetics, University of Rochester, Rochester NY 14627, USA\\
miguel.alonso@fresnel.fr}

\maketitle

\begin{abstract}
This tutorial provides an overview of the local description of polarization for nonparaxial light, for which all Cartesian components of the electric field are significant. The polarization of light at each point is characterized by a $3$ component vector in the case of full polarization or by a $3\times3$ polarization matrix for partial polarization. Standard concepts for paraxial polarization like the degree of polarization, the Stokes parameters and the Poincar\'e sphere then have generalizations for nonparaxial light that are either not unique or not trivial. This work aims to clarify some of these discrepancies, present some new concepts, and provide a framework that highlights the similarities and differences with the description for the paraxial regimes. Particular emphasis is placed on geometric interpretations. 
\end{abstract}

\section{Introduction}

The polarization of electromagnetic waves refers to the local 
geometric behavior of the oscillations of the electric (or sometimes the magnetic) field vector. The study of optical polarization and the implementation of techniques for measuring it have been largely restricted until fairly recently to light with a well-defined direction of propagation. This restriction is valid in many common situations, such as when the light source is distant and subtends a small range of angles at the point of observation, or when a collimated laser beam is considered. The transversality of the electric and magnetic fields then means that their component in the main direction of propagation is much smaller than those normal to this direction and hence has a negligible effect on measurements. These longitudinal field components can therefore be ignored for most practical purposes.

In recent years, however, there has been growing interest within areas such as nano-optics, plasmonics and microscopy, in the characterization of the polarization of light in cases where all three Cartesian field components can be significant. In these situations, the polarization properties often vary considerably within length scales of the order of the wavelength, which explains why some of the early work on the subject was for electromagnetic waves at low frequencies \cite{Samson1973,Holm1988,LiIEEE2016}, and why measurements within the optical spectrum were challenging until fairly recently. Optical measurements of nonparaxial polarization typically imply the interaction of the field with a known small probe, such as a metallic nanoparticle, placed at the point where the polarization is to be measured \cite{Lindfors2005,Lindfors2007,Deutsch2010,Bauer2015}. The field scattered by this particle is collected over a high numerical aperture by a microscope objective that collimates it, so that standard polarimetric techniques can be used to characterize the light distribution in the Fourier (or angular spectrum) domain. The polarization of the nonparaxial field at the point can be inferred from these measurements. These techniques have lead to the experimental verification of interesting local polarization effects such as transverse spin \cite{BliokhPR2015,Aiello2015,BliokhNP2015,Lodahl2017,Eismann2020}. Further, by scanning the probe, the spatial distribution of polarization can be detected, hence allowing the observation of (extended) topological features such as M\"obius bands formed by the directions of largest electric field component at all points over a loop \cite{Bauer2015}, as predicted by Freund \cite{FreundOC2010,FreundOL2010} and Dennis \cite{Dennis2011}, knotted structures \cite{Larocque2018,Maucher2019,Sugic2019} and skyrmionic distributions \cite{Donati2016,Tsesses2018,Du2019,Gutierrez-Cuevas2021,Sugic2021,Marco2022}.

Another application for which nonparaxial measures of polarization are of interest is fluorescence microscopy, where the nanoparticles in question are not elastic scatterers but fluorescent molecules that behave as sources. Therefore, rather than the particle allowing the retrieval of local information about the field, the measured emitted field reveals information about the particle \cite{Foreman2008,Aguet2009,Backlund2012,Backer2015,Zhang2018,Zhang2019,CHIDO,Backlund2018,Hulleman2021,Ding2021,Wu2022}. In particular, the measured nonparaxial polarization of the field emitted by each fluorophore provides information about its orientation and even whether it is static or ``wobbling''. This information is encoded in a $3\times3$ matrix, referred to in this context as the second moment matrix, that essentially corresponds to the polarization matrix of the emitted field, but that is usually assumed to be real due to the fact that the fluorophores typically emit as linear dipoles. Some of the techniques used in this context seek to recover simultaneously the information of the $3\times3$ correlation matrix for several molecules whose positions are also being estimated. Therefore, in order to resolve them the measurement is often performed not in the Fourier plane but in the image plane, but after some appropriate filtering within the Fourier plane is performed to encode information about the molecule's orientation (i.e. the polarization of the emitted light) in the shape of the molecules' point spread function \cite{Foreman2008,Aguet2009,Backlund2012,Backer2015,Zhang2018,CHIDO,Hulleman2021,Ding2021,Wu2022}.

The aim of the current work is to present a unified description of different theoretical aspects of the polarization of nonparaxial light as extensions to the standard treatment in the paraxial regime. Please note that Brosseau and Dogariu provided a first excellent extended review on this topic \cite{BrosseauProgOpt2006}. The emphasis of the treatment on the present article is on geometric interpretations. The goal is to summarize many recent results on this topic, supplemented by concepts that to the knowledge of the author are introduced here, in order to present a coherent description of the geometry of nonparaxial polarization. The treatment in this article avoids as much as possible relying on group-theoretical terminology, for the benefit of readers not familiar with this formalism. To contextualize the presentation, standard concepts used in the paraxial regime, such as the degree of polarization, the Stokes parameters and the Poincar\'e sphere, are summarized in Section~\ref{sec:par}. Section~\ref{sec:FP3D} is devoted to nonparaxial full polarization, particularly its representation in terms of two points over a unit sphere. The discussion of partial polarization for nonparaxial light begins in Section~\ref{sec:NPPPL}, where the $3\times3$ polarization matrix is introduced, as well as its geometric interpretation in physical space. Section~\ref{sec:DOP} presents a discussion of the several nonparaxial generalizations of the degree of polarization and their physical and geometric interpretations. Section~\ref{sec:Stokes} focuses on the generalization of the Stokes parameters for nonparaxial polarization, and the inequalities that constrain them. A representation of partial polarization as a collection of points inside a unit sphere is proposed in Section~\ref{threepointrep}. Finally, some concluding remarks and outlooks are presented in Section~\ref{sec:CR}.

\section{Summary of paraxial polarization}
\label{sec:par}
Let us start by giving a brief summary of some aspects of the theory of polarization for paraxial fields, in order to provide a context for its extension into the nonparaxial regime in the remainder of the article. More complete summaries are provided in appropriate textbooks \cite{Brosseau1998,GilOssikovski2016}. Also discussed in this section is the convention of signs and terminology that will be used.

\subsection{Monochromatic beams, full polarization, and the Poincar\'e sphere}
Consider a paraxial monochromatic beam with temporal frequency $\omega$ propagating in the positive $z$ direction in an isotropic medium of refractive index $n$:
\begin{align}
\vec{\cal E}={\rm Re}\left[{\bf E}\exp(\ui knz-\ui\omega t)\right],
\label{eq:paraxialplanewave}
\end{align}
where $k$ is the wavenumber and ${\bf E}=(E_x,E_y,E_z)^{\rm T}$ (with $^{\rm T}$ denoting a transpose) is a complex vector independent of time. When the field is a plane wave, ${\bf E}$ is constant, while for structured beams this vector is a (slowly-varying over the scale of the wavelength) function of the spatial coordinates that satisfies the paraxial wave equation. In either case, given the transversality of the electric field, only the $x$ and $y$ components of ${\bf E}$ can take significant values, and the $z$ component can be ignored. The {\it Jones vector} ${\bf E}_{\rm 2D}=(E_x,E_y)^{\rm T}$ is then defined as the two-vector in which the $z$ component is dropped. Since at any given spatial location its two components are arbitrary complex numbers, each with a real and an imaginary part, the Jones vector has four independent degrees of freedom and hence requires the specification of four real parameters. 

\begin{figure}
\centering
\includegraphics[scale=0.4]{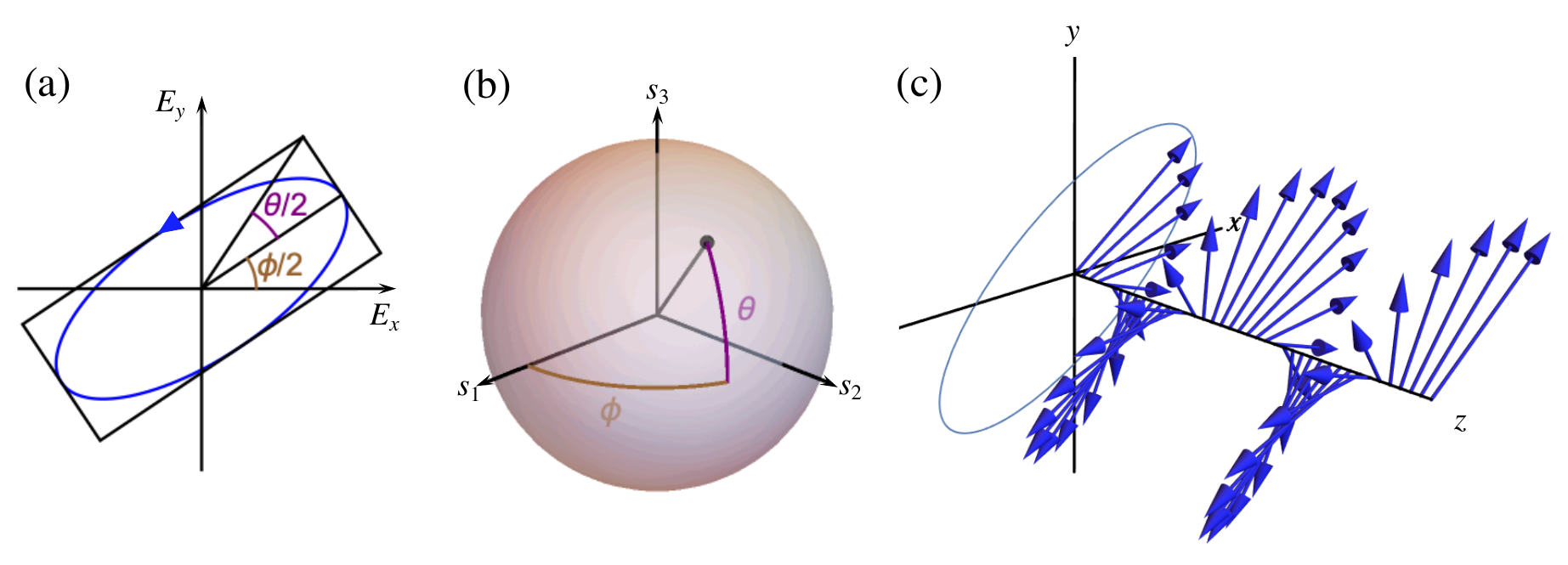}
\caption{(a) Polarization ellipse traced by the electric field at a point for monochromatic paraxial light. The major axis is at an angle $\phi/2$ from the $x$ direction, and the ellipticity is characterized by $\theta/2$, corresponding to the angle between the major axis and a corner of a rectangle that contains the ellipse and is aligned with it. The sense in which the electric field traces this ellipse is encoded in the sign of $\theta\in[-\pi/2,\pi/2]$. The case shown in the figure corresponds to counterclockwise rotation, which is referred to here as left-handed. Note that, following the right-hand rule, left-handed circulation corresponds to spin pointing in the positive $z$ direction (out of the plane). The length of the rectangle's half diagonal (from the origin to the corner) corresponds to the norm of the Jones vector, which equals $A$ if the field is not normalized and unity if it is normalized. (b) The Poincar\'e sphere as an abstract compact space for representing polarization. Each possible orientation, ellipticity and handedness of the polarization ellipse is represented by the position of a point over the sphere's surface, where $\phi$ equals longitude and $\theta$ latitude. The three Cartesian coordinates correspond to the normalized Stokes parameters defined in Section~\ref{normstokes}. (c) Instantaneous distribution of the electric field vector along the propagation axis for the same case as in (a) and (b), corresponding to a left-handed helix.}
\label{fig:fullpol2D}
\end{figure}

There are several ways to parametrize this vector, but the following one in terms of the four parameters $A,\Phi,\theta,\phi$ highlights the link to the geometry of the electric field oscillations:
\begin{align}
{\bf E}_{\rm 2D}=\left(\begin{array}{c}E_x\\E_y\end{array}\right)&=A\exp(\ui\Phi)\left(\begin{array}{cc}\cos\frac\phi2-\sin\frac\phi2\\\sin\frac\phi2+\cos\frac\phi2\end{array}\right)\left(\begin{array}{c}\cos\frac\theta2\\\ui\sin\frac\theta2\end{array}\right)\nonumber\\
&=A\exp(\ui\Phi)\left(\begin{array}{c}\cos\frac\theta2\cos\frac\phi2-\ui\sin\frac\theta2\sin\frac\phi2\\\cos\frac\theta2\sin\frac\phi2+\ui\sin\frac\theta2\cos\frac\phi2\end{array}\right).
\label{Jones}
\end{align}
The form at the end of the first line of this equation simplifies the geometric interpretation, since each of the four parameters appears alone in a different factor. For a fixed spatial position, consider the path traced over the transverse plane by the electric field, corresponding to the parametric graph of ${\rm Re}[(E_x,E_y)\exp(-\ui\omega t)]$ as a function of $t$. As shown in Fig.~\ref{fig:fullpol2D}(a), this path is an ellipse centered at the origin. The global amplitude $A$ provides the scale of the ellipse, and the global phase $\Phi$ has no influence on the global shape but only on the time at which each value of the electric field takes place. The shape and orientation of the oscillations, which is what we refer to as polarization, are determined by the remaining two parameters, $\phi$ and $\theta$. 
For $\phi=0$ and $\theta\in[-\pi/2,\pi/2]$, the electric field traces an ellipse whose major semi-axes are aligned with the $x$ direction and have magnitude $A\cos\theta/2$, while the minor semi-axes are in the $y$ direction and have magnitude $A|\sin\theta/2|$. The ellipticity is controlled by $\theta$, so that the ellipse goes then from a line for $\theta=0$ to a circle for $\theta=\pm\pi/2$. The handedness of the circulation of the electric field around this ellipse depends on the sign of $\theta$. 
Different authors adopt different conventions, but here we use the convention in which ``left-handed'' oscillations correspond to $\theta>0$, while ``right-handed'' ones correspond to $\theta<0$. The reason for this choice is that, if we fix time and consider the path traced by the electric field as a function of propagation distance $z$ following Eq.~(\ref{eq:paraxialplanewave}), this path is a helix (with elliptic projection) with the corresponding handedness, as shown in Fig.~\ref{fig:fullpol2D}(c).

The factor including $\phi$ within the first line of Eq.~(\ref{Jones}) is simply a rotation matrix, and therefore for $\phi\neq0$ the ellipse is rotated by $\phi/2$. As shown in Fig.~\ref{fig:fullpol2D}(a), $\phi$ is one half of the angle between the $x$ direction and the major axis of the ellipse, and $\theta$ is a measure of the ellipticity given by the angle (bisected by the major axis) between two corners of a rectangle boxing the ellipse. 
Note that, from the point of view of polarization, $\phi$ is periodic with period $2\pi$, while $\theta$ is constrained to the interval $[-\pi/2,\pi/2]$, with $\phi$ becoming irrelevant for the extreme values of $\theta=\pm\pi/2$. This makes $\phi$ and $\theta$ similar to the longitude and latitude spherical angles, respectively. Polarization can then be represented as a coordinate over the surface of a unit sphere, known as the \emph{Poincar\'e sphere}, shown in Fig.~\ref{fig:fullpol2D}(b), where the two poles correspond to the two circular polarizations, with left-handed circular at the north pole and right-handed circular at the south pole. Points along the equator correspond to linear polarizations with different orientations, and the rest of the sphere's surface corresponds to elliptical shapes with different ellipticity, handedness and orientation. The convention adopted here of placing left-handed (rather than right-handed) polarization at the northern hemisphere is perhaps not the most common. However, it is a direct consequence of naming the handedness of the polarization according to the handedness of the polarization helix in space (as described in Fig.~\ref{fig:fullpol2D}(c)), and making the sign of the vertical coordinate of the Poincar\'e sphere the same as that of the spin density of the field along the positive $z$ direction according to the right-hand rule.  

This article focuses on the electric field, whose interaction with detectors is typically dominant. Notice, however, that for paraxial light where there is a well-defined wavevector pointing in the $z$ direction, Maxwell's equations dictate that the Jones vector for the magnetic field (and hence the ellipse this field traces) is identical to that for the electric field except for a rotation by $\pi/2$ around the propagation axis, and for a factor of the inverse of the speed of light. 

\subsection{$2\times2$ polarization matrix and Stokes parameters}
When the field is not purely monochromatic, the shape traced by the electric field is in general not periodic and is much more complex than an ellipse. However, the details of the oscillation are typically over a time scale that is inaccessible to detectors, and what can be measured are averages over the detector's integration time of quantities that have a quadratic dependence in the field. If we make the assumption that the fields are statistically stationary (namely that the measured averages are independent of when the measurement is made), 
the field's polarization at a given point is well described by the $2\times2$ autocorrelation matrix of the field components, known as the polarization (or coherency) matrix:
\begin{align}
\mathbf{\Gamma}_{\rm 2D}=\langle{\bf E}_{\rm 2D}{\bf E}_{\rm 2D}^\dagger\rangle=\left(\begin{array}{cc}\langle E^*_xE_x\rangle&\langle E^*_yE_x\rangle\\\langle E^*_xE_y\rangle&\langle E^*_yE_y\rangle\end{array}\right),
\end{align}
where $^\dagger$ denotes a transpose conjugate and 
$\langle\cdot\rangle$ denotes an average. 
Because this $2\times2$ matrix is explicitly Hermitian, it contains four degrees of freedom (e.g. the two real diagonal elements and the real and imaginary parts of one of the off-diagonal elements). One choice for these four parameters, associated with simple combinations of measurable quantities, was proposed by Gabriel Stokes in 1852. Again, different conventions exist, but here these \emph{Stokes parameters} are defined as
\begin{align}
S_0&={\rm Tr}(\mathbf{\Gamma}_{\rm 2D})=\langle|{\bf E}_{\rm 2D}|^2\rangle=\langle|E_x|^2\rangle+\langle|E_y|^2\rangle,\\
S_1&=\langle|E_x|^2\rangle-\langle|E_y|^2\rangle,\\
S_2&=2\,{\rm Re}\langle E_x^*E_y\rangle=\langle|E_{\rm p}|^2\rangle-\langle|E_{\rm m}|^2\rangle,\\
S_3&=2\,{\rm Im}\langle E_x^*E_y\rangle=\langle|E_{\rm l}|^2\rangle-\langle|E_{\rm r}|^2\rangle,
\end{align}
where $E_{\rm p,m}=(E_x\pm E_y)/\sqrt{2}$ are the field components in a Cartesian frame rotated by $\pi/4$ with respect to the $x,y$ axes, and $E_{\rm r,l}=(E_x\pm\ui E_y)/\sqrt{2}$ are the right/left circular components. The quantities $\langle|E_i|^2\rangle$, for $i=x,y,{\rm p},{\rm m},{\rm r},{\rm l}$, are directly measurable through the appropriate use of polarizers and quarter-wave plates prior to the detector \cite{Brosseau1998,GilOssikovski2016}. Written in terms of the Stokes parameters, the polarization matrix becomes
\begin{align}
\mathbf{\Gamma}_{\rm 2D}=\frac12\left(\begin{array}{cc}S_0+S_1&S_2-\ui S_3\\S_2+\ui S_3&S_0-S_1\end{array}\right).\label{Gamma2D}
\end{align}

Surprisingly, the four Stokes parameters correspond to the coefficients of the decomposition of the polarization matrix into a complete orthonormal basis of $2\times2$ Hermitian matrices known as the Pauli matrices, proposed by Wolfgang Pauli in 1927 (three quarters of a century after the Stokes parameters) for the quantum study of electrons \cite{Pauli1927}. 
Note that each Pauli matrix can be recovered from Eq.~(\ref{Gamma2D}) by setting to 2 (to remove the prefactor of $1/2$) the corresponding parameter $S_n$ while setting to zero the remaining parameters. The Pauli matrices were proposed in a different physical context to that of polarization, and hence a different labeling scheme is often used. Here we use a labeling scheme and sign convention consistent with the convention for the Stokes parameters. More details on the Pauli matrices can be found in Appendix~\ref{appA}.

\subsection{Ellipse of inertia and spin}
\begin{figure}
\centering
\includegraphics[scale=0.4]{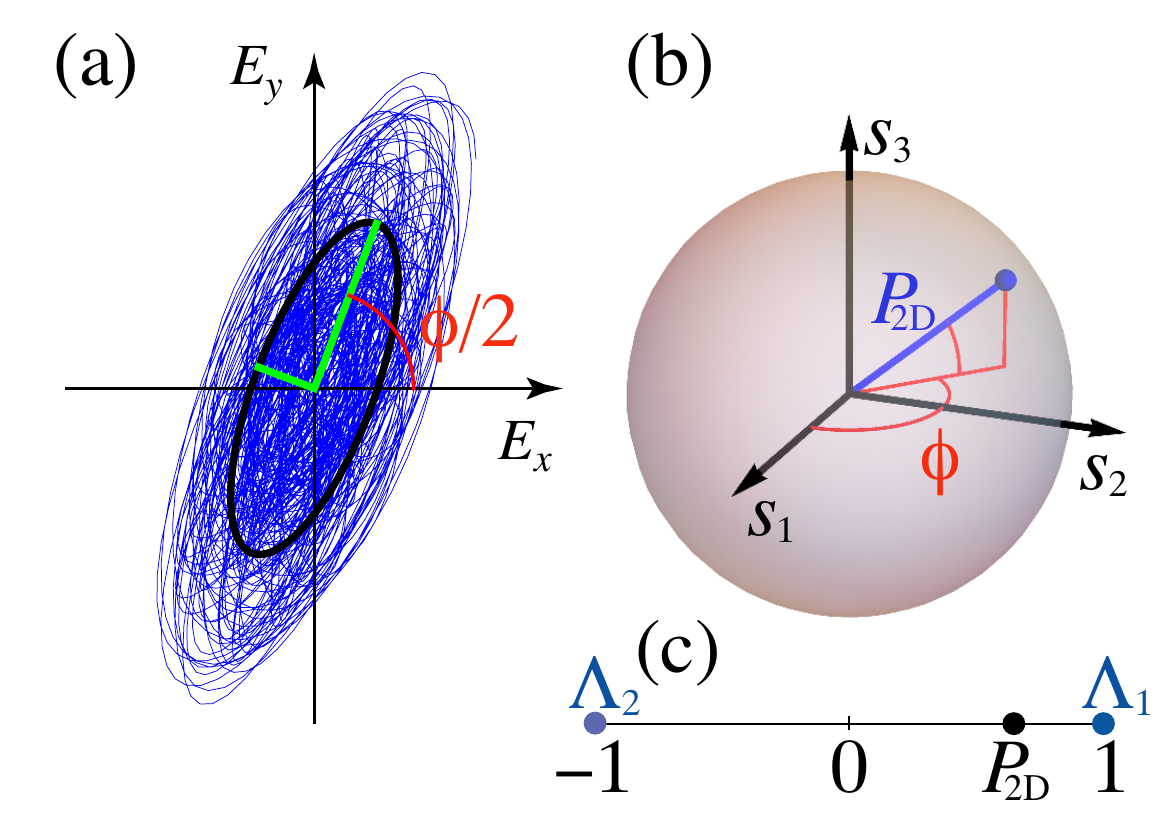}
\caption{(a) Path traced by the electric field (blue) over some time. The black ellipse denotes the ellipse of inertia, which denotes a region where the field is constrained a significant part of the time. Its semi-axes (green) are given by the square roots of the eigenvalues of ${\rm Re}(\mathbf{\Gamma})$. (b) The state of polarization is described by a point $\vec s_{\rm 2D}$ constrained inside the unit Poincar\'e sphere. The distance $P_{\rm 2D}$ of this point from the origin corresponds to the degree of polarization. (c) Interpretation of $P_{\rm 2D}$ as the coordinate (or distance from the origin) of the center of mass of two point masses at $\pm1$ of magnitudes $\Lambda_{1,2}$.}
\label{fig:2D}
\end{figure}

As mentioned earlier, for monochromatic light the electric field vector at a given point traces repeatedly over time an ellipse, following the equation ${\rm Re}[(E_x,E_y)\exp(-\ui\omega t)]$. Such a field is therefore said to be fully polarized. On the other hand, when light is  not strictly monochromatic (nor fully polarized), 
the electric field vector traces 
a more complicated oscillation that, over a sufficiently long time, explores a region within the plane of (real) field components, according to a probability density with an elliptical cross-section, as shown in Fig.~\ref{fig:2D}(a). The orientation of this cross-section, the {\it ellipse of inertia}, is given by the eigenvectors of the real part of the matrix in Eq.~(\ref{Gamma2D}) (that is, with $S_3$ being ignored), and its semi-axes correspond to the square roots of the corresponding eigenvalues. (If the probability density is Gaussian, this elliptical cross-section corresponds to the contour at which the probability drops by ${\rm e}^{-1/2}$, or equivalently, it encloses a region which the field occupies $1-{\rm e}^{-1/2}\approx40\%$ or the time.) The area enclosed by the ellipse of inertia equals 
\begin{align}
\pi\sqrt{\det[{\rm Re}(\mathbf{\Gamma}_{\rm 2D})]}=\frac{\pi}2\sqrt{S_0^2-S_1^2-S_2^2}.\label{area}
\end{align}
While this ellipse describes the average shape traced by the field, it does not distinguish whether the oscillation involves more rotations in the left-handed or the right-handed sense. This is precisely the role of the parameter $S_3$, which then complements this simple second-order statistical/geometrical description of the oscillations. The matrix $\mathbf{\Gamma}_{\rm 2D}$ is not only explicitly Hermitian but also non-negative-definite, and therefore its determinant must be non-negative, which straightforwardly gives the condition $S_0^2\ge S_1^2+S_2^2+S_3^2$. This constraint, combined with the expression in Eq.~(\ref{area}) implies that $|S_3|$ is constrained to be at most equal to $2/\pi$ times the area enclosed by the ellipse of inertia.

\subsection{Normalized Stokes parameters, degree of polarization and the Poincar\'e sphere's radial coordinate}
\label{normstokes}
The Stokes parameter $S_0$ describes the intensity of the field and not its polarization (namely the shape of the elliptical profile just described and the dominance of one handedness over the other). It is then useful for the purpose of characterizing polarization to define the three {\it normalized Stokes parameters} $s_n=S_n/S_0$ for $n=1,2,3$. Given the relation $S_0^2\ge S_1^2+S_2^2+S_3^2$, the normalized Stokes vector $\vec s_{\rm 2D}=(s_1,s_2,s_3)$ is constrained to the interior and surface of a unit sphere (i.e. a unit 2-ball). It is easy to show that, for a monochromatic field with Jones vector as given in Eq.~(\ref{Jones}), this vector gives $\vec s_{\rm 2D}=(\sin\theta\cos\phi,\sin\theta\sin\phi,\cos\theta)$, so that this sphere is precisely the Poincar\'e sphere mentioned earlier. For a general polarization matrix, the relation $|\vec s_{\rm 2D}|\le1$ indicates that the whole interior of the sphere is inhabitable, as illustrated in Fig.~\ref{fig:2D}(b), and the sphere's surface (a 2D manifold) separates the accessible and inaccessible regions, and corresponds to fully polarized fields, for which indeed only two parameters are needed. 
Partial polarization makes it necessary to introduce a third parameter, the magnitude of $\vec s_{\rm 2D}$, which is a measure of how polarized the field is at the location in question. This radial coordinate in the Poincar\'e sphere is referred to as the \emph{degree of polarization}, which can be written in the equivalent forms:
\begin{align}
P_{\rm 2D}=|\vec s_{\rm 2D}|=\sqrt{\sum_{m}s_{m}^2}=\sqrt{\frac{2\,{\rm Tr}\mathbf{\Gamma}_{\rm 2D}^2}{({\rm Tr}\mathbf{\Gamma}_{\rm 2D})^2}-1}=\sqrt{1-4\frac{\det\mathbf{\Gamma}_{\rm 2D}}{({\rm Tr}\mathbf{\Gamma}_{\rm 2D})^2}}.\label{DoP2D}
\end{align}
The equivalence of the last two forms to the first can be easily verified by substituting into them the form of the polarization matrix in Eq.~(\ref{Gamma2D}). 

Because the polarization matrix is Hermitian and non-negative definite, it has two normalized eigenvectors ${\bf e}_i$ with corresponding real, non-negative eigenvalues $\Lambda_i>0$ such that $\mathbf{\Gamma}_{\rm 2D}\cdot{\bf e}_i=\Lambda_n{\bf e}_i$, for $i=1,2$. Without loss of generality, we can order these so that $\Lambda_1\ge\Lambda_2$. Note that the polarization matrix can also be written in terms of these quantities as
\begin{align}
\mathbf{\Gamma}_{\rm 2D}&=\Lambda_1\,{\bf e}_1{\bf e}_1^\dagger+\Lambda_2\,{\bf e}_2{\bf e}_2^\dagger
=(\Lambda_1-\Lambda_2)\,{\bf e}_1{\bf e}_1^\dagger+\Lambda_2\,\left(\begin{array}{cc}1&0\\0&1\end{array}\right)
,\label{polandnotpol}
\end{align}
where in the last step we used the fact that the eigenvectors ${\bf e}_i$ form an orthonormal basis, and therefore ${\bf e}_1{\bf e}_1^\dagger+{\bf e}_2{\bf e}_2^\dagger$ equals the $2\times2$ identity. The first term in this expression, factorizable as an outer product of a vector with its complex conjugate, can be interpreted as the ``polarized part of the field'' because alone it would have a degree of polarization of unity. The second term, proportional to the identity matrix, can be interpreted instead as the ``unpolarized part of the field'', because on its own it would have a degree of polarization of zero. It is trivial to see that the degree of polarization of the complete matrix can be written in terms of the two eigenvalues $\Lambda_1\ge\Lambda_2\ge0$ of the polarization matrix as
\begin{align}
P_{\rm 2D}=\frac{\Lambda_1-\Lambda_2}{\Lambda_1+\Lambda_2}.\label{DoP2Devals}
\end{align}
This means that the degree of polarization can be given a physical interpretation as the fraction of the optical power that is fully polarized (the total power being proportional to $\Lambda_1+\Lambda_2$). 

Equation~(\ref{DoP2Devals}) allows also a simple geometric picture \cite{CoM} for the degree of polarization, illustrated in Fig.~\ref{fig:2D}(c): consider two point masses along a line, at unit distances from the origin. Let the magnitude of the mass at $+1$ be $\Lambda_1$ and that at $-1$ be $\Lambda_2$. The coordinate for the center of mass, that is, its distance to the origin, is then precisely $P_{\rm 2D}$. The conceptual value of this simple picture for the degree of polarization will become apparent in the discussion of nonparaxial polarization.


\subsection{Some properties of the Poincar\'e sphere as a space for polarization}
\label{properties}

Let us finish this review of 2D polarization by listing a few of the properties of the Poincar\'e sphere as a suitable abstract space for the description of paraxial polarization, as well as some considerations.

\subsubsection{Unitary transformations}
\begin{figure}
\centering
\includegraphics[scale=0.4]{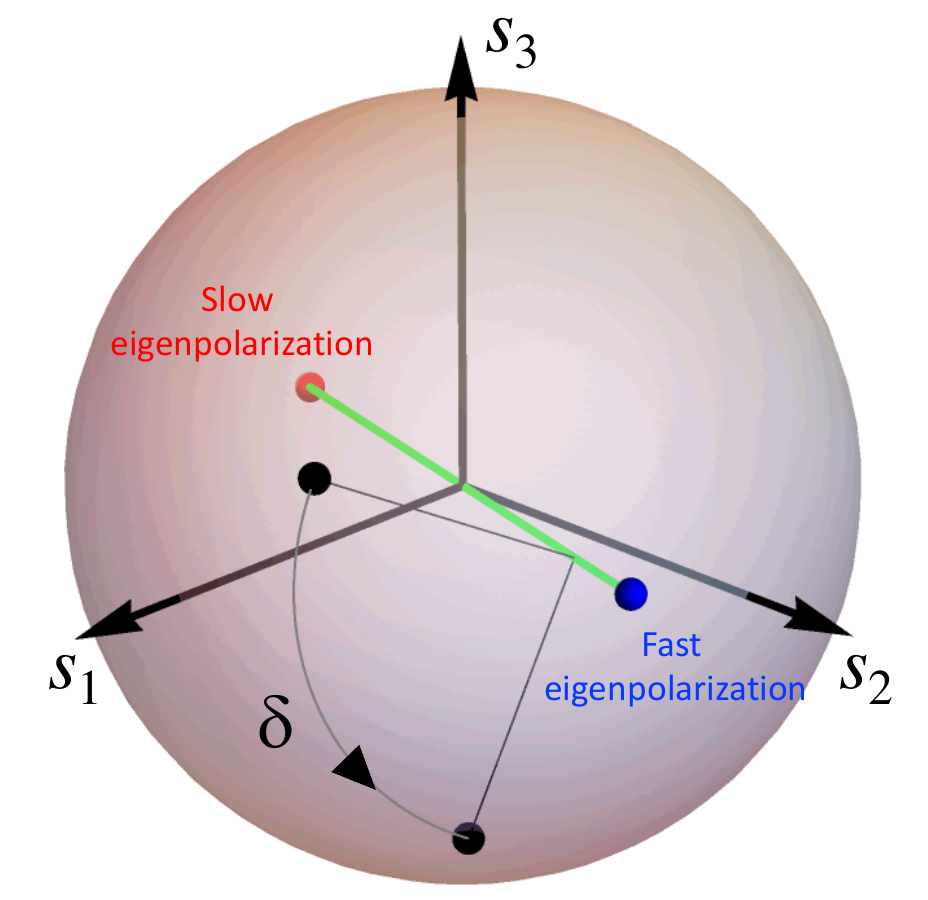}
\caption{A unitary transformation on the Jones vector corresponds to the effect the beam passing through an optical element consisting of one or a succession of transparent birefringent materials. These optical elements have two orthogonal eigenpolarizations, which are the two states of full polarization that emerge from the element unchanged except for a phase factor. A general polarization state can be decomposed into these two eigenpolarizations, and after traversing the element, one of them, referred to as the slow eigenpolarization, is dephased by $\delta$ with respect to the other, so-called fast eigenpolarization. Over the Poincar\'e sphere representation, this effect corresponds to a rigid rotation of $\vec s_{\rm 2D}$ (which can be either on the surface or the interior of the sphere) by an angle $\delta$ around the axis corresponding to the line joining the points representing the two eigenpolarizations over the Poincar\'e sphere.}
\label{fig:rotations}
\end{figure}
The fact that the inhabitable region in the abstract space $\vec s_{\rm 2D}=(s_1,s_2,s_3)$ is a sphere reveals the natural symmetries 
inherent to paraxial polarization. Lossless polarization transformations performed by transparent birefringent or optically active materials, for which a phase difference is applied to two orthogonal polarization components of the field without the loss or gain of light, correspond to unitary transformations acting on the Jones vector. As illustrated in Fig.~\ref{fig:rotations}, these transformations translate simply into rigid rotations of $\vec s_{\rm 2D}$, and hence preserve the degree of polarization and the shape of the parameter space.

\subsubsection{Geometric phase} 
The Poincar\'e sphere provides a beautiful geometric interpretation for the phenomenon known as the Pancharatnam-Berry geometric phase \cite{Pancharatnam1956,Berry1984,Berry1987,Bliokh2019}, which is the accumulation of an extra phase by a beam following a sequence of transformations of polarization that correspond to a closed path over the Poincar\'e sphere. When each segment of this path obeys what is referred to as {\it parallel transport}, the geometric phase equals one half of the enclosed solid angle over the Poincar\'e sphere. In paraxial optical systems, parallel transport is guaranteed when the path is piecewise geodesic, such as for changes of polarization enacted by polarizers, or by wave retarders in which the input and output polarizations are at $90^\circ$ from the eigenpolarizations over the Poincar\'e sphere. Even when the transformations do not obey parallel transport, the Poincar\'e sphere construction allows geometric interpretations for the resulting phases \cite{Courtial1999,Kurzynowski2011,Vella2018}. 

\subsubsection{Meaning of the latitude angle for partially polarized light}
For fully polarized light, the angular variables $\theta$ and   $ \phi$, which describe the orientation and ellipticity of the polarization ellipse, correspond to spherical coordinates (latitude and longitude, respectively) over the Poincar\'e sphere. On the other hand, the normalized Stokes parameters, which are simple linear combinations of measurement results (up to a normalization) correspond to the Cartesian coordinates in the Poincar\'e space. For partially polarized light, corresponding to the interior of the sphere, latitude and longitude are then supplemented by a radial coordinate, given by the degree of polarization $P_{\rm 2D}$. Notice, though, that in this case one cannot talk of a polarization ellipse, and perhaps the closest concept to it is the ellipse of inertia. While the orientation of the major axis of the ellipse of inertia is still given by $\phi/2$ as shown in Fig.~\ref{fig:2D}(a), its ellipticity depends not only on $\theta$ but also on $P_{\rm 2D}$: the ratio between the minor and major semi-axes of the ellipse of inertia is not equal to $\tan\theta/2$ (as for the polarization ellipse of a fully polarized field), but to $\tan\vartheta/2$, where $\vartheta$ is defined such that $\cos\vartheta=P_{\rm 2D}\cos\theta$.  
 
\subsubsection{Relation between two polarizations}
\label{sec:similarity}
\begin{figure}
\centering
\includegraphics[scale=0.3]{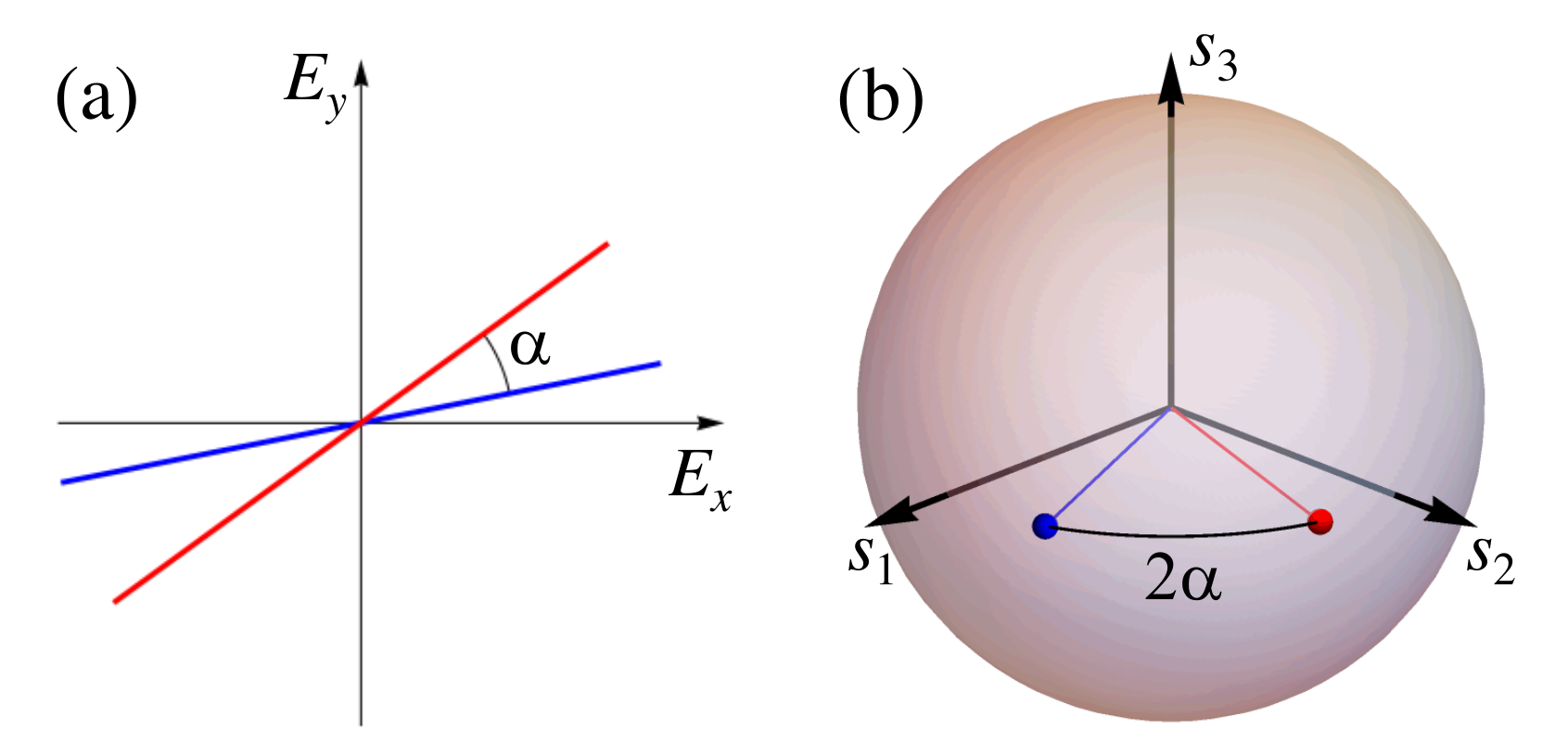}
\caption{(a) Angle $\alpha$ between two linearly polarized fields. (b) The corresponding angle between the two points representing them over the Poincar\'e sphere is $2\alpha$.}
\label{fig:angles}
\end{figure}

Let us now consider how the relation between two polarizations translates into the geometrical relation between their corresponding points in the Poincar\'e space. 
Consider two fully polarized fields with Jones vectors ${\bf E}_{\rm 2D}^{\rm (I)}$ and ${\bf E}_{\rm 2D}^{\rm (II)}$. The similarity of their polarizations can be characterized by the angle $\alpha$ defined as
\begin{align}
\cos^2\alpha=\frac{|{\bf E}_{\rm 2D}^{\rm (I)*}\cdot{\bf E}_{\rm 2D}^{\rm (II)}|^2}{|{\bf E}_{\rm 2D}^{\rm (I)}|^2|{\bf E}_{\rm 2D}^{\rm (II)}|^2}.
\label{alphasim}
\end{align}
If the two fields are mutually proportional, the right-hand side of this equation is unity and $\alpha$ can be chosen as zero. On the other hand, when the fields are orthonormal, the right-hand side of the equation vanishes, so $\alpha$ can be chosen as $\pi/2$. If both fields happen to be linearly polarized, $\alpha$ corresponds to the angle between them, as shown in Fig.~\ref{fig:angles}(a).
This expression can be written in terms of the polarization matrices of both fields, $\mathbf{\Gamma}_{\rm 2D}^{\rm (I)}$ and $\mathbf{\Gamma}_{\rm 2D}^{\rm (II)}$, as
\begin{align}
\cos^2\alpha=\frac{{\rm Tr}\left[\mathbf{\Gamma}_{\rm 2D}^{\rm (I)}\mathbf{\Gamma}_{\rm 2D}^{\rm (II)}\right]}{{\rm Tr}\left[\mathbf{\Gamma}_{\rm 2D}^{\rm (I)}\right]{\rm Tr}\left[\mathbf{\Gamma}_{\rm 2D}^{\rm (II)}\right]}. 
\end{align}
By now writing each of the polarization matrices in terms of its Stokes parameters as in Eq.~(\ref{Gamma2D}) we arrive at the expression
\begin{align}
\cos^2\alpha=\frac{\sum_{n=0}^3S_n^{\rm (I)}S_n^{\rm (II)}}{2\,S_0^{\rm (I)}S_0^{\rm (II)}}=\frac{1+\vec s^{\rm (I)}\cdot\vec s^{\rm (II)}}2.
\end{align}
By using the property $\cos^2\alpha=(1+\cos2\alpha)/2$ we can simplify this expression to
\begin{align}
\cos2\alpha=\vec s^{\rm (I)}\cdot\vec s^{\rm (II)}.
\end{align}
In the case in which both fields are fully polarized and therefore $\vec s^{\rm (I)}$ and $\vec s^{\rm (II)}$ are unit vectors, the right-hand side of this equation is simply the cosine of the angle between them, which then equals $2\alpha$. That is, the angle in the Poincar\'e space between the normalized Stokes vectors is twice the angle $\alpha$ characterizing the similarity between two Jones vectors. In particular, for any pair of orthogonal states, $\alpha=\pi/2$ and therefore the corresponding normalized Stokes vectors are antiparallel, corresponding to antipodal points over the Poincar\'e sphere.

\subsubsection{Statistical properties for random fields}
\begin{figure}
\centering
\includegraphics[scale=0.35]{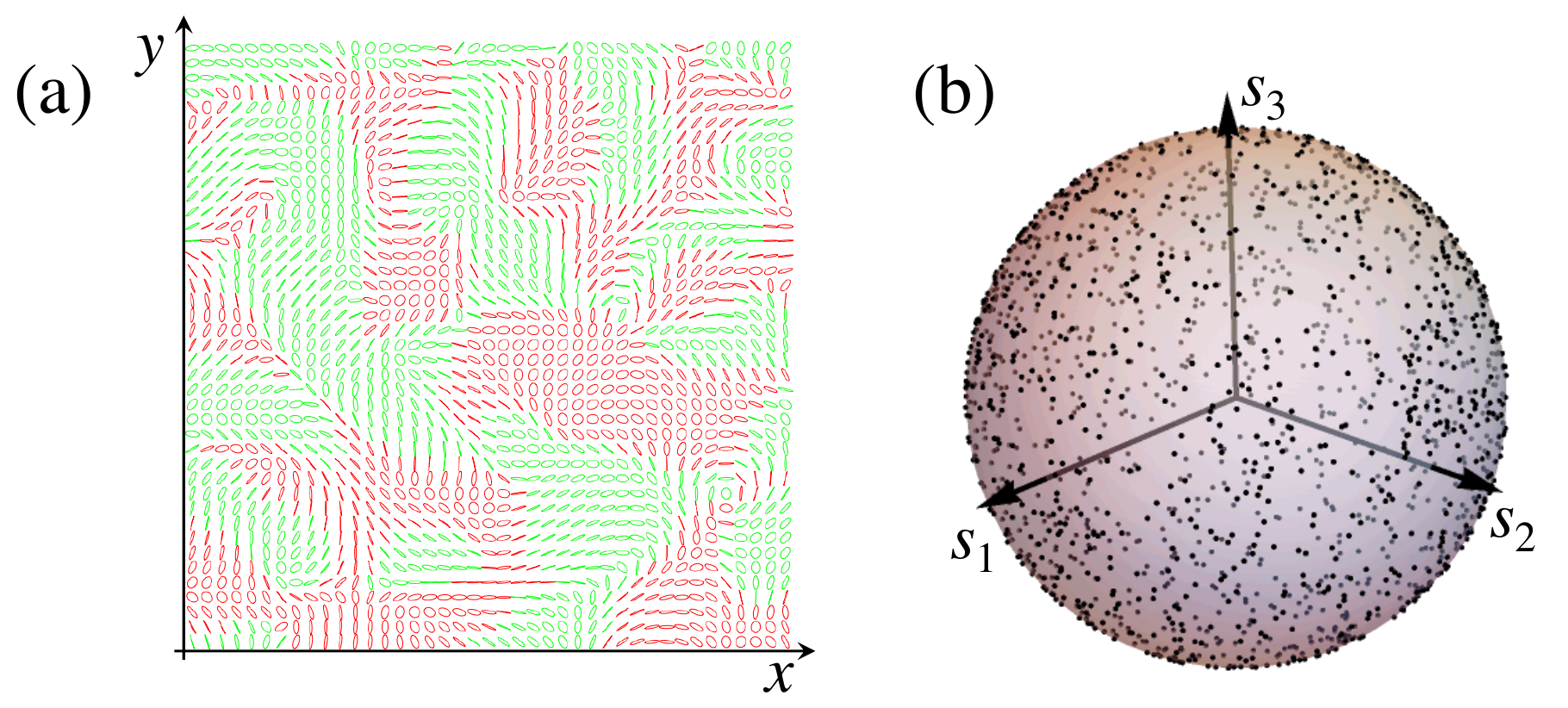}
\caption{(a) Distribution of polarization over a sample of points for a paraxial field composed of a random superposition of paraxial plane waves, where green represents left-handed ellipses and red represents right-handed ones. (b) Representation over the PoincarŽ sphere of each polarization ellipse shown in (a). These points are statistically uniformly distributed over the sphere.}
\label{fig:speckle}
\end{figure}
Finally, we consider the statistics of the polarization state for a field composed of a random superposition of a large number of monochromatic paraxial plane waves whose relative phases and polarizations are uncorrelated. Since we assume all waves are monochromatic and of the same frequency, the superposition still yields a monochromatic, fully polarized field. This field, however, varies spatially and corresponds to a speckle pattern, where the intensity, phase and polarization change from point to point. An example of a polarization distribution of one such field over the $xy$ plane is shown in Fig.~\ref{fig:speckle}(a). It is shown in Appendix~\ref{appB} that for such a field, the statistical coverage of the Poincar\'e sphere is uniform. That is, any two subsets of polarizations, corresponding to two patches over the Poincar\'e sphere subtending equal solid angles, are equally probable. This is illustrated in Fig.~\ref{fig:speckle}(b) where the points over the Poincar\'e sphere corresponding to the ellipses shown in Fig.~\ref{fig:speckle}(a) are seen to be uniformly distributed. This would not be the case if polarization were parametrized over some other abstract space that is not a sphere.

\section{Nonparaxial fully polarized fields}
\label{sec:FP3D}
We now begin the discussion of fields that do not propagate in a preferential direction. We start by considering in this section perfectly monochromatic, and hence fully polarized, fields. The electric field at a point must then be described by a three-component complex vector ${\bf E}=(E_x,E_y,E_z)^{\rm T}$. This complex field is independent of time, but it is a function of position that is ruled by the time-harmonic version of Maxwell's equations. As mentioned in the introduction, the spatial distribution of polarization can be topologically very rich, but this is not the main subject of this work; here we focus on the description of polarization at each point. This complex field ${\bf E}$ is then treated as a constant. 


\subsection{Polarization ellipse and spin density}
\begin{figure}
\centering
\includegraphics[scale=0.3]{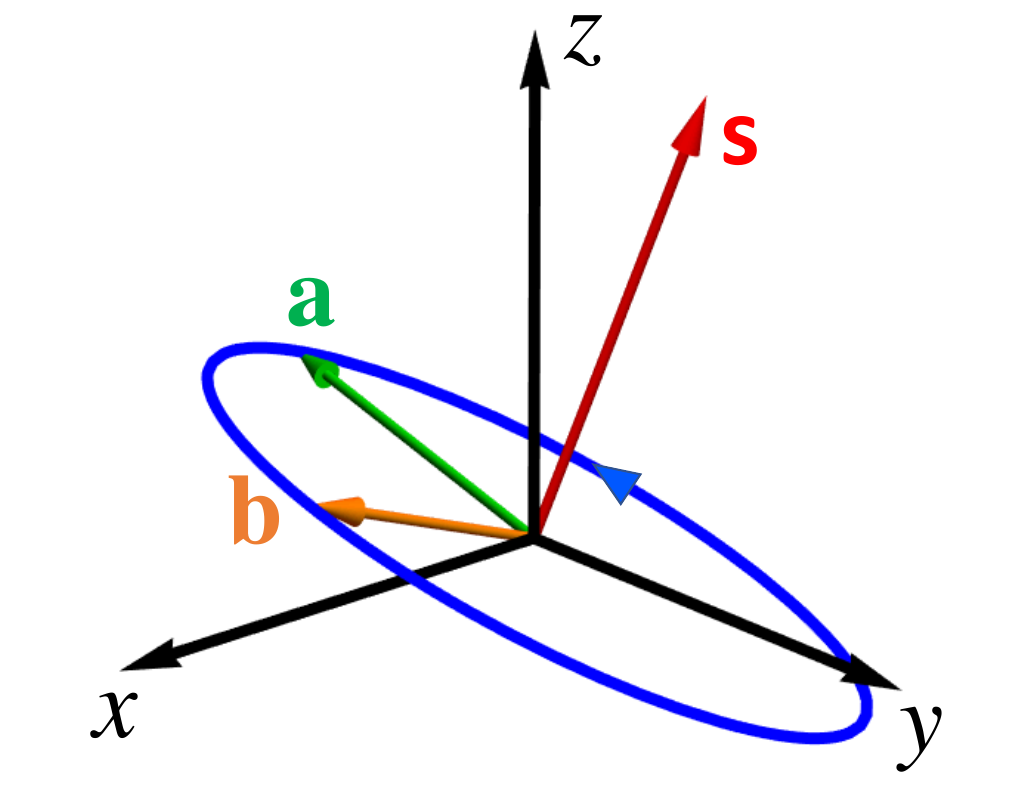}
\caption{Normalized nonparaxial polarization ellipse, with major semi-axis ${\bf a}$, minor semi-axis ${\bf b}$, and normalized spin density $\pmb{\mathsf s}$. These three vectors form a right-handed orthogonal set. Note that reversing the sign of both ${\bf a}$ and ${\bf b}$ does not change the polarization ellipse nor the normalized spin density.}
\label{fig:3Dellipse}
\end{figure}

The real field as a function of time is calculated from the complex field ${\bf E}$ according to 
\begin{align}
\vec {\cal E}(t)={\rm Re}[{\bf E}\exp(-\ui\omega t)].
\label{eq:RealE3D}
\end{align}
It is easy to see that, like in the paraxial case, this field traces an ellipse, although now the ellipse is not necessarily constrained to the $xy$ plane. To see this, we can adopt the notation \cite{Berry2001} of writing the complex field as
\begin{align}
{\bf E}=A\exp(\ui\Phi)({\bf a}+\ui{\bf b}),
\label{eq:ab}\end{align}
where $A=|{\bf E}|$ is the global amplitude, $\Phi$ is a global phase, and the dimensionless vectors ${\bf a}$ and ${\bf b}$ are chosen to be purely real, mutually orthogonal (${\bf a}\cdot{\bf \ b}=0$) with $|{\bf a}|\ge|{\bf b}|$, and satisfying the normalization condition $|{\bf a}|^2+|{\bf b}|^2=1$. This decomposition is unique up to a global sign for ${\bf a}$ and ${\bf b}$, except in the degenerate case in which both vectors have the same magnitude. 
By substituting the form in Eq.~(\ref{eq:ab}) into Eq.~(\ref{eq:RealE3D}), we get
\begin{align}
\vec {\cal E}(t)=A\,[{\bf a}\cos(\omega t-\Phi)+{\bf b}\sin(\omega t-\Phi)].
\label{eq:3Dellipse}
\end{align}
This is the parametric equation for an ellipse whose major and minor axes are aligned with ${\bf a}$ and ${\bf b}$, respectively. This polarization ellipse can have an arbitrary orientation in 3D, as shown in Fig.~\ref{fig:3Dellipse}. Unlike in the paraxial case, the plane of oscillation is not linked to a main direction of propagation. 

Determining $A$, $\Phi$, ${\bf a}$ and ${\bf b}$ from the complex field ${\bf E}$ is relatively simple. As mentioned earlier, $A$ is simply the norm of the complex field:
\begin{subequations}
\label{eqs:Eparts}
\begin{align}
A=|{\bf E}|=\sqrt{{\bf E}^*\cdot{\bf E}}\ge0.
\end{align} 
Let us now define de scalar complex quadratic field \cite{Berry2001} as ${\bf E}\cdot{\bf E}=A^2\exp(2\ui\Phi)(|{\bf a}|^2-|{\bf b}|^2)$, where we used Eq.~(\ref{eq:ab}) in the last step. Since the factors $A^2$ and $(|{\bf a}|^2-|{\bf b}|^2)$ are real and non-negative, we can find the global phase to within an integer multiple of $\pi$ as
\begin{align}
\Phi=\frac12{\rm Arg}({\bf E}\cdot{\bf E}).
\end{align}
This expression becomes indeterminate when ${\bf E}\cdot{\bf E}=0$, namely when $|{\bf a}|^2=|{\bf b}|^2$, which corresponds to a circular polarization ellipse. The points where this is true are referred to as {\it c-points} \cite{Berry2001} and they are regarded as singularities of $\Phi$. In general, because the condition ${\bf E}\cdot{\bf E}=0$ constitutes two constraints (both the real and imaginary parts must vanish), c-points in 3D space form curves, known as {\it c-lines}. For any point that is not a c-point, the global phase $\Phi$ is well determined modulo $\pi$. With this, we can define the normalized field ${\bf f}$ as
\begin{align}
{\bf f}=\frac{{\bf E}}{A}\exp(-\ui\Phi),
\end{align}
so that
\begin{align}
{\bf a}={\rm Re}({\bf f}),\,\,\,{\bf b}={\rm Im}({\bf f}).
\end{align}
\end{subequations}
Notice that the fact that $\Phi$ is only defined modulo $\pi$ is consistent with the fact that ${\bf a}$ and ${\bf b}$ are defined to within a global sign.

A quantity that will be used in what follows is the {\it spin density}, which for a fully polarized field is defined as
\begin{align}
\pmb{\mathsf S}={\rm Im}({\bf E}^*\times{\bf E})=2A^2{\bf a}\times{\bf b}
\end{align}
We also define the {\it normalized spin density}, which equals $\pmb{\mathsf S}$ divided by the intensity,
\begin{align}
\pmb{\mathsf s}=\frac{{\rm Im}({\bf E}^*\times{\bf E})}{|{\bf E}|^2}=2\,{\bf a}\times{\bf b}.
\end{align}
These quantities represent vectors that point perpendicularly to the plane containing the ellipse, following the right-hand rule. Their length is proportional to the area enclosed by the polarization ellipse. The normalized spin density $\pmb{\mathsf s}$ takes a maximum value of unity when the ellipse is a circle, and vanishes when the ellipse is a line. In the paraxial limit in which the ellipse is constrained to the $xy$ plane, the $x$ and $y$ components of these vectors vanish, and their $z$ component equal the third Stokes parameter and its normalized version, respectively, that is ${\mathsf S}_z=S_3$ and ${\mathsf s}_z=s_3$.

\subsection{Geometric representations using two points over a unit sphere}
The complex field vector has three complex components and hence involves six degrees of freedom. However, two of these degrees of freedom can be made to correspond to a global amplitude which determines the magnitude and not the shape of the polarization ellipse, and a global phase which has no effect on the shape of the ellipse. This means that the shape and orientation of the ellipse involves four degrees of freedom. There are many ways of choosing these four quantities. For example, one could specify the three components of the major axis vector ${\bf a}$, which would fix the magnitude of ${\bf b}$ and constrain it to a plane; the fourth degree of freedom would then be the orientation of ${\bf b}$ within this plane. One could instead start by specifying the three components of ${\bf b}$, or of the normalized spin density $\pmb{\mathsf s}$ and then provide an orientation angle for one of the other two vectors. 

The goal of this section is to describe geometric representations in which the four degrees of freedom of a nonparaxial polarization ellipse are given in a more democratic way, inheriting some of the desirable properties that the Poincar\'e sphere representation has for paraxial light. It is then tempting to think that points over the surface of a unit sphere are a good option and, since in this case four degrees of freedom are involved, two points will be required rather than just one. A set of constructions of this type are now described, each having its own desirable properties. For all these representations, the three coordinate axes of the ambient space where the sphere is defined are simply the three Cartesian directions of physical space, instead of abstract quantities such as the Stokes parameters for the Poincar\'e sphere.
 
\subsubsection{Hannay-Majorana construction}
\begin{figure}
\centering
\includegraphics[scale=0.3]{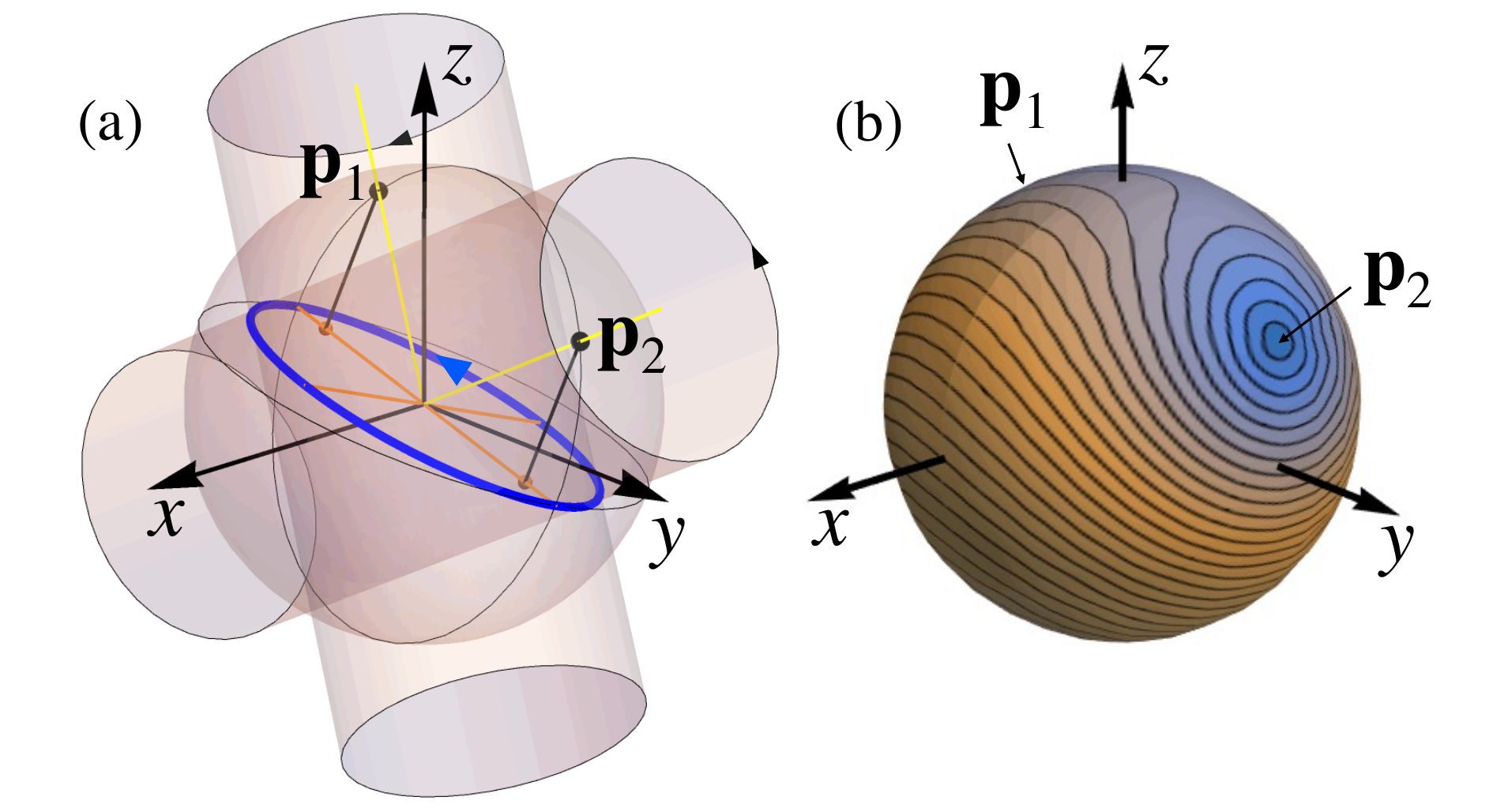}
\caption{(a) The two Hannay-Majorana points ${\bf p}_{1,2}$ are contained within the great circle corresponding to the intersection of the sphere with the plane that contains both the ellipse's major axis and the normal to the ellipse. They correspond to the two directions in which the projection of the polarization ellipse is a (right-handed) circle, as illustrated by the fact that this ellipse coincides with the intersection of two cylinders whose axes are the (yellow) lines joining the origin and each ${\bf p}_{1,2}$. Another interpretation for ${\bf p}_{1,2}$ results from scaling the polarization ellipse so that its major semi-axis is unity: ${\bf p}_{1,2}$ are then the intersection with the sphere of normals to the plane of the ellipse that contain the ellipse's foci (orange dots). (b) Husimi distribution over the sphere, where the two zeros (in blue, one hidden behind the sphere) correspond to ${\bf p}_{1,2}$.}
\label{fig:Hannay}
\end{figure}

Hannay \cite{Hannay1998} proposed a representation of nonparaxial polarization in terms of two points over a unit sphere, based on Majorana's construction for spin systems \cite{Majorana1932} and its geometric description by Penrose \cite{Penrose1989}. 
The shape and orientation of the polarization ellipse in 3D is fully characterized by the two points (or unit vectors) ${\bf p}_{1,2}$, which correspond to the two directions in which this ellipse projects onto a circle, as shown in Fig.~\ref{fig:Hannay}(a), and in the sense for which circulation follows the right-hand rule. Clearly the bisector of the two points is normal to the plane containing the ellipse and hence points in the direction of the spin density $\pmb{\mathsf s}$, and the line joining the two points is parallel to the major axis of the ellipse and hence in the direction of ${\bf a}$. The separation of the two points encodes the ellipticity of the polarization: the points get closer together until they coincide as the ellipse tends towards a circle, while on the other hand they separate until they become antipodal as the ellipse tends to a line. 
Note that the two points are {\it indistinguishable} in the sense that exchanging them has no effect on the polarization ellipse. As explained in the caption of Fig.~\ref{fig:Hannay}(a), the separation of the two Hannay-Majorana points also has the property that the line segment joining them is identical in direction and length to that joining the two foci of the ellipse, if the ellipse is scaled so that its major semi-axis is unity. 

The interpretation of ${\bf p}_{1,2}$ as the two directions over which the ellipse projects onto a circle following the right-hand rule has both experimental and mathematical consequences. As mentioned in the Introduction, one of the standard experimental techniques for measuring the polarization at a point is to place a small scatterer (whose dimensions are much smaller than the wavelength) at the point in question and observe the scattered field distribution at the far field. In the Rayleigh approximation, the two directions in which this scattered field is left-circular coincide with  ${\bf p}_1$ and ${\bf p}_2$ (and similarly the scattered field is right-circular in the directions $-{\bf p}_1$ and $-{\bf p}_2$). Mathematically, the two points can then be interpreted as the two zeros of a Husimi distribution over the sphere of unit vectors ${\bf u}$, in this case defined as
\begin{align}
H({\bf u})=|\hat{\bf r}^*({\bf u}) \cdot{\bf E}|^2,
\end{align}
where $\hat{\bf r}({\bf u})$ is a normalized right-circular 3D polarization vector for a plane wave propagating in the direction of the unit vector ${\bf u}$, which can be defined, for example, as
\begin{align}
\hat{\bf r}({\bf u})=\frac{{\bf c}\times{\bf u}-\ui\,{\bf u}\times{\bf c}\times{\bf u}}{|{\bf c}\times{\bf u}|},
\end{align}
where ${\bf c}$ is some constant nonzero real vector whose choice has no influence on the values of $H$. This Husimi distribution is illustrated in Fig.~\ref{fig:Hannay}(b).


\subsubsection{Poincarana construction}
\begin{figure}
\centering
\includegraphics[scale=0.3]{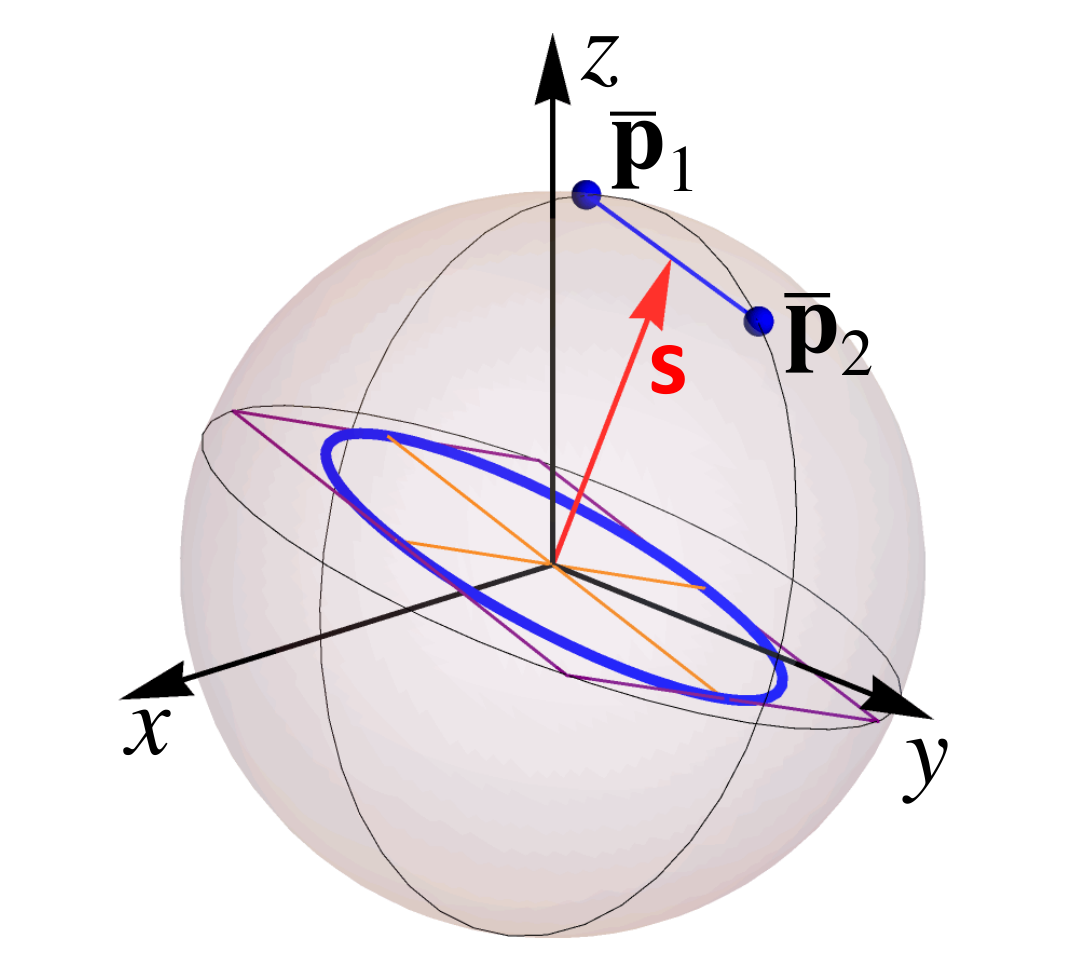}
\caption{The two Poincarana points $\overline{\bf p}_{1,2}$ are along the same great circle as the Hannay-Majorana points ${\bf p}_{1,2}$. Their distance from the plane of the ellipse is precisely the magnitude of ${\bf s}_3$. That is, ${\bf s}_3$ is the mid point between ${\bf p}_1$ and ${\bf p}_2$.}
\label{fig:Poincarana}
\end{figure}

A similar construction was proposed more recently \cite{Bliokh2019}, which was referred to as the \emph{Poincarana} representation since it incorporates aspects from both Hannay's Majorana-based construction as well as from the Poincar\'e sphere. The Poincarana representation also characterizes polarization by using two points over the sphere. These points, $\overline{\bf p}_{1,2}$, are also along the great circle normal to the plane of the ellipse and aligned with the major axis, and have the same bisector as the points ${\bf p}_{1,2}$ in the Hannay-Majorana construction. The only difference is how the angular separation of the two points encodes ellipticity. For the Poincarana construction, this separation is chosen such that the midpoint, $(\overline{\bf p}_1+\overline{\bf p}_2)/2$, corresponds exactly to the normalized spin density $\pmb{\mathsf s}$, as shown in Fig.~\ref{fig:Poincarana}. This property of the Poincarana construction emerges from the fact that it was defined to be directly connected to geometric phase \cite{Bliokh2019}. Consider a continuous transformation of the polarization ellipse due to the evolution of some parameter $\tau$. The geometric phase corresponding to a smooth transformation of this type that is cyclic (i.e. where the final polarization state is identical to the initial one) can then be written as
\begin{align}
\Phi_{\rm G}={\rm Im}\left(\oint\frac{{\bf E}^*\cdot\partial_\tau{\bf E}}{|{\bf E}|^2}\,\ud\tau\right).
\end{align}
This evolution over a cycle corresponds to closed trajectories traced by the points $\overline{\bf p}_{1,2}$. Given the indistinguishability of the points, two scenarios are possible \cite{Hannay1998,Bliokh2019}: either each point traces a closed loop, or one ends where the other began so that together they trace one loop. The Poincarana construction is such that, in both cases, the accumulated geometric phase corresponds directly to one half of the solid angle enclosed by the two points, as shown in \cite{Bliokh2019}, similarly to what happens in the paraxial case when using the Poincar\'e representation to describe geometric phase under parallel transport. This geometric phase includes both transformations of the polarization ellipse within one plane, as in the standard Pancharatnam phase, or due to changes in the plane containing the polarization ellipse, as in the redirection geometric phase.

\subsubsection{Relation between the Hannay-Majorana and the Poincarana constructions}
\begin{figure}
\centering
\includegraphics[scale=0.7]{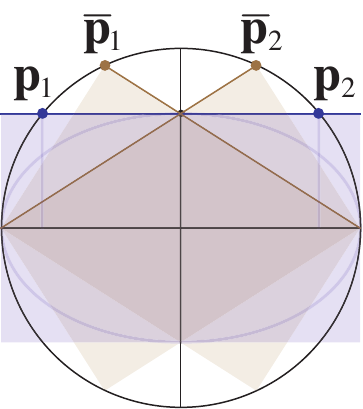}
\caption{Geometric relation between the Hannay-Majorana and Poincarana points over the section of the sphere corresponding to the plane containing the ellipse's major axis and ${\bf s}_3$. The three rectangles (two in pale brown, one in pale blue) have the same aspect ratio as each other and as the polarization ellipse. The horizontal black line through the middle of the circle corresponds to the intersection with the plane of the polarization ellipse (specifically with its major axis). The Hannay-Majorana points ${\bf p}_{1,2}$ are at the intersections with the unit disk of the top side of the horizontal rectangle, whose long side equals 2. The separation between the two points equals the distance between the foci of an ellipse inscribed in the rectangle. The Poincarana points $\overline{\bf p}_{1,2}$ correspond to the top corners of the two tilted rectangles inscribed in the circle, whose diagonals are horizontal and of length 2. Note that the top long sides of the three rectangles always intersect at a point (marked with a black dot), so that the Poincarana points can be easily found from the Hannay-Majorana ones and vice-versa. This diagram shows that the Poincarana points $\overline{\bf p}_{1,2}$ are always closer to each other (and further away from the plane of the ellipse) than the Hannay-Majorana points ${\bf p}_{1,2}$.}
\label{fig:relation}
\end{figure}

It turns out that a simple geometrical relation exists between the two points for the Hannay-Majorana representation and those for the Poincarana representation. To understand this relation, it is sufficient to look at the circular cross-section of the sphere that contains the points, as shown in Fig.~\ref{fig:relation} where the horizontal axis corresponds to the direction of the major axis of the polarization ellipse and the vertical axis corresponds to the direction of the spin density. Consider the height of these pairs of points within this plane. The angle between the vertical axis and each of the two Hannay-Majorana points ${\bf p}_{1,2}$ is $\arccos(|{\bf b}|/|{\bf a}|)$ in order for the ellipse to project onto a circle. The height of the two points on this plane is therefore the cosine of this angle, namely $|{\bf b}|/|{\bf a}|$, and by using the relations $|{\bf a}|^2+|{\bf b}|^2=1$ and $|\pmb{\mathsf s}|=2|{\bf a}||{\bf b}|$ this height can be written purely in terms of the normalized spin density as $|\pmb{\mathsf s}|/(1+\sqrt{1-|\pmb{\mathsf s}|^2})$. For the Poincarana points, on the other hand, the height is simply $|\pmb{\mathsf s}|$, and the half-separation between the two points is then $\sqrt{1-|\pmb{\mathsf s}|^2}$. It is easy to see then that, if one draws a straight line from one of the Poincarana points to the intersection of the horizontal axis with the edge of the sphere that is most distant to this point, this line will cross the vertical axis at a height $|\pmb{\mathsf s}|/(1+\sqrt{1-|\pmb{\mathsf s}|^2})$, which is the height of the Hannay-Majorana points, as illustrated in Fig.~\ref{fig:relation}.

The Hannay-Majorana and Poincarana points coincide only in the two limiting situations: i) for circular polarization in 3D (for which ${\bf E}\cdot{\bf E}=0$) in which a case all points coincide, namely ${\bf p}_1={\bf p}_2=\overline{\bf p}_1=\overline{\bf p}_2=\pmb{\mathsf s}$, and ii) for linear polarization in 3D (for which ${\rm Im}(E_x^*E_y)={\rm Im}(E_y^*E_z)={\rm Im}(E_z^*E_x)=0$), where for each of the representations the two points are antipodal and define the direction of oscillation of the field, namely ${\bf p}_1=-{\bf p}_2=\overline{\bf p}_1=-\overline{\bf p}_2\propto{\bf E}$. For all other cases, the angle between the Poincarana points is always smaller than that between the Hannay-Majorana points.

\subsubsection{Relation of the Hannay-Majorana and Poincarana representations in the paraxial case with the Poincar\'e sphere}
It is useful to consider the relation between the Hannay-Majorana and Poincarana representations with the Poincar\'e representation in the limiting case of paraxial light traveling in the positive $z$ direction. In this case, the ellipse traced by the field is contained within the $xy$ plane, and the normalized spin density vector $\pmb{\mathsf s}$ is constrained to the $z$ direction. Let us start by considering the Poincarana representation. The points $\overline{\bf p}_{1,2}$ are bisected by the $z$ axis, and then have the same height as each other. Because this height is, by definition, the normalized spin density, it coincides with the height $s_3$ of the point $\vec s_{\rm 2D}$ for the Poincar\'e sphere. Further, let the angles with respect to the $x$ axis of the projections of the Poincarana points $\overline{\bf p}_{1,2}$ onto the $xy$ plane be referred as $\xi_{1,2}=\arg(\{\overline{\bf p}_{1,2}\}_x+\ui\{\overline{\bf p}_{1,2}\}_y)$. Since the two points are bisected by the $z$ axis, these two angles differ (modulo $2\pi$) by $\pi$. These angles correspond to the angle between the $x$ axis and the major semi-axes of the polarization ellipse. The corresponding angle $\phi=\arg(s_1+\ui s_2)$ for the Poincar\'e sphere is then given, modulo $2\pi$, by $\phi=2\xi_1=2\xi_2$. That is, the two Poincarana points result from rotating around the vertical axis ($s_3$ for Poincar\'e, $z$ for Poincarana) the Poincar\'e point so that its angles with respect to the $s_1$ axis, both clockwise and anti-clockwise, are halved. 

The corresponding transformation from Poincar\'e to Hannay-Majorana is equivalent, except that it also involves a change in height according to $\{\overline{\bf p}_{1,2}\}_z=s_3/(1+\sqrt{1-s_3^2})$. It turns out that this extra change in height makes the mapping between the spheres conformal, a property that is important in the definition of the Majorana representation for any number of dimensions \cite{Majorana1932}. 

\subsubsection{A third two-point construction motivated by statistically uncorrelated light}
Both two-point representations for nonparaxial polarization ellipses described earlier obey the following simple rules: 
\begin{itemize}
\item the two points are indistinguishable; 
\item their bisector is parallel to the spin density; 
\item the line joining them is parallel to the major axis of the polarization ellipse; 
\item their separation uniquely and continuously encodes ellipticity such that circular polarization corresponds to the two points coinciding and linear polarization corresponds to the two points being antipodal. 
\end{itemize}
What distinguishes these representations is simply how ellipticity is encoded as point separation or, equivalently, what the relation is between the magnitudes of the centroid of the two points and the normalized spin density: for the Hannay-Majorana construction we have $|{\bf p}_1+{\bf p}_2|/2=|\pmb{\mathsf s}|/(1+\sqrt{1-|\pmb{\mathsf s}|^2})$, while for the Poincarana representation we have the simpler relation $|\overline{\bf p}_1+\overline{\bf p}_2|/2=|\pmb{\mathsf s}|$. These definitions convey each construction with different desirable properties: interpretation in terms of circular projection of the ellipse for Hannay-Majorana, connection with the geometric phase for Poincarana.

Here we propose a third option that is motivated by a statistical argument. Consider the superposition of a large number of monochromatic plane waves of the same temporal frequency, whose propagation directions are uniformly distributed over the whole sphere of directions, and whose polarizations and phases are random and statistically uncorrelated \cite{Berry2001}. 
At any point in space this field is fully polarized, but the polarization changes significantly from point to point within the scale of a wavelength. If we sample the polarization over a large number of spatial points, the direction of the spin density would be statistically uniformly distributed over the sphere. As shown in Appendix~\ref{appC}, the magnitude of the normalized spin density, $|\pmb{\mathsf s}|$, turns out to follow a probability density that is constant over its allowed range of values $|\pmb{\mathsf s}|\in[0,1]$. For such a statistically isotropic monochromatic field, each point of a two-point representation would be able to access the complete unit sphere with uniform probability distribution. However, the statistical distribution of the angle between the two points would depend on which representation we are using. Dennis \cite{DennisPersonal} calculated this distribution for the Hannay-Majorana representation. 

Consider a generic two-point representation, where the two points over the sphere are $\widetilde{\bf p}_{1,2}$. One could naively expect that the isotropy and lack of correlation inherent to a fully random onmidirectional plane-wave superposition would result in the positions of these two points being mutually completely uncorrelated. That is, if we identify a large number of polarizations in this field in which, say, $\widetilde{\bf p}_1$ is at a given location over the sphere, the probability of finding $\widetilde{\bf p}_2$ anywhere over the sphere would be uniform. However, such property would require a very specific statistical distribution of the angle between the points that is not that for the Hannay-Majorana nor for the Poincarana constructions. Nevertheless, a two-point representation that shows this property can be found that has a fairly simple expression. 

Full decorrelation of the locations of the two points over the sphere implies a statistically uniform distribution of $\cos\nu$, where $\nu$ is the angle between the two points. This follows from the fact that, if we change reference frames so that, say, $\widetilde{\bf p}_1$ is aligned with the vertical axis, uniform coverage of the sphere by $\widetilde{\bf p}_2$ implies that the vertical coordinate of this second point, which corresponds to $\cos\nu$, is uniformly distributed. By using trigonometric properties, we can write this uniformly-distributed quantity as $\cos\nu=2\cos^2(\nu/2)-1$. However, the magnitude of the centroid of the two points is given by $|\widetilde{\bf p}_1+\widetilde{\bf p}_2|/2=\cos(\nu/2)$. Therefore, for the location of the two points to be statistically uncorrelated, we must choose $|\widetilde{\bf p}_1+\widetilde{\bf p}_2|/2=|\pmb{\mathsf s}|^{1/2}$, so that $\cos\nu=2|\pmb{\mathsf s}|-1$ is uniformly distributed in the interval $[-1,1]$ for a random omnidirectional plane-wave superposition. 

This statistically-motivated construction then represents a nonparaxial polarization ellipse in terms of the two indistinguishable points $\widetilde{\bf p}_{1,2}$ that are closer to each other than the corresponding pairs of points for the Hannay-Majorana and Poincarana constructions, except in the limits of circular and linear polarization. When applied to a random field, any statistical correlation between $\widetilde{\bf p}_1$ and $\widetilde{\bf p}_2$ when sampling the field would signal either a lack of omnidirectionality for the field or a correlation in the polarization states or phases of the constitutive plane waves.

\subsubsection{Expressions for the two-point constructions in terms of the field}
To facilitate their computation, let us give simple expressions for the two points of any of the three two-point representations discussed so far. The coordinates of the two points can be written as
\begin{align}
{\bf P}_{1,2}=C\,\frac{\pmb{\mathsf s}}{|\pmb{\mathsf s}|}\pm\sqrt{1-C^2}\,\frac{{\bf a}}{|{\bf a}| },
\label{eq:pointsfield}
\end{align}
where $\pmb{\mathsf s}$ and ${\bf a}$ can be found by using Eqs.~(\ref{eqs:Eparts}) and $C$ is a function of the magnitude of the spin density corresponding to the cosine of the angle between the points and their bisector, which for the three cases discussed earlier is given by the values in Table~1.
\begin{table}
\centering
\caption{Functions $C$ for different two-point representations of full nonparaxial polarization.}
\begin{tabular}{c|c|c}
Representation& Points ${\bf P}_i$&$C$\\
\hline\hline
Hannay-Majorana&${\bf p}_i$&$\frac{|\pmb{\mathsf s}|}{1+\sqrt{1-|\pmb{\mathsf s}|^2}}$\\
\hline
Poincarana&$\overline{\bf p}_i$&$|\pmb{\mathsf s}|$\\
\hline
Statistical&$\widetilde{\bf p}_i$&$\sqrt{|\pmb{\mathsf s}|}$
\end{tabular}
\end{table}

\begin{figure}
\centering
\includegraphics[scale=0.4]{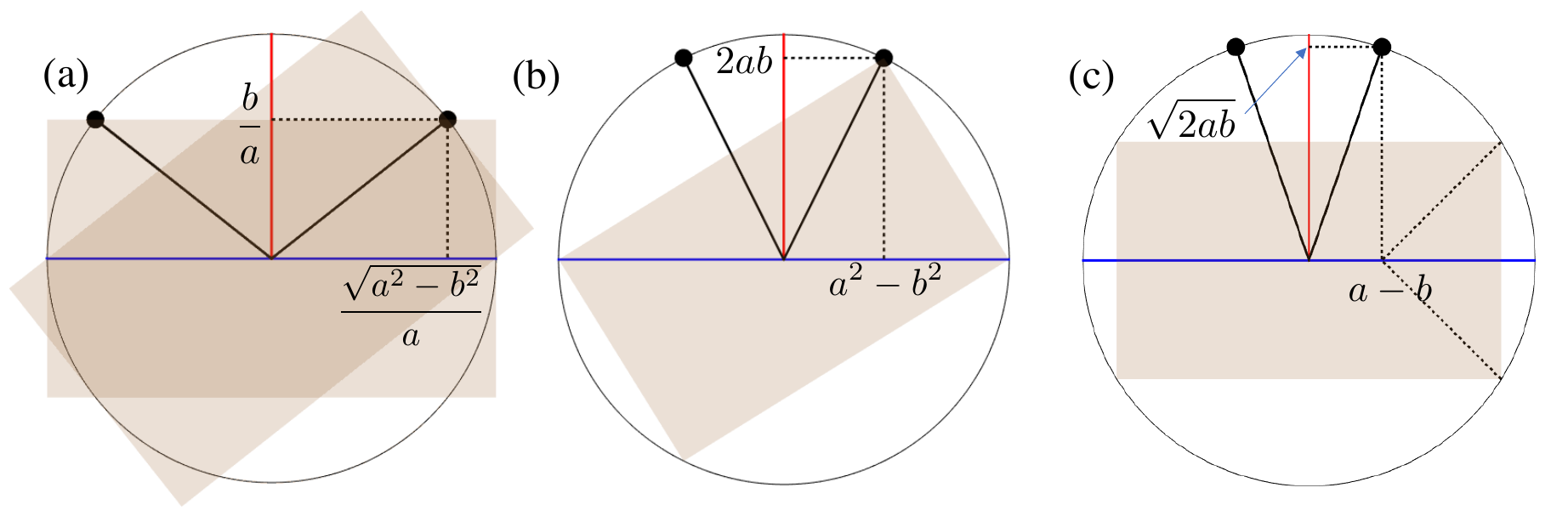}
\caption{Coefficients $C$ (vertical coordinate) and $\sqrt{1-C^2}$ (horizontal coordinate) in terms of the major and minor semiaxis lengths, $a=|{\bf a}|$ and $b=|{\bf b}|$, respectively, of the normalized polarization ellipse, for (a) the Hannay-Majorana, (b) the Poincarana, and (c) the statistically motivated representations of nonparaxial full polarization. In all cases the two points are indicated as the two black dots, the circle over which they lay represents the cross-section of the unit sphere that contains them, with the horizontal blue line representing the direction of the ellipse's major axis and the vertical red line the direction of the spin density. The brown rectangles shown all have the same aspect ratio as the polarization ellipse: those in (a) being slightly larger, having dimensions of $2$ times $2b/a$, while those in (b,c) have the same dimensions as the normalized polarization ellipse, namely $2a$ times $2b$. These rectangles are included to illustrate the geometrical relation in each case between the points and the aspect ratio of the ellipse.}
\label{fig:allthree}
\end{figure}

It might be useful to present also the expressions for the coefficients $C$ and $\sqrt{1-C^2}$ in Eq.~(\ref{eq:pointsfield}) in terms of the lengths of the major and minor axes of the normalized polarization ellipse, namely $|{\bf a}|$ and $|{\bf b}|$. These are presented in Fig.~\ref{fig:allthree}, along with some geometric relations for the three constructions.

\subsubsection{Orthogonality of polarizations}
\begin{figure}
\centering
\includegraphics[scale=0.4]{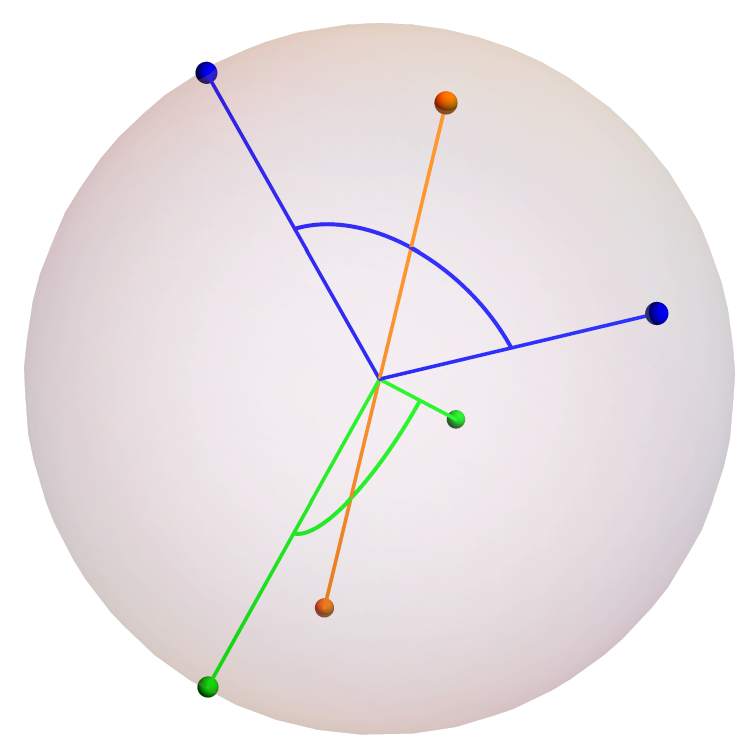}
\caption{For a given polarization represented by the two blue dots, to orthogonal polarizations are those for which i) the coplanar counter-rotating ellipse for which the two points (green) are at the same angle around the same bisecting line as the original polarization, but are in the opposite hemisphere and rotated around the bisector by $\pi/2$, and ii) the linear polarization represented by two antipodal points at the ends of the bisector of the original polarization. }
\label{fig:orthogonality}
\end{figure}

In the paraxial regime, for any given polarization ellipse there is a unique orthogonal polarization (given that a global phase and amplitude are ignored), and the orthogonality of two polarizations is obvious from their representations as points over the Poincar\'e sphere, which must be antipodal. For nonparaxial polarization, on the other hand, each polarization state has not one but a two-parameter set of orthogonal polarization states, since for given ${\bf E}$, the complex equality ${\bf E}^*\cdot{\bf E}'$ only imposes two constraints on the four degrees of freedom of the polarization of ${\bf E}'$. For any of these three constructions, it is in general not easy to identify directly if two polarizations are orthogonal given their pairs of points, with the exception of two cases. Suppose that we have a given polarization state whose normalized complex field vector is ${\bf f}={\bf a}+\ui{\bf b}$ and whose normalized spin density is $\pmb{\mathsf s}$. The two simple orthogonal states, with complex field vector ${\bf f}'={\bf a}'+\ui{\bf b}'$ and normalized spin density $\pmb{\mathsf s}'$, are:
\begin{subequations}
\begin{itemize}
\item the coplanar ellipse with the opposite spin density and whose major and minor axes have the same magnitudes but their directions are exchanged, namely,
\begin{align}
{\bf a}'=\frac{|{\bf a}|}{|{\bf b}|}{\bf b},\,\,\,\,\,\,\,\,{\bf b}'=\frac{|{\bf b}|}{|{\bf a}|}{\bf a},\,\,\,\,\,\,\,\,\pmb{\mathsf s}'=-\pmb{\mathsf s};
\end{align}
\item the lineal polarization oriented perpendicularly to the plane of the polarization ellipse, so that
\begin{align}
{\bf a}'=\frac{\pmb{\mathsf s}}{|\pmb{\mathsf s}|},\,\,\,\,\,\,\,\,{\bf b}'={\bf 0},\,\,\,\,\,\,\,\,\pmb{\mathsf s}'={\bf 0}.
\end{align}
\end{itemize}
\end{subequations}
The two-point coordinates of these polarizations are illustrated in Fig.~\ref{fig:orthogonality}. Any other orthogonal polarization can be constructed as a complex linear combination of these two. For these other states, the coordinates of the points will depend on the specifics of the chosen two-point representation. Of course, the two special cases just described become degenerate if the initial polarization is either circular (the two points coincide) or linear (the two points are antipodal).

\subsubsection{Topology of two-point constructions and polarization M\"obius strips}
Any of the two-point constructions described in this section can be used for visualizing topological properties of a cyclic polarization evolution. For example, consider the variation of polarization along a closed contour in a monochromatic field  \cite{Bliokh2019}. The polarization at each stage is represented by two points over the sphere, so that for the complete loop, these points trace closed curves over the sphere. Since the separation between the two points is parallel to the direction of the polarization ellipse's major axis, the two points exchange roles when completing the loop if and only if the major axis describes a M\"obius strip over the loop \cite{Bauer2015,FreundOC2010,FreundOL2010,Dennis2011}. That is, for a loop over which the field's major axis describes a M\"obius strip, there is a single closed curve over the sphere, where each point traces a segment of this curve. On the contrary, if the field's major axis does not describe a M\"obius strip, each point traces a separate closed curve, so there are two closed curves over the sphere.

\subsubsection{Unitary transformations}
One of the key properties of the Poincar\'e sphere representation is that any linear unitary transformation of the Jones vector corresponds simply to a rigid rotation of the point over the sphere. This is important because linear unitary transformations correspond to the physical effect of light passing through transparent materials that cause a retardation of one specific polarization component with respect to the orthogonal one. The fact that such physical components correspond to unitary transformations (independent of the polarization of the incident field) relies on the paraxial approximation, in which the field is known to propagate in a given direction. 

In the nonparaxial regime, knowledge of the local polarization is largely independent of the range of propagation directions that compose the field, and therefore the effect of a transparent optical element cannot generally be associated with a unitary transformation acting on the three-component complex vector. Therefore, the fact that general linear unitary transformations of the complex field do not correspond to simple geometric transformations of the two-point representations poses no problems to their physical usefulness. The only unitary transformations with direct physical relevance in the nonparaxial regime are rigid rotations (associated, for example, with changes of coordinate reference frame) and inversions (resulting, for example, from reflection by a perfect flat mirror). Since the coordinate axes for the two-point representations are precisely the directions of the physical space, invariance to these transformations is guaranteed.

\section{Nonparaxial partially polarized light}
\label{sec:NPPPL}

In this section we introduce the basic elements for the description of fields that are not purely monochromatic but that are statistically stationary, and hence can be regarded as partially polarized. 
%

\subsection{$3\times3$ polarization matrix, ellipsoid of inertia and spin density}
\label{section:ellipsoid}
For a nonparaxial, partially polarized field, polarization at a given point is described by the $3\times3$ polarization matrix:
\begin{align}
\mathbf{\Gamma}=\langle{\bf E}{\bf E}^\dagger\rangle=\left(\begin{array}{ccc}\langle E^*_xE_x\rangle&\langle E^*_yE_x\rangle&\langle E^*_zE_x\rangle\\\langle E^*_xE_y\rangle&\langle E^*_yE_y\rangle&\langle E^*_zE_y\rangle\\\langle E^*_xE_z\rangle&\langle E^*_yE_z\rangle&\langle E^*_zE_z\rangle\end{array}\right),
\end{align}
For simplicity of notation, we do not use the subindex 3D to label either this matrix nor the measures of polarization that follow. This matrix is Hermitian and non-negative definite, and therefore it contains nine degrees of freedom. Like in the paraxial case, however, one of these degrees of freedom can be associated with the total intensity, and hence can be removed through normalization, leaving eight real parameters needed to fully determine the state of polarization.
 
\subsection{Ellipsoid of inertia and spin density}
\begin{figure}
\centering
\includegraphics[scale=0.5]{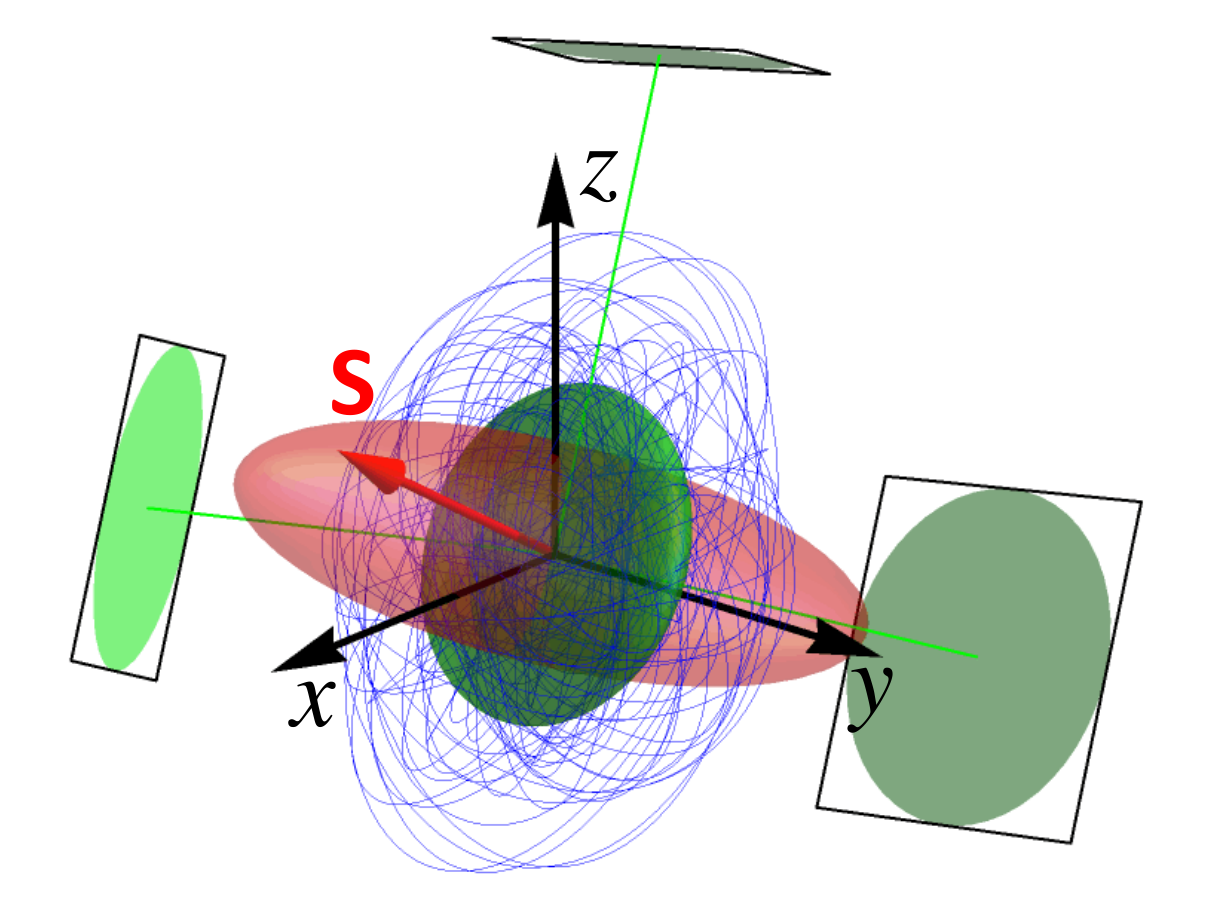}
\caption{Path traced by the electric field (blue) over several optical cycles. The green ellipsoid is the ellipsoid of inertia, which characterizes the average shape described by the oscillations of the electric field. Its semi-axes are given by the square roots of the eigenvalues of ${\rm Re}(\mathbf{\Gamma})$ and are aligned with the corresponding eigenvectors (the direction of the green lines). The vector $\pmb{\mathsf S}$ (red arrow) describes the direction around which the field spins on average, and the amount of this spin. This vector is constrained to the interior of a dual ellipsoid (red), whose semi-axes are equal to $2/\pi$ times the areas of the corresponding projections of the green ellipsoid in the directions of the eigenvectors. Note that different elements of this figure have different units (field for the blue curve, the green ellipsoid and its projections, field squared for $\pmb{\mathsf S}$ and the red ellipsoid), and hence an arbitrary scaling between them was used in the figure.}
\label{fig:Mark}
\end{figure}

Dennis \cite{Dennis2004} proposed an intuitive geometric interpretation for this $3\times3$ polarization matrix, which generalizes the one shown in Fig.~\ref{fig:2D}(a) for paraxial light. The oscillations of the electric field vector follow a volumetric probability density with ellipsoidal cross-section, and to second order, this ellipsoidal shape is characterized by the {\it ellipsoid of inertia}, whose semi-axes are aligned with the eigenvectors of ${\rm Re}(\mathbf{\Gamma})$ and have lengths equal to the square roots of the corresponding eigenvalues. Recall that its paraxial analogue, the ellipse of inertia, is supplemented with the (scalar) spin density $S_3$, given by the imaginary part of the off-diagonal matrix elements, whose value is constrained by the area of the ellipse. Similarly, for nonparaxial fields the ellipsoid of inertia is supplemented with the spin density vector, which for partially polarized light includes an average: $\pmb{\mathsf S}=2{\rm Im}(\Gamma_{yz},\Gamma_{zx},\Gamma_{xy})={\rm Im}\langle{\bf E}^*\times{\bf E}\rangle$. 
This averaged spin density quantifies both the preferential axis and sense of rotation of the time-dependent electric field vector. This description is illustrated in Fig.~\ref{fig:Mark}. 

It is easy to show that, given the Hermiticity of the polarization matrix, its determinant can be written as the difference of two contributions, where only one of them depends on the imaginary parts of the matrix components, encoded in the vector $\pmb{\mathsf S}$:
\begin{align}
\det(\mathbf{\Gamma})=\det[{\rm Re}(\mathbf{\Gamma})]-\frac14\pmb{\mathsf S}\cdot{\rm Re}(\mathbf{\Gamma})\cdot\pmb{\mathsf S}.
\end{align}
The non-negative definiteness of the polarization matrix implies that this determinant must be equal to or greater than zero, and this fact imposes the following restriction for the spin vector:
\begin{align}
\pmb{\mathsf S}\cdot{\rm Re}(\mathbf{\Gamma})\cdot\pmb{\mathsf S}\le4\det[{\rm Re}(\mathbf{\Gamma})],\label{ellipsoidforS3}
\end{align}
where the equality holds only if at least one of the eigenvalues of the polarization matrix vanishes. This inequality restricts the spin vector to the interior of a {\it dual} ellipsoid \cite{Dennis2004}, whose semi-axes are aligned with those of the first ellipsoid, but where the length of each semi-axis of the dual ellipsoid equals $2/\pi$ times the area subtended by the projection of the first ellipsoid in the direction of the corresponding semi-axis, as shown in Fig.~\ref{fig:Mark}.

\section{Measures of polarization and their geometric interpretations}
\label{sec:DOP}
Several measures have been proposed that seek to generalize the concept of degree of polarization as defined in Eq.~(\ref{DoP2D}) to the nonparaxial regime, based on its different interpretations. Several reviews of this topic and comparisons between these measures exist in the literature \cite{BrosseauProgOpt2006,Petruccelli2010,Sheppard2012,Gamel2012,Aunon2013,Gamel2014}. Most of these measures were defined to be invariant under general unitary transformations of the electric field, following the example of the paraxial definition. A consequence of this invariance is that it is possible to express these measures exclusively in terms of the three eigenvalues $\Lambda_1\ge\Lambda_2\ge\Lambda_3\ge0$ of the polarization matrix, independently from the eigenvectors. Sheppard \cite{Sheppard2011,Sheppard2020,Sheppard2022} found that these eigenvalues and eigenvectors can be calculated analytically from the elements of the matrix through fairly simple expressions.

\subsection{The two most common definitions and their companions}
The first measure of degree of polarization discussed here was proposed by Samson \cite{Samson1973} and later by others \cite{Barakat1977,SetalaPRL2002,SetalaPRE2002}. This measure can be written in the equivalent forms
\begin{align}
P_{\rm I}&=\sqrt{\frac{3\,{\rm Tr}\mathbf{\Gamma}^2}{2\,({\rm Tr}\mathbf{\Gamma})^2}-\frac12}\nonumber\\
&=\sqrt{\lambda_1^2+\lambda_2^2+\lambda_3^2-\lambda_1\lambda_2-\lambda_2\lambda_3-\lambda_3\lambda_1}\nonumber\\
&=\sqrt{\frac{(\lambda_1-\lambda_2)^2+(\lambda_2-\lambda_3)^2+(\lambda_1-\lambda_3)^2}2}.\label{DoPSampson}
\end{align}
where $\lambda_i=\Lambda_i/\sum_{j=1}^3\Lambda_j$ are the normalized eigenvalues. 
The measure $P_{\rm I}$ is monotonically linked to measures used commonly in quantum physics and linear algebra, such as the purity \cite{Moya2008}, the Schmidt index \cite{Qian2011}, and the trace distance of the polarization matrix to the identity matrix \cite{Luis2005}. According to this measure, a field is fully polarized if and only if two eigenvalues vanish, and fully unpolarized only if all three eigenvalues are equal.

An alternative measure was proposed by Gil \cite{Gil2004} and independently by Ellis and colaborators \cite{EllisWolfOC2005,EllisOC2005,EllisPRL2005}, based on the interpretation of the degree of polarization as the fraction of the optical power that is fully polarized. To understand this measure, it is convenient to write the polarization matrix in terms of its three orthonormal eigenvectors in a fashion similar to that in Eq.~(\ref{polandnotpol}):
\begin{align}
\mathbf{\Gamma}&=\Lambda_1\,{\bf e}_1{\bf e}_1^\dagger+\Lambda_2\,{\bf e}_2{\bf e}_2^\dagger+\Lambda_3\,{\bf e}_3{\bf e}_3^\dagger\nonumber\\
&=(\Lambda_1-\Lambda_2)\,{\bf e}_1{\bf e}_1^\dagger+(\Lambda_2-\Lambda_3)\,({\bf e}_1{\bf e}_1\dagger+{\bf e}_2{\bf e}_2^\dagger)+\Lambda_3\,\left(\begin{array}{ccc}1&0&0\\0&1&0\\0&0&1\end{array}\right)
,\label{polhalfnotpol}
\end{align}
where the eigenvectors ${\bf e}_i$ have now three components. Notice that in the last step a fraction of each of the first two terms in the first line was transferred to the terms to its right, so that the last term is proportional to the $3\times3$ identity composed as ${\bf e}_1{\bf e}_1^\dagger+{\bf e}_2{\bf e}_2^\dagger+{\bf e}_3{\bf e}_3^\dagger$. Like in Eq.~(\ref{polandnotpol}), the first term is fully factorizable and therefore on its own would correspond to a fully polarized field, while the last is proportional to the ($3\times3$) identity and alone would correspond to a fully unpolarized field. However there is an extra term, proportional to $\Lambda_2-\Lambda_3$ that is neither fully polarized nor fully unpolarized. That is, in general a $3\times3$ polarization matrix cannot be expressed as the sum of two parts that are respectively fully polarized and fully unpolarized.
The degree of polarization in question is then the ratio of the power of the fully polarized part to the total, namely 
\begin{align}
P_{\rm II}=\frac{\Lambda_1-\Lambda_2}{\Lambda_1+\Lambda_2+\Lambda_3}=\lambda_1-\lambda_2,\label{DoPE}
\end{align}
It is tempting to interpret the second term on the right-hand side of Eq.~(\ref{polhalfnotpol}) as a ``2D-unpolarized'' component, since, if the eigenpolarizations ${\bf e}_1$ and ${\bf e}_2$ were contained in the same plane, there would be a reference frame in which this term would be proportional to the matrix ${\rm Diag}(1,1,0)$, giving a contribution similar to that of unpolarized light in the paraxial sense. In general, however, the polarization ellipses for ${\bf e}_1$ and ${\bf e}_2$ are not contained in the same plane, a case referred to by Gil and collaborators as polarimetric nonregularity \cite{SanJose2011,Gil2017,GilPRA2018,GilEPJ2018,GilOL2018,GilNJP2021} which has consequences e.g. on the distribution of spin amongst the first and second terms. 
Note that according to the measure $P_{\rm II}$, for a field to be fully polarized two eigenvalues must vanish, but (unlike for $P_{\rm I}$) a completely unpolarized field is one for which the two largest eigenvalues are equal, regardless of the third. One could say that $P_{\rm II}=0$ can be interpreted as meaning that the field has no fully polarized component, rather than it being fully unpolarized.

The measures $P_{\rm I}$ and $P_{\rm II}$ (and those related univocally to each of them) have been used to characterize with a single quantity the level of polarization of the matrix. However, the polarization matrix has three eigenvalues whose sum gives the total intensity (which is not relevant to polarization). In other words, only two normalized eigenvalues $\lambda_i$ are independent since $\lambda_1+\lambda_2+\lambda_3=1$. Therefore, the full characterization of polarization from the eigenvalue point of view requires the specification of two numbers, and so each of the two measures above can be supplemented with a second measure. For example, Barakat \cite{Barakat1977} proposed a second measure to supplement $P_{\rm I}$, referred to here as $Q_{\rm I}$ and given by
\begin{align}
Q_{\rm I}&=\sqrt{1-27 \frac{\det\mathbf{\Gamma}}{({\rm Tr}\mathbf{\Gamma})^3}}=\sqrt{1-27\lambda_1\lambda_2\lambda_3}.\label{DoPBarakat}
\end{align}
Note that $\sqrt{1-Q_{\rm I}^2}$ has been referred to as a degree of isotropy \cite{Hioe2006}. It has been shown that for fields with Gaussian statistics, $P_{\rm I}$ and $Q_{\rm I}$ are related to the Shannon and Renyi entropies \cite{Refregier2006}. Similarly, measures that supplement $P_{\rm II}$ have been proposed that like it are linear combinations of the normalized eigenvalues, such as \cite{Gil2004}
\begin{align}
Q_{\rm II}&=\lambda_1+\lambda_2-2\lambda_3.
\end{align}

\subsection{Wobbling fluorophores and rotational constraint}
\label{microscopy}
A definition that is mathematically similar to $P_{\rm II}$ was proposed within the context of fluorescence microscopy to quantify the amount of vibration (often called wobble) of a fluorophore \cite{Zhang2019}. This measure, referred to as the {\it rotational constraint}, results from a slightly different decomposition of the matrix into three parts, according to
\begin{align}
\mathbf{\Gamma}=\left(\Lambda_1-\frac{\Lambda_2+\Lambda_3}2\right)\,{\bf e}_1{\bf e}_1^\dagger+\frac{\Lambda_2-\Lambda_3}2\,({\bf e}_2{\bf e}_2^\dagger-{\bf e}_3{\bf e}_3^\dagger)+\frac{\Lambda_2+\Lambda_3}2\,\left(\begin{array}{ccc}1&0&0\\0&1&0\\0&0&1\end{array}\right).
\label{formovility}
\end{align}
Notice that the second term cannot be interpreted on its own as a valid polarization matrix because it is not non-negative definite. The motivation for this type of separation comes from its physical context: we consider light emitted by a linear dipole that wobbles at a time scale much larger than the optical period. The resulting light has essentially no spin, so that the three eigenvectors can be chosen as real and point in orthogonal directions. The eigenvector ${\bf e}_1$ corresponds then to the main direction of the dipole, and if, say, the wobbling were within an isotropic cone (a common assumption in this context), the two smaller eigenvalues would coincide. The second term in Eq.~(\ref{formovility}) therefore accounts for possible rotational asymmetry of the wobbling around the main direction ${\bf e}_1$. Like $P_{\rm II}$, the rotational constraint used to quantify wobble is defined as the ratio of the factorizable part to the total:
\begin{align}
\gamma=\lambda_1-\frac{\lambda_2+\lambda_3}2.
\end{align}
Notice that, if it were to be considered as a measure of polarization, $\gamma$ would agree with $P_{\rm I}$ in defining what states correspond to full and null polarization. 

\begin{figure}
\centering
\includegraphics[scale=0.35]{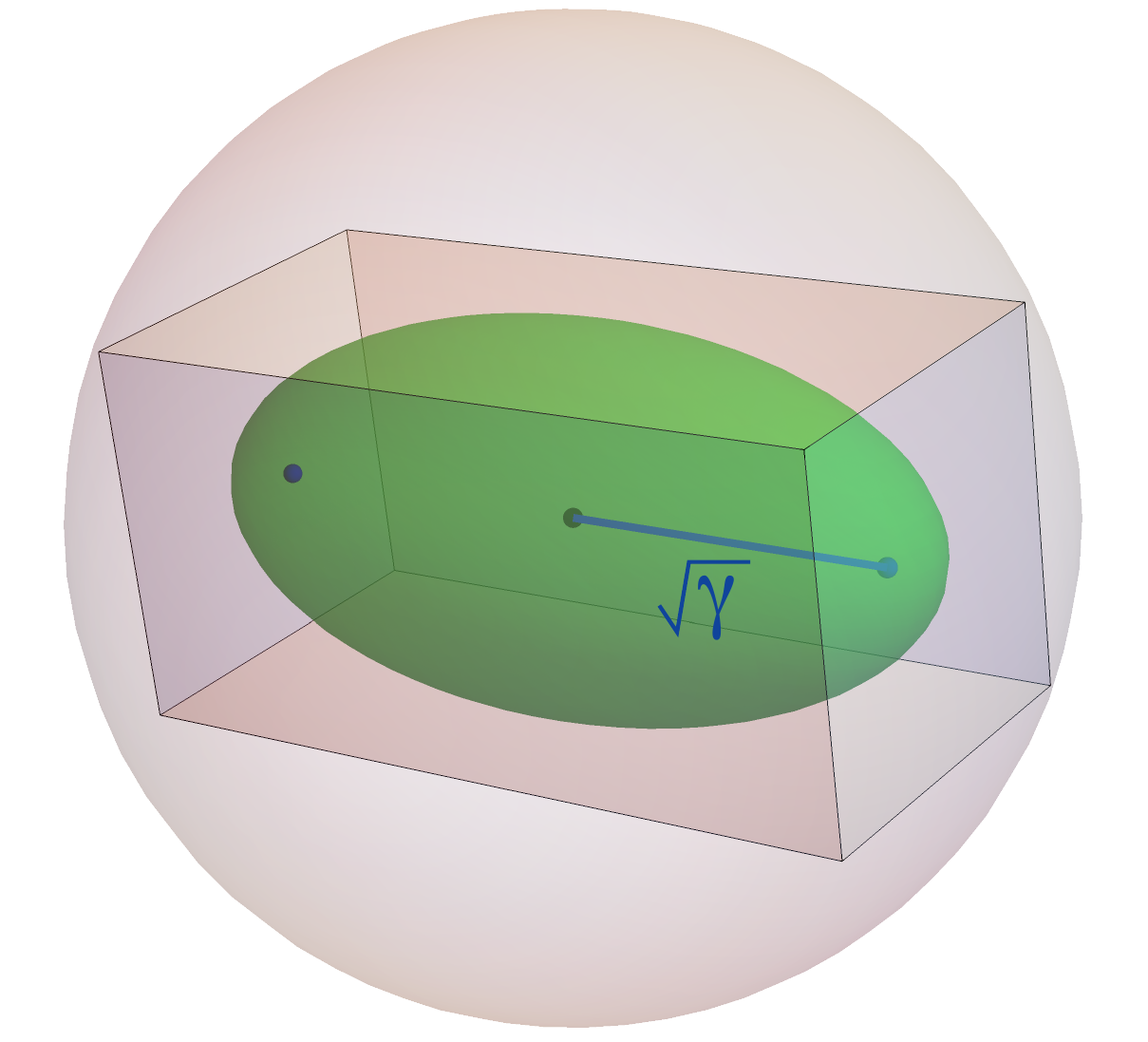}
\caption{Interpretation of the rotational mobility $\gamma$ in the case of symmetric wobble ($\Lambda_2=\Lambda_3$) as the square root of the distance from the center to the foci of the normalized ellipsoid of inertia (green), which in this case is a prolate spheroid. Here, normalization implies that the ellipsoid must be inscribed in a box that is itself inscribed in a unit sphere. It is assumed in this context that $\pmb{\mathsf s}={\bf 0}$.}
\label{fig:wobbleellipse}
\end{figure}

For the common assumption of symmetric wobble, where $\lambda_2=\lambda_3$, the three measures actually coincide, namely $P_{\rm I}=P_{\rm II}=\gamma$, and they have a geometric interpretation. As mentioned earlier, the spin density $\pmb{\mathsf S}$ vanishes for (static or wobbling) linear dipole emitters, so $\mathbf{\Gamma}$ is fully represented graphically by the ellipsoid of inertia, whose semi-axes are the square roots of the eigenvalues. Let us consider the dimensionless version of this ellipsoid normalized by the intensity, whose semi-axes are the square roots of the normalized eigenvalues $\lambda_i$. As shown in Fig.~\ref{fig:wobbleellipse}, this normalized ellipsoid is inscribed in a box that is itself inscribed in a unit sphere. The assumption of symmetric wobble means that this ellipsoid is a prolate spheroid, and therefore it has two focal points. The distance from the center to the foci is given by $\sqrt{\gamma}$.

\subsection{Barycentric interpretation}
\begin{figure}
\centering
\includegraphics[scale=0.8]{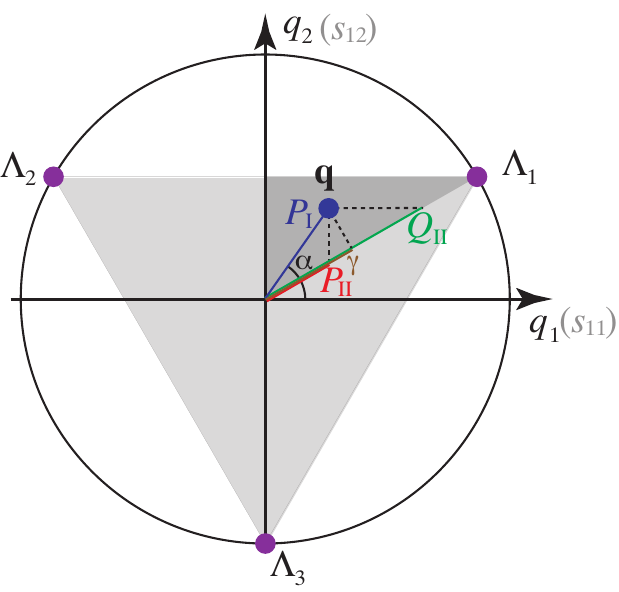}
\caption{Geometric interpretation of the measures of polarization $P_{\rm I}$, $P_{\rm II}$, $Q_{\rm II}$, $\gamma$, and the angle $\alpha$ in terms of the center of mass (blue dot) of three point masses (purple dots) whose magnitudes are $\Lambda_i$, the eigenvalues of $\mathbf{\Gamma}$. Note that $P_{\rm II}$ and $Q_{\rm II}$ are proportional to the coordinates $q_1$ and $q_2$, respectively, and that the proportionality factors can be accounted for geometrically by looking at the distance from the origin of the intersections of the lines of constant $q_1$ and $q_2$ with a radial line at $30^\circ$ from the $q_1$ axis. The measure $Q_{\rm I}$ is given by $P_{\rm I}^2 (3-2P_{\rm I}\sin3\alpha)$. The rotational mobility $\gamma$ is directly the projection onto this radial line. The equilateral triangle corresponds also to the inhabitable region for the Stokes-Gell-Mann sub-vector ${\bf s}_1$.}
\label{fig:CoM}
\end{figure}

To provide an interpretation to these quantities, we use a simple geometric construction \cite{CoM} like that described at the end of Section 2 for paraxial polarization. Consider three point masses within a plane, at equal distances from each other and at a unit distance from the origin. That is, these three masses are at the corners of an equilateral triangle inscribed in the unit circle, as shown in Fig.~\ref{fig:CoM}. Let the magnitudes of each of these masses be one of the eigenvalues $\Lambda_i$ (or their normalized versions $\lambda_i$). The point ${\bf q}$ corresponding to the center of mass 
of the three masses is necessarily inside the equilateral triangle, and given the chosen ordering of the eigenvalues, it is further constrained to one sixth of this triangle, as shown in Fig.~\ref{fig:CoM}. The measures discussed so far are associated with coordinates for this center of mass. For example, it is easy to show that $P_{\rm I}$ and $Q_{\rm I}$ are simply related to the polar coordinates of ${\bf q}$: the first is directly the radial coordinate or distance to the origin, $P_{\rm I}=|{\bf q}|$, while the second depends on the angular coordinate, $Q_{\rm I}=|{\bf q}|^2 (3-2|{\bf q}|\sin3\alpha)$, where $\alpha\in[\pi/6,\pi/2]$ is the angle between the $q_1$ axis and ${\bf q}$. On the other hand, it is easy to see that $P_{\rm II}=q_1\sqrt{3}/2$ and $Q_{\rm II}=q_2/2$, so these measures are just scaled versions of the Cartesian coordinates of ${\bf q}$. The rotational constraint $\gamma$ \cite{Zhang2019} also corresponds to a Cartesian coordinate along a rotated coordinate axis aligned with the line joining the origin and the point mass $\Lambda_1$; its corresponding second measure, which would be the complementary Cartesian coordinate, would be proportional to $\lambda_2-\lambda_3$, which characterizes the rotational asymmetry of the wobble. 
This barycentric construction illustrates why many authors have chosen to represent the space of nonparaxial polarization in terms of equilateral triangles or segments of them \cite{Sheppard2012,Saastamoinen2004,Bjork2014,Bosyk2017,Greffet2007}. Other authors have used spheres \cite{Sheppard2012,Qian2011}, because triangles can be mapped onto octants of the sphere.
 
\subsection{Other measures inspired by measurements}
We conclude this section by noting that other measures of polarization have been proposed that are inspired by thought experiments. We briefly describe two of these:

\subsubsection{Measured inspired by interferometry}
A measure of polarization was defined \cite{Bjork2014} that is physically inspired by the maximization of visibility in interferometric setups. Mathematically, this measure relies on the concept of ``distance'' between two matrices. There are several definitions of this type of distance, but a simple and intuitive one is the Hilbert-Schmidt distance, given by the square root of one half of the trace of the square of the difference of the two matrices:
\begin{align}
D_{\rm HS}(\mathbf{\Gamma},\mathbf{\Gamma}')=\sqrt{\frac{{\rm Tr}[(\mathbf{\Gamma}-\mathbf{\Gamma}')^2]}2}.
\end{align}
The degree of polarization is then defined as the maximum distance between the polarization matrix and any meaningful transformation of it:
\begin{align}
P_{\rm HS}=\max_gD_{\rm HS}(\mathbf{\Gamma},\mathbf{R}_g\mathbf{\Gamma}\mathbf{R}_g^\dagger),
\end{align}
where $\mathbf{R}_g$ are matrices that define meaningful transformations. In particular, the authors consider all possible unitary transformations, which allow a simplification of the result. Incidentally, the resulting measure coincides with $P_{\rm I}$ when $\lambda_1=\lambda_2$, for which they both take the value $\lambda_1-\lambda_3$ (which happens to be twice the value of $\gamma$ and half the value of $Q_{\rm II}$, and for which $P_{\rm II}$ vanishes). 
 
\subsubsection{Measure based on averaged projections through Rayleigh scattering}
A different measure of polarization was proposed based on the idea of Rayleigh scattering \cite{Dennis2007}. Suppose that a small scatterer is placed at the point where polarization is to be measured and the far field scattered by the particle (in the Rayleigh regime) is collected and polarimetrically characterized. For each scattering direction, the measured field is paraxial with respect to that direction, so $P_{\rm 2D}$ can be used to characterize the degree of polarization in the corresponding direction. These directional degrees of polarization can then be averaged over all directions, weighted by their corresponding radiant intensity. It turns out that the directional integrals can be evaluated in closed form if one uses instead the square root of the directional average of the square of the product of the 2D degree of polarization and the radiant intensity. The resulting measure then takes the form
\begin{align}
P_{\rm RS}=\sqrt{\frac{11{\rm Tr}\mathbf{\Gamma}^2-4({\rm Tr}\mathbf{\Gamma})^2+{\rm Tr}(\mathbf{\Gamma}\mathbf{\Gamma}^*)}{{\rm Tr}\mathbf{\Gamma}^2+6({\rm Tr}\mathbf{\Gamma})^2+{\rm Tr}(\mathbf{\Gamma}\mathbf{\Gamma}^*)}},
\end{align}
Unlike all other measures discussed in this section, $P_{\rm RS}$ cannot be expressed purely in terms of the eigenvalues $\lambda_i$, because it is not invariant to unitary transformations.  As mentioned already, though, this lack of invariance to general unitary transformations should not be seen as problematic, since such invariance does not carry the physical importance for nonparaxial fields as it does for paraxial beams, because general unitary transformations cannot be associated with the action of simple optical elements. Let us stress that this is not a current technological limitation, but a fundamental one, because the field in general does not have a well-defined direction of propagation: 
a given local polarization matrix can be achieved through extremely different combinations of (traveling and/or evanescent) plane waves, and it is hard to envision a device that would cause the same local unitary transformation independently of the more global behavior of the field. %
Despite this qualitative difference, $P_{\rm RS}$ has been shown \cite{Petruccelli2010} to take very similar numerical values as $P_{\rm I}$, since the following inequality is always satisfied:
\begin{align}
P_{\rm I}\le P_{\rm RS}\le\sqrt{\frac65}\,P_{\rm I}=1.095\,P_{\rm I},
\end{align}
so these two measures never differ by more than $10\%$.

\section{Stokes-Gell-Mann parameters}
\label{sec:Stokes}
The natural extension of the Pauli matrices to the $3\times3$ case are the Gell-Mann matrices, which were defined within the context of particle physics \cite{GellMann}. These eight matrices, 
supplemented by the $3\times3$ identity, constitute a complete orthonormal basis (under the trace of the product) for $3\times3$ Hermitian matrices. They have been used in the context of optical polarization to decompose the polarization matrix, therefore providing a generalization for the concept of the Stokes parameters into the nonparaxial regime, where nine parameters are needed \cite{Holm1988,Brosseau1998,GilOssikovski2016,Carozzi2000,Ramachandran1980,SheppardPRA2014}. (The corresponding generalization of Mueller's calculus for describing polarization transformations then requires $9\times9$ Mueller matrices \cite{SheppardJOSAA2016,Krouglov2019}). The goal of this section is to not only review these parameters but also to propose an intuitive convention for them within this context. The numbering scheme and sign conventions used here for the Gell-Mann matrices and the resulting parameters are then different to those in other publications, in order to stress the connections with the paraxial case. 

\subsection{Definition of the parameters}
Rather than writing here each Gell-Mann matrix separately, we directly write the $3\times3$ polarization matrix $\mathbf{\Gamma}$ as a linear combination of these matrices:
\begin{align}
\mathbf{\Gamma}=\frac12\left(\begin{array}{ccc}\frac23S_0+S_{11}+\frac{S_{12}}{\sqrt{3}}&S_{23}-\ui S_{33}&S_{22}+\ui S_{32}\\S_{23}+\ui S_{33}&\frac23S_0-S_{11}+\frac{S_{12}}{\sqrt{3}}&S_{21}-\ui S_{31}\\ S_{22}-\ui S_{32}&S_{21}+\ui S_{31}&\frac23S_0-2 \frac{S_{12}}{\sqrt{3}}\end{array}\right),\label{Gamma}
\end{align}
where the parameters $S_0$ and $S_{mn}$ are the nine nonparaxial analogs of the Stokes parameters, referred to here as the Stokes-Gell-Mann parameters. To extract the Gell-Mann matrix associated with each parameter, we simply set this parameter to 2 and the others to zero in the expression above. The explicit form for these matrices as well as some of their properties are given in Appendix~\ref{appE}. It is worth noting a difference between the Pauli and the Gell-mann matrices: while the Pauli matrices have the same norm (defined as the square root of the trace of their square) as the $2\times2$ identity matrix used to complete the set, namely $\sqrt{2}$, the same is not true for the Gell-mann matrices (with norm $\sqrt{2}$) and the $3\times3$ identity (with norm $\sqrt{3}$). There are therefore different possible conventions for the normalization factors of $S_0$ with respect to the others; the reason for the choice used here will become apparent in what follows. 

We now describe the different subsets of parameters. 
First, as in the paraxial case, the parameter $S_0$ 
equals the local intensity:
\begin{align}
S_0=
{\rm Tr}\mathbf{\Gamma}=
\langle|{\bf E}|^2\rangle;
\end{align}
the parameters $S_{1m}$ characterize discrepancies amongst the diagonal terms of the polarization matrix:
\begin{subequations}
\begin{align}
S_{11}&=\langle |E_x|^2\rangle-\langle|E_y|^2\rangle,\\
S_{12}&=\frac{\langle |E_x|^2\rangle+\langle|E_y|^2\rangle-2\langle|E_z|^2\rangle}{\sqrt{3} };
\end{align}
\end{subequations}
the parameters $S_{2m}$ characterize the real parts of the correlations between the different Cartesian components:
\begin{subequations}
\begin{align}
S_{21}&=2\,{\rm Re}\langle E_y^*E_z\rangle,\\
S_{22}&=2\,{\rm Re}\langle E_z^*E_x\rangle,\\
S_{23}&=2\,{\rm Re}\langle E_x^*E_y\rangle;
\end{align}
\end{subequations}
and the parameters $S_{3m}$ characterize the imaginary parts of the correlations between the different Cartesian components:
\begin{subequations}
\begin{align}
S_{31}&=2\,{\rm Im}\langle E_y^*E_z\rangle,\\
S_{32}&=2\,{\rm Im}\langle E_z^*E_x\rangle,\\
S_{33}&=2\,{\rm Im}\langle E_x^*E_y\rangle.
\end{align}
\end{subequations}

Let us make a few observations about these definitions:
\begin{itemize}
\item Let us start with the two elements $S_{1m}$. In the paraxial treatment where the matrix is $2\times2$, there is a natural choice for the measure of discrepancy between the two diagonal elements, corresponding to their difference, the only marginally ``nondemocratic'' choice being that of which diagonal element is subtracted from which in the definition of $S_1$. For $3\times3$ matrices, on the other hand, there is no natural choice of two parameters that treats the three diagonal elements equally: we can see that the third diagonal element in Eq.~(\ref{Gamma}) has a different form than the other two. The form for the diagonal elements can be made to look more natural by grouping the two Stokes-Gell-Mann parameters $S_{1m}$ in a two-vector ${\bf S}_1=(S_{11},S_{12})$; the three diagonal elements of Eq.~(\ref{Gamma}) can then be written concisely as $S_0/3-{\bf u}_m\cdot{\bf S}_1/\sqrt{3}$ where ${\bf u}_m=(\sin\theta_m,\cos\theta_m)$ with $\theta_m =-m\,2\pi/3$ for $m=1,2,3$. Note that we could have chosen any other set of three unit vectors ${\bf u}_m$ that are equally spaced angularly, so that their sum vanishes and the trace of the matrix is $S_0$. The choice that is implicit in the definition of the Gell-Mann matrices is the alignment of the vector ${\bf u}_3$, corresponding to the matrix element $\mathbf{\Gamma}_{zz}$, with one of the axes within the plane of ${\bf S}_1$. From the point of view of optical fields, this arbitrary choice can be justified by the fact that the $z$ axis is often associated with the main direction of propagation and is hence perhaps special. In other words, this choice lets $S_{11}$ take the same form as the paraxial Stokes parameter $S_1$.
\item The three Stokes-Gell-Mann parameters $S_{2m}$ are a measure of the misalignment between the chosen Cartesian coordinate axes and the natural axes of the ellipsoid of inertia. It is convenient to group these elements in a three-vector ${\bf S}_2=(S_{21},S_{22},S_{23})$, even though it must be stressed that this is a vector in an abstract space, not in the physical 3D space.
\item The last three Stokes-Gell-Mann parameters, $S_{3m}$, can also be grouped in a vector as ${\bf S}_3=(S_{31},S_{32},S_{33})$. However, notice that this is precisely the local spin density vector of the field shown in Fig.~\ref{fig:Mark}, namely ${\bf S}_3={\rm Im}\langle{\bf E}^*\times{\bf E}\rangle=\pmb{\mathsf S}$. Therefore (unlike ${\bf S}_1$ and ${\bf S}_2$), ${\bf S}_3$ is truly a (pseudo)vector in the physical coordinate system. 
\item Note that, if only the $x$ and $y$ components of the field are significant, the Stokes-Gell-Mann parameters $S_0,S_{11},S_{23},S_{33}$ reduce to the standard Stokes parameters for paraxial fields, while the parameter $S_{12}$ becomes redundant with $S_0$ and the remaining ones vanish.
\end{itemize}

\subsection{Normalized parameters and eight-dimensional polarization space}
As in the paraxial case, we define a normalized set of parameters as $s_{nm}=(\sqrt{3}/2)
S_{nm}/S_0$, which are then independent of the intensity. The reason for the extra numerical factor will become apparent in what follows. These eight normalized parameters can be used to define a polarization vector in an eight-dimensional abstract space:
\begin{align}
\vec s=(s_{11},s_{12},s_{21},s_{22},s_{23},s_{31},s_{32},s_{33}).
\end{align}
It can be shown that the Euclidean magnitude of the eight-component vector $\vec s$ corresponds precisely to the measure of degree of polarization in Eq.~(\ref{DoPSampson}):
\begin{align}
P_{\rm I}
=\sqrt{\sum_{nm}s_{nm}^2}=|\vec s|.\label{DoP}
\end{align}
Because $P_{\rm I}$ is constrained to the interval $[0,1]$, ${\vec s}$ is constrained to the interior and hypersurface of a unit hypersphere in eight dimensions (a 7-ball). Full polarization then corresponds to the hypersurface of this 8D hypervolume, namely to a 7D manifold. This suggest that the local description of a fully polarized field requires the specification of seven parameters. This is not the case, however, as it was established in Section~\ref{sec:FP3D} that only four parameters are required to describe full polarization. 
Therefore, not all points inside the unit 7-ball are inhabitable, so several constraints limit the true accessible hypervolume \cite{BrosseauProgOpt2006,Gamel2012,Ramachandran1980,Bloore1976,Kimura2003} which is inscribed in the 7-ball. The inequalities that shape the inhabitable region are described later in this section. 

Before determining the shape of the space, however, let us illustrate another difference with the paraxial case by extending the discussion in Section~\ref{sec:similarity} to the nonparaxial regime. Let us consider two fully polarized fields given by the complex three-vectors ${\bf E}^{\rm (I)}$ and ${\bf E}^{\rm (II)}$. We again use the angle $\alpha$ to characterize their similarity by using the definition in Eq.~(\ref{alphasim}): 
\begin{align}
\cos^2\alpha=\frac{|{\bf E}^{\rm (I)*}\cdot{\bf E}^{\rm (II)}|^2}{|{\bf E}^{\rm (I)}|^2|{\bf E}^{\rm (II)}|^2}=\frac{{\rm Tr}\left[\mathbf{\Gamma}^{\rm (I)}\mathbf{\Gamma}^{\rm (II)}\right]}{{\rm Tr}\left[\mathbf{\Gamma}^{\rm (I)}\right]{\rm Tr}\left[\mathbf{\Gamma}^{\rm (II)}\right]}.
\end{align}
Again, if both fields are linearly polarized, $\alpha$ represents the angle between these polarizations. By writing the polarization matrices in terms of the normalized Stokes-Gell-Mann parameters and simplifying, we get the relation
\begin{align}
\cos^2\alpha=\frac{1+2\vec s^{\rm (I)}\cdot\vec s^{\rm (II)}}3,
\end{align}
A few observations can be made from this result. First, unlike for the paraxial case, the relation between $\alpha$ and the angle between the normalized Stokes-Gell-Mann vectors is not linear. Second, while $\alpha=0$ indeed corresponds to parallel normalized Stokes-Gell-Mann vectors, orthogonal polarizations ($\alpha=\pi/2$) do not correspond to antiparallel normalized Stokes-Gell-Mann vectors, but to $\vec s^{\rm (I)}\cdot\vec s^{\rm (II)}=-1/2$. Because for fully polarized fields the norm of the normalized Stokes-Gell-Mann vectors is unity, two orthogonal polarization states have Stokes-Gell-Mann vectors that are at $120^\circ$ from each other. The fact that orthogonal polarizations do not correspond to antiparallel (or antipodal) normalized vectors is consistent with the fact that each polarization has not a unique orthogonal polarization state but a two-parameter continuum of them. Further, even for partially-polarized fields the inequality $1\ge\vec s^{\rm (I)}\cdot\vec s^{\rm (II)}\ge-1/2$ holds, showing that indeed not all regions of the hypersphere's interior are inhabitable. In particular, if a normalized Stokes-Gell-Mann vector $\vec s$ with unit magnitude represents a physical fully polarized state, a large segment of the hypersphere's interior and surface surrounding its antipode $-\vec s$ does not correspond to physical states.

\subsection{Some inequalities constraining the normalized Stokes-Gell-Mann parameters}
For paraxial light all three standard normalized Stokes parameters play very similar roles, but this is not true for the normalized Stokes-Gell-Mann parameters. It is convenient to separate these into the three normalized Stokes-Gell-Mann sub-vectors ${\bf s}_1=(s_{11},s_{12})$, ${\bf s}_2=(s_{21},s_{22},s_{23})$ and ${\bf s}_3=(s_{31},s_{32},s_{33})$, the latter being proportional to the normalized spin density, ${\bf s}_3=\sqrt{3}\,\pmb{\mathsf s}/2$. 

Let us start by considering the diagonal elements of Eq.~(\ref{Gamma}), which limit the values of the sub-vector ${\bf s}_1$. Since the polarization matrix is non-negative definite, these elements must be equal to or greater than zero, leading to the three inequalities
\begin{align}
{\bf u}_m\cdot{\bf s}_1\le\frac12,\,\,\,\,m=1,2,3,
\label{inplane}
\end{align}
where as before ${\bf u}_m=(\sin\theta_m,\cos\theta_m)$ with $\theta_m =-m\,2\pi/3$. 
These inequalities imply that the sub-vector ${\bf s}_1=(s_{11},s_{12})$ is constrained to an equilateral triangle inscribed within the unit disk \cite{Bloore1976}. This restriction to a triangle should no longer be surprising: if the other two Stokes-Gell-Mann subvectors vanished (${\bf s}_2={\bf s}_3={\bf 0}$), the polarization matrix would be diagonal so that its three diagonal elements would correspond to the eigenvalues $\Lambda_i$. The center-of mass interpretation would then give directly ${\bf s}_1={\bf q}$, but with the diagonal elements (or eigenvalues) not necessarily being ordered from largest to smallest, so that the whole equilateral triangle in Fig.~\ref{fig:CoM} would be inhabitable.

\begin{figure}
\centering
\includegraphics[scale=0.5]{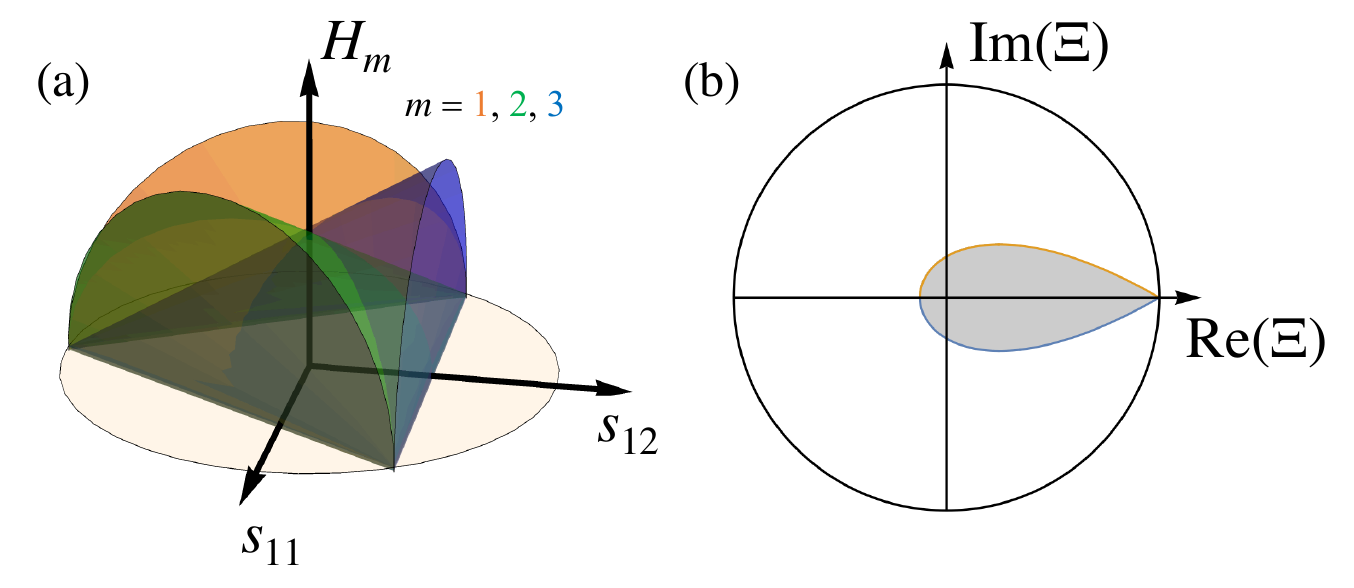}
\caption{(a) Inhabitable hypervolumes (hypercones) over the subspaces $s_{11},s_{12},s_{2m},s_{3m}$ for $m=1,2,3$: for a given coordinate $(s_{11},s_{12})$, each magnitude $\sqrt{s_{2m}^2+s_{3m}^2}$ must be equal to or smaller than the height $H_m$ of the corresponding cone at that point. (b) Inhabitable region for the complex quantity $\Xi$ within the unit complex disk.}
\label{fig:3D}
\end{figure}

We now consider inequalities that apply to each non-diagonal element of the matrix. From the Cauchy-Bunyakowsky-Schwarz inequality it follows directly that correlation matrices satisfy $|\Gamma_{ij}|^2\le\Gamma_{ii}\Gamma_{jj}$ for $i,j=x,y,z$, the equality holding only when $E_i$ and $E_j$ are fully correlated. The resulting three inequalities can be written concisely as
\begin{align}
({\bf v}_m\cdot{\bf s}_1)^2+s_{2m}^2+s_{3m}^2\le\frac{(1+{\bf u}_m\cdot{\bf s}_1)^2}3,
\label{cones}
\end{align}
where ${\bf v}_m=(\cos\theta_m,-\sin\theta_m)$. Each of these relations implies a restriction in a 4D subspace $(s_{11},s_{12},s_{2m},s_{3m})$ to the interior of a section of a hypercone, as represented in Fig.~\ref{fig:3D}(a). These three hypervolumes inhabit different subspaces, but they intersect at the plane $(s_{11},s_{12})$ where they all have a cross-section corresponding to the equilateral triangle. That is, these three inequalities restrict $(s_{11},s_{12})$ to the same region as those in (\ref{inplane}), but provide stronger limitations involving also other Stokes-Gell-Mann parameters. It is easy to see that the sum of the three inequalities in (\ref{cones}) gives, after some rearrangement, $\sum_{mn}s_{mn}^2\le1$, so that these restrictions are sufficient to constrain $\vec s$ to a region that is fully inside the unit 7-ball. The constraints in (\ref{cones}) can also be written as
\begin{align}
|s_{2m}+\ui s_{3m}|\le H_m({\bf s}_1),\,\,\,\,\,\,\,\,H_m({\bf s}_1)=\sqrt{(1+{\bf u}_m\cdot{\bf s}_1)^2/3-({\bf v}_m\cdot{\bf s}_1)^2},
\label{cones2}
\end{align}
where the three functions $H_m({\bf s}_1)$ are the heights of each of the cones at each point $(s_{11},s_{12})$, as shown in Fig.~\ref{fig:3D}(a). 

The three constraints in relations (\ref{cones}) or (\ref{cones2}) already limit significantly the hypervolume inhabitable by $\vec s$ to $9\sqrt{3}\/\pi^3/1120$, which is about $10.6\%$ of the unit 7-ball's interior of $\pi^4/24$. These inequalities can be supplemented with a fourth (non-tight) inequality that restricts the phases of the off-diagonal elements. By using a result found in Appendix~\ref{appE}, we find
\begin{align}
27[{\rm Im}(\Xi)]^2\le[1-{\rm Re}(\Xi)]^2[1+8\,{\rm Re}(\Xi)],\label{phase}
\end{align}
where the complex quantity $\Xi$ is defined as
\begin{align}
\Xi=\frac{\Gamma_{yx}\Gamma_{zy}\Gamma_{xz}}{\Gamma_{xx}\Gamma_{yy}\Gamma_{zz}}=\prod_{m=1}^3\frac{(s_{2m}+\ui s_{3m})}{H_m({\bf s}_1)}.\label{Xi}
\end{align}
This range of possible values for $\Xi$ is shown in Fig.~\ref{fig:3D}(b). Note that ${\rm Re}(\Xi)\in[-1/8,1]$, ${\rm Im}(\Xi)\in[-1/4,1/4]$. 
Let us define $h_m=\sqrt{s_{2m}^2+s_{3m}^2}$ (namely, the heights in Fig.~\ref{fig:3D}(a)) and $\phi_m={\rm arg}(s_{2m}+\ui s_{3m})$. 
We can then rewrite $\Xi$ as
\begin{align}
\Xi&=\frac{h_1h_2h_3}{H_1H_2H_3}\exp[\ui(\phi_1+\phi_2+\phi_3)].
\label{Xi1}
\end{align}
The inequality in (\ref{phase}) can then be expressed as a constraint on the phases of the off-diagonal matrix elements:
\begin{align}
\cos(\phi_1+\phi_2+\phi_3)\ge{\rm Max}\left[-1,\frac{3|\Xi|^{2/3}-1}{2|\Xi|}\right].\label{phase2}
\end{align}
That is, the value of $\phi_1+\phi_2+\phi_3$ is constrained only for $1/8<|\Xi|\le1$. 

\subsection{Rigorous relations for the inhabitable region}
\label{subsRI}
\begin{figure}
\centering
\includegraphics[scale=0.25]{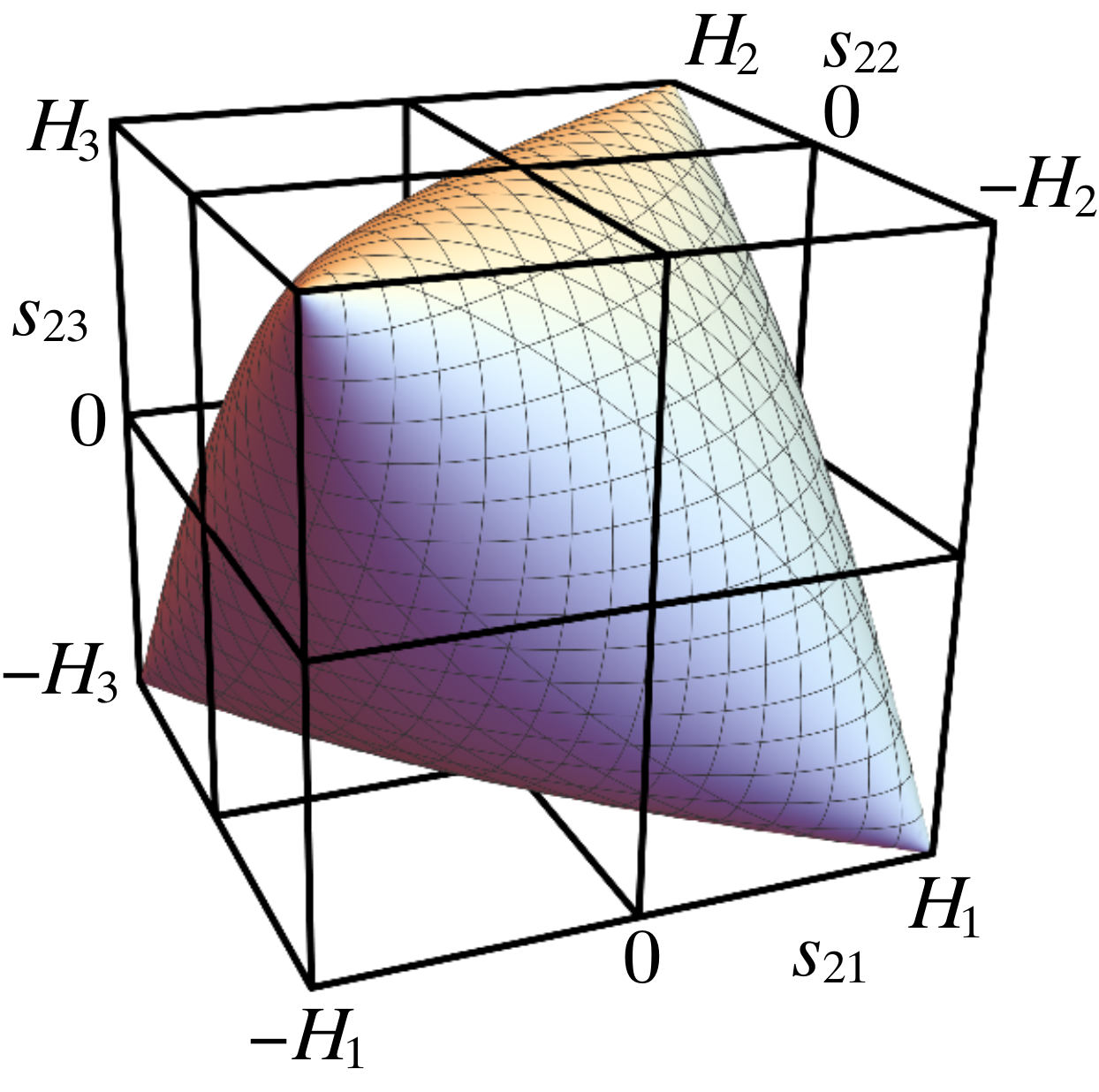}
\caption{Inhabitable region for the Stokes-Gell-Mann sub-vector ${\bf s}_2$, according to relation (\ref{fatth}).}
\label{fig:fatth}
\end{figure}

It turns out that the three constraints in relations (\ref{cones2}) plus the one in (\ref{phase2}) are sufficient to reduce from eight to four the number of free parameters in the limit of full polarization, since fully polarized fields must be at the four boundaries. Surprisingly, however, away from this limit these inequalities are not sufficiently strong to provide the true shape of the accessible hypervolume for $\vec s$. The rigorous inequalities result from making sure that the three eigenvalues of $\mathbf{\Gamma}$ are non-negative. 
Note that, while $\Lambda_i\ge0$ for $i=1,2,3$ implies that $\det(\mathbf{\Gamma})\ge0$, the converse is not necessarily true. Therefore, the inequality in relation (\ref{ellipsoidforS3}), which results from enforcing $\det(\mathbf{\Gamma})\ge0$, is in itself not sufficient. 
Nevertheless, the physical interpretation for relation (\ref{ellipsoidforS3}) in Section 3 provides a useful hint: the eigenvalues of ${\rm Re}(\mathbf{\Gamma})$ must also be non-negative so that the ellipsoid of inertia is well defined. If the diagonal elements of the matrix are guaranteed to be non-negative by constraining ${\bf s}_1$ to the triangle in Fig.~\ref{fig:CoM}, then the eigenvalues of ${\rm Re}(\mathbf{\Gamma})$ are non-negative as long as the following inequality is satisfied:
\begin{align}
\sum_{m=1}^3\frac{s_{2m}^2}{H_m^2({\bf s}_1)}-2\prod_{m=1}^3\frac{s_{2m}}{H_m({\bf s}_1)}\le1.\label{fatth}
\end{align}
This relation constrains ${\bf s}_2$ to the surface and interior of shape shown in Fig.~\ref{fig:fatth}, described by Bloore as an over-inflated tetrapack \cite{Bloore1976}. Note that all cross-sections of this shape in which one of the parameters $s_{2m}$ is fixed correspond to ellipses in the remaining two parameters. This volume is inscribed in a box defined by $|s_{2m}|\le H_m({\bf s}_1)$ implied by the inequalities in relation~(\ref{cones2}). 

An ordered way to determine the true boundaries of the space of the normalized Stokes-Gell-Mann parameters is the following: 
\begin{itemize}
\item The subvector ${\bf s}_1$ is constrained to the triangle in Fig.~\ref{fig:CoM} following relation (\ref{inplane});
\item The subvector ${\bf s}_2$ is constrained to the 3D shape described in relation (\ref{fatth}) and shown in Fig.~\ref{fig:fatth}, whose dimensions are fixed by ${\bf s}_1$ through the functions $H_m({\bf s}_1)$ (that correspond to the heights of the three cones in Fig.~\ref{fig:3D}(a) at the corresponding point); 
\item The subvector ${\bf s}_3$ is constrained to the normalized version of the dual ellipsoid in Fig.~\ref{fig:Mark} following the inequality in relation (\ref{ellipsoidforS3}) with $\pmb{\mathsf S}=2S_0{\bf s}_3/\sqrt{3}$, whose shape is determined by both ${\bf s}_1$ and ${\bf s}_2$ through the construction of ${\rm Re}(\mathbf{\Gamma})$ following Eq.~(\ref{Gamma}). 
\end{itemize}
These three constraints imply that $\vec s$ is restricted to a 8D hyper-volume much smaller than $\pi^4/24$, the interior of the unit hypersphere. By using the volume of the ellipsoid inhabitable by ${\bf s}_3$ (which depends on ${\bf s}_1$ and ${\bf s}_2$), integrating it in ${\bf s}_2$ over the volume in Fig.~\ref{fig:fatth} (which depends on ${\bf s}_1$), and integrating the result in ${\bf s}_1$ over the surface of the triangle in Fig.~\ref{fig:CoM}, a closed form result of $567\pi^3/163840$ is found, which constitutes only $2.64\%$ of the interior of the unit hypersphere. This analytic result was corroborated by numerical Monte-Carlo integration. 

Note that none of the constraints described in this section impose a restriction on the normalized Stokes-Gell-Mann vector $\vec s$ within the inner region $|\vec s|\le1/2$. A simple proof of this fact follows from using the formula found by Sheppard \cite{Sheppard2011,Sheppard2012} for the eigenvalues $\lambda_i$ in terms of the measures $P_{\rm I}=|\vec s|$, which can be written as
\begin{align}
\lambda_i=\frac13\left[1+2P_{\rm I}\cos\left(\psi+\frac{2\pi i}3\right)\right],
\end{align}
where $\sin\psi=(\sqrt{3}/2)P_{\rm II}/P_{\rm I}$, with $P_{\rm II}$ as defined in Eq.~(\ref{DoPE}). It is clear that for $P_{\rm I}=|\vec s|\le1/2$ all eigenvalues are non-negative. Therefore, the restrictions found here enforce constraints only in the region of the 8D space for which $1/2<|\vec s|\le1$, i.e., for states with considerable polarization. Even for the approximate inequalities in relations (\ref{inplane}) and (\ref{cones}) it is clear from their geometric interpretation in Figs.~\ref{fig:3D}(a) that a sphere of radius $1/2$ centered at the origin fits completely within all the restricted volumes. While less evident, this is also true for the inequality in relation (\ref{phase2}), as shown in Appendix~\ref{appF}. That is, a part of the $2.64\%$ of the interior of the 8D hypersphere inhabitable by physical polarization states corresponds to the $1/2^8\approx0.4\%$ occupied by the central hypersphere of radius $1/2$ that represents significantly unpolarized light.

\subsection{Limit of full polarization}
Let us finish this section by discussing the limit of full polarization, corresponding to completely deterministic fields. 
The fact that fully polarized fields involve only four degrees of freedom could naively be interpreted as the result of imposing four constraints to the eight parameters: the vector $\vec s$ must lie at the hypersurface $|\vec s|=1$, and each of the three sub-vectors ${\bf s}_n$ must lie at the boundary of its inhabitable regions. It turns out, however, that the three constraints for the sub-vectors are sufficient to guarantee $|\vec s|\le1$. That is, these three constraints define a region in the 8D space that does not cross the unit hypersphere but simply touches it; this contact region defines the four-parameter subspace of full polarization. This limit can be better understood in the following way: let ${\bf s}_1$ be at any point within its allowed triangular area, and let ${\bf s}_2$ be at any point over the surface of its allowed volume shown in Fig.~\ref{fig:fatth}; it is then clear that $|\vec s|^2=|{\bf s}_1|^2+|{\bf s}_2|^2+|{\bf s}_3|^2$ achieves its maximum value of unity when $|{\bf s}_3|^2$ is maximal given its constraints, so it is not enough that ${\bf s}_3$ be at the surface of the dual ellipsoid, but it must be at one of its two vertices, corresponding to the intersection of the dual ellipsoid with its major axis. All other points over the surface of the dual ellipsoid constitute boundaries between physical and unphysical states but do not correspond to full polarization. The four degrees of freedom that parametrize states of full polarization can then be thought of as the two coordinates ${\bf s}_1$ and the position of ${\bf s}_2$ over the surface of its allowed region; ${\bf s}_3$ is then fully determined to within a sign, each choice corresponding to a sense of circulation of the same geometric ellipse.

\section{Representation of 3D partial polarization as a collection of vectors/points}
\label{threepointrep}
The inequalities just described illustrate the similarities and the important differences between the characterizations of partial polarization in the paraxial and nonparaxial regimes. For the paraxial case, the state of polarization can be fully represented by a single point within a spherical region of an abstract three-dimensional space, where the surface of the region corresponds to states of full polarization. Further, physically meaningful transformations of the state of polarization that preserve the amount of light correspond to simple rotations of the Poincar\'e sphere. For the nonparaxial case, on the other hand, the normalized Stokes-Gell-Mann parameters provide a description that is considerably more challenging to visualize: the state of polarization is represented as a single point in an abstract eight-dimensional space with a complex irregular shape, and where only a subset of the edges of the region correspond to states of full polarization.  
Additionally, the physically meaningful unitary transformations for nonparaxial light are rotations and inversions of the coordinate system (instead of all complex unitary transformations), and while these transformations correspond to linear transformations of the 8D vector $\vec s$, these are not particularly intuitive. 

The goal in this section is to propose a complete representation of partial polarization in a three-dimensional space in order to facilitate visualization. Further, by making the three axes in this space correspond to the physical Cartesian coordinates, the effect of rotations and inversions of the coordinate system becomes trivial. Note that the interpretation \cite{Dennis2004} in terms of the ellipsoid of inertia supplemented by the spin density vector fits within this category, as discussed in Section~\ref{section:ellipsoid}. However, the idea here is to use instead a collection of vectors or points that allow an interpretation that is independent of the coordinate system, similarly to the two-point representations for full polarization. Because partial polarization corresponds to eight parameters, at least three points or vectors are required, with at least one constraint. There are probably several ways to choose one such representation, but the construction proposed in what follows aims to minimize the number of points (three, with one constraint), to avoid discontinuous jumps for the coordinates of these points under small perturbations of the field, and to make the position of the points unique (other than perhaps for a sign ambiguity).

Note that the normalized spin density $\pmb{\mathsf s}$ can be chosen as the coordinates of one of the three points of the proposed description, since it fully encodes the imaginary part of the polarization matrix and its components are associated with the spatial coordinate axes, as desired. Further, this vector is by construction constrained to a unit ball, its magnitude taking the maximum value of unity only for fully polarized circular polarization. 

\subsection{Vectors for representing the ellipsoid of inertia}
The remaining two points or vectors must represent the real part of the polarization matrix, ${\rm Re}(\mathbf{\Gamma})$, whose normalized version depends on five degrees of freedom. This matrix is fully characterized by its eigenvalues $\bar{\Lambda}_n$ and eigenvectors $\bar{\bf e}_n$, defined by ${\rm Re}(\mathbf{\Gamma})\bar{\bf e}_n=\bar{\Lambda}_n\bar{\bf e}_n$ with $\bar{\Lambda}_1\ge\bar{\Lambda}_2\ge\bar{\Lambda}_3$. Note that the eigenvalues $\bar{\Lambda}_n$ are in general not the same as $\Lambda_n$, the eigenvalues of $\mathbf{\Gamma}$. However, because the imaginary components of the polarization matrix are all within its nondiagonal elements, the traces of $\mathbf{\Gamma}$ and ${\rm Re}(\mathbf{\Gamma})$ are the same, and therefore $\bar{\Lambda}_1+\bar{\Lambda}_2+\bar{\Lambda}_3=\Lambda_1+\Lambda_2+\Lambda_3$. We then use the normalized eigenvalues $\bar{\lambda}_n=\bar{\Lambda}_n/(\Lambda_1+\Lambda_2+\Lambda_3)$, whose sum is unity. The eigenvectors $\bar{\bf e}_n$ are chosen to be real and of unit magnitude, so they are uniquely defined up to a global sign. (Strictly speaking they are then not vectors but {\it directors}, as they specify a direction but not a sense.) They are mutually orthogonal and correspond to the directions of the semi-axes of the ellipsoid of inertia illustrated in Fig.~\ref{fig:Mark}, where the lengths of the corresponding semi-axes are $\bar{\Lambda}_n^{1/2}$. 

\begin{figure}
\centering
\includegraphics[scale=0.4]{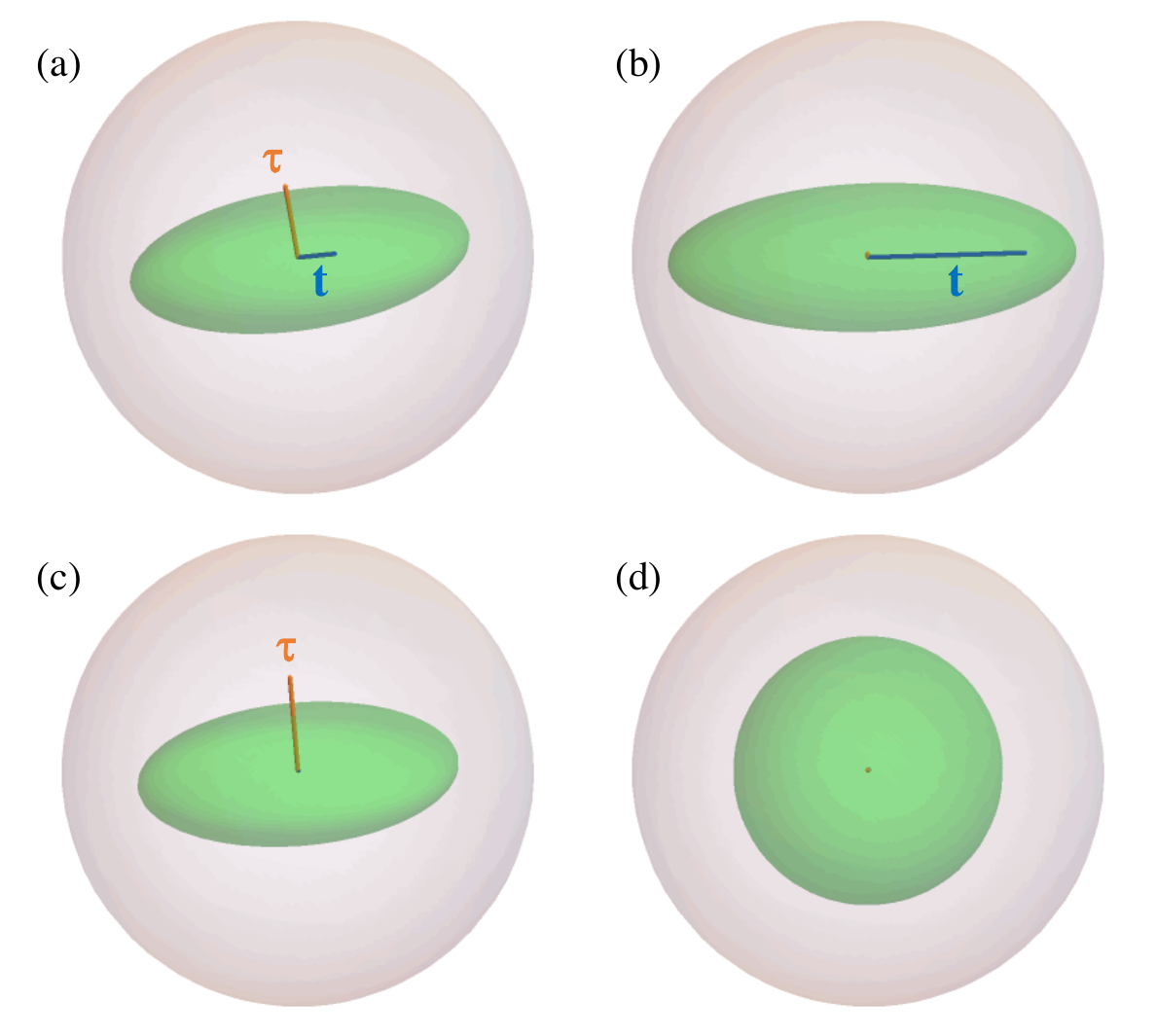}
\caption{Vectors ${\bf t}$ (blue) and $\pmb{\tau}$ (orange) for four different shapes of the ellipsoid of inertia (green): (a) a generic ellipsoid, (b) a prolate spheroid (for which $\pmb{\tau}$ vanishes), (c) an oblate spheroid (for which ${\bf t}$ vanishes), and (d) a sphere (for which both vectors vanish). The ellipoids of inertia in this figure are normalized, so their size is compared to that of the unit sphere (white).}
\label{fig:ellipsoids}
\end{figure}

It is natural to chose the coordinates of the two remaining points as proportional to two of the eigenvectors. This choice would already include a restriction since these vectors are orthogonal. The lengths of the two vectors must encode the normalized eigenvalues. Note that the choice of the eigenvectors becomes non-unique in cases of degeneracy, namely when $\bar{\lambda}_1=\bar{\lambda}_2$ (in which a case the choice of $\bar{\bf e}_1$ and $\bar{\bf e}_2$ is not unique) or when $\bar{\lambda}_2=\bar{\lambda}_3$ (in which a case the choice of $\bar{\bf e}_2$ and $\bar{\bf e}_3$ is not unique). These possible ambiguities can be avoided by choosing the length of each of the vectors to vanish when a degeneracy would cause its direction not to be well defined, and their directions to also remain well defined in case of degeneracy.
The two vectors are then chosen as
\begin{subequations}
\begin{align}
{\bf t}&=(\bar{\lambda}_1-\bar{\lambda}_2)\bar{\bf e}_1,\\
\pmb{\tau}&=2(\bar{\lambda}_2-\bar{\lambda}_3)\bar{\bf e}_3.
\end{align}
\label{eq:tvectors}\end{subequations}
Note that the direction of $\bar{\bf e}_2$ is not used since this eigenvector becomes undefined for both types of degenerate situations. To understand the connection of this construction with the ellipsoid of inertia, let us consider the different types of ellipsoid, as shown in Fig.~\ref{fig:ellipsoids}: (a) for a general non-degenerate ellipsoid ($\bar{\lambda}_1>\bar{\lambda}_2>\bar{\lambda}_3$), ${\bf t}$ points in the direction of the major axis and $\pmb{\tau}$ in the direction of the minor axis; (b) for a prolate spheroid ($\bar{\lambda}_1>\bar{\lambda}_2=\bar{\lambda}_3$), $\pmb{\tau}$ vanishes and ${\bf t}$ points in the direction of the axis of rotational symmetry, and in the limit of a very elongated ellipsoid $|{\bf t}|$ tends to unity; (c) for an oblate spheroid ($\bar{\lambda}_1=\bar{\lambda}_2>\bar{\lambda}_3$), ${\bf t}$ vanishes and $\pmb{\tau}$ points in the direction of the axis of rotational symmetry, and in the limit of a very flat ellipsoid $|\pmb{\tau}|$ tends to unity; (d) for a sphere ($\bar{\lambda}_1=\bar{\lambda}_2=\bar{\lambda}_3$), both vectors vanish. 

\subsection{Inequalities for the three vectors}
A complete description of the state of polarization is then given by the two orthogonal vectors ${\bf t}$ and $\pmb{\tau}$ (each of which is uniquely defined up to a sign) and the spin density vector ${\bf s}$. None of these vectors can be larger than unity in magnitude, and their lengths are in fact significantly more restricted: for the two vectors that characterize the real part of the polarization matrix we have
\begin{subequations}
\label{eq:restrictions}
\begin{align}
0&\le|{\bf t}|\le1,\\
0&\le|\pmb{\tau}|\le1-|{\bf t}|,
\end{align}
while for ${\bf s}$ the expression follows from that in relation~(\ref{ellipsoidforS3}), and can be written as
\begin{align}
\sum_{n=1}^3\bar{\lambda}_n\,({\bf s}\cdot\bar{\bf e}_n)^2\le 4\bar{\lambda}_1\bar{\lambda}_2\bar{\lambda}_3.
\label{eq:ellipsoid2}
\end{align}
\end{subequations}

Further, it is easy to verify from Eqs.~(\ref{eq:tvectors}) that the following relation holds:
\begin{align}
|{\bf t}|^2+\frac{|\pmb{\tau}|^2}4+\frac{|{\bf t}||\pmb{\tau}|}2=\bar{\lambda}_1^2+\bar{\lambda}_2^2+\bar{\lambda}_3^2-\bar{\lambda}_1\bar{\lambda}_2-\bar{\lambda}_2\bar{\lambda}_3-\bar{\lambda}_3\bar{\lambda}_1=|{\bf s}_1|^2+|{\bf s}_2|^2.
\end{align}
The final part of this expression follows from the fact that, if the spin density vanishes, this expression coincides with the square of the degree of polarization in Eq.~(\ref{DoPSampson}) with $\lambda_i=\bar{\lambda}_i$, namely the squared norm of $\vec s$ following Eq.~(\ref{DoP}). Since the contribution of spin to the square of this degree of polarization is additive, we can write the following general expression for this degree of polarization in terms of the three vectors:
\begin{align}
P_{\rm I}&=\sqrt{|{\bf t}|^2+\frac{|\pmb{\tau}|^2}4+\frac{|{\bf t}||\pmb{\tau}|}2+|{\bf s}_3|^2}
=\sqrt{\frac{(|{\bf t}|+|\pmb{\tau}|)^2+3(|\pmb{\mathsf s}|^2+|{\bf t}|^2)}4}\le1,
\label{eq:Ppts}
\end{align}
where in the second step we used the fact that ${\bf s}_3=\sqrt{3}\,\pmb{\mathsf s}/2$. In the special case in which $|\pmb{\mathsf s}|$ vanishes and $\bar{\lambda}_2=\bar{\lambda}_3$, $P_{\rm I}=|{\bf t}|$, while if $|\pmb{\mathsf s}|$ vanishes and $\bar{\lambda}_1=\bar{\lambda}_2$, $P_{\rm I}=|\pmb{\tau}|/2$.

We now show that the following simple relationship between ${\bf t}$ and $\pmb{\mathsf s}$ also holds:
\begin{align}
|\pmb{\mathsf s}\pm{\bf t}|\le1,
\label{eq:tpluss}
\end{align}
with the equality holding only in the limit of full polarization, in which $\pmb{\mathsf s}$ and ${\bf t}$ are perpendicular. To prove this inequality, we take the extreme values of $\pmb{\mathsf s}$ from relation (\ref{eq:ellipsoid2}), parametrized as an ellipsoid, add this vector to ${\bf t}$, and take the norm squared. This norm squared is maximized over the ellipsoid, giving the expression $\bar{\lambda}_1[(1-\bar{\lambda}_3)^2-4\bar{\lambda}_2\bar{\lambda}_3]/(\bar{\lambda}_1-\bar{\lambda}_3)$, which takes a maximum value of unity when $\bar{\lambda}_3=0$. In this limit, $\pmb{\mathsf s}=\pm2\sqrt{\bar{\lambda}_1\bar{\lambda}_2}\,\bar{\bf e}_3$, which indeed maximizes relation (\ref{eq:ellipsoid2}) and hence guarantees full polarization. The corresponding fully polarized field is proportional to $\sqrt{\bar{\lambda}_1}\,\bar{\bf e}_1\pm\ui\sqrt{\bar{\lambda}_2}\,\bar{\bf e}_2$. Further, in this limit $\pmb{\tau}$ becomes superfluous since its direction is that of $\pmb{\mathsf s}$ and its magnitude reduces to $2\bar{\lambda}_2$ which is also equal to $1-|{\bf t}|$. Notice that this is consistent with relation (\ref{eq:Ppts}) (where in this case both $P_{\rm I}$ and $|\pmb{\mathsf s}|^2+|{\bf t}|^2$ equal unity).

\subsection{Construction in terms of two indistinguishable points and a director.}
\begin{figure}
\centering
\includegraphics[scale=0.4]{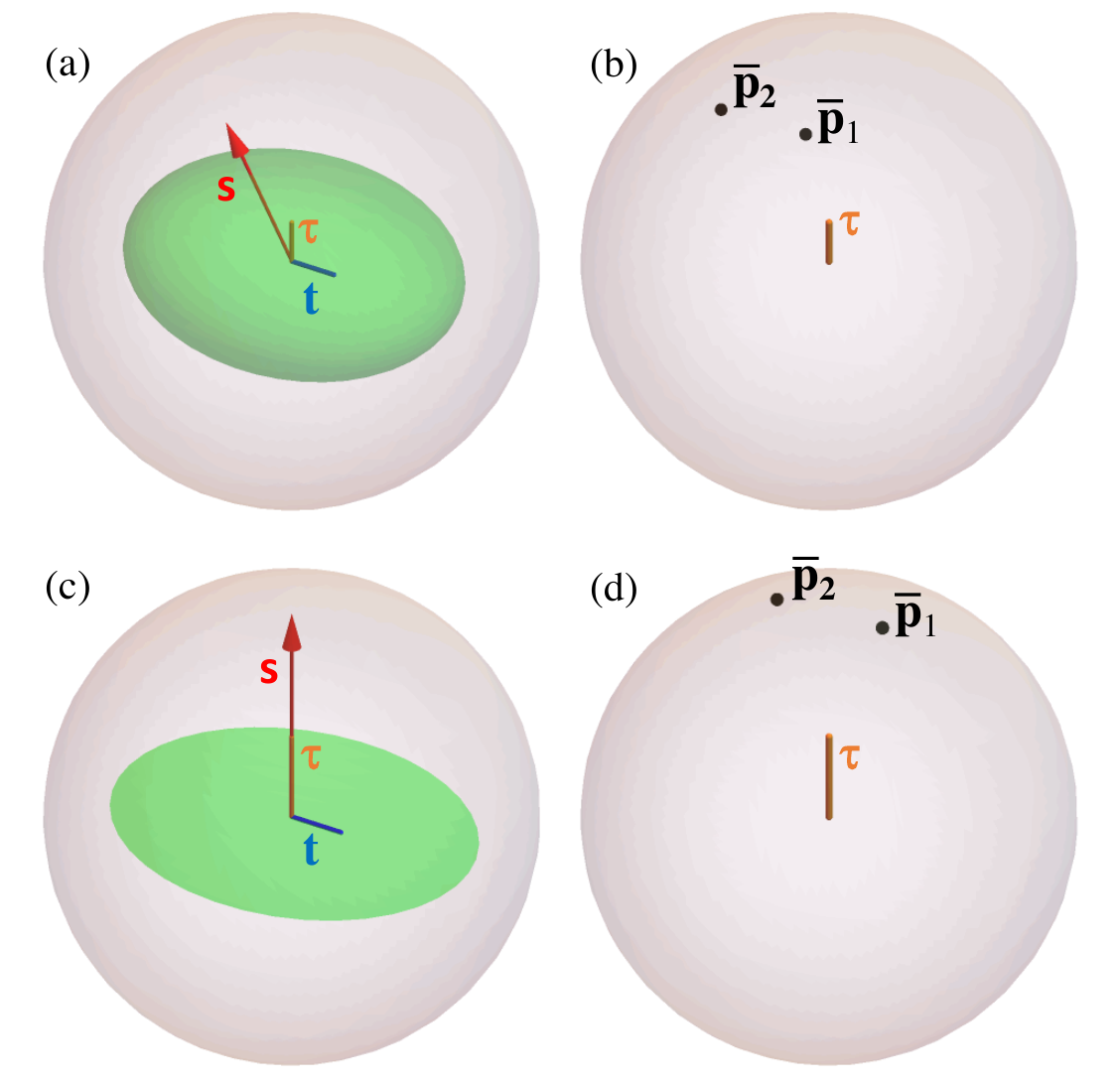}
\caption{
(a,b) two representations of a partially polarized state. (a) the normalized ellipsoid of inertia (green) and normalized spin density $\pmb{\mathsf s}$ (red) as well as the two perpendicular directors ${\bf t}$ (blue) and $\pmb{\tau}$ (orange). (b) The same state represented in terms of the two indistinguishable points $\overline{\bf p}_1$ and $\overline{\bf p}_2$ (black) and the director $\pmb{\tau}$ (orange), which is perpendicular to the line joining $\overline{\bf p}_1$ and $\overline{\bf p}_2$. All three are inside the unit sphere. Notice that, to distinguish it from $\overline{\bf p}_1$ and $\overline{\bf p}_2$, $\pmb{\tau}$ is represented by a line from the origin, and recall that it is sign-ambiguous. (c,d) Same as for (a,b) but for a state corresponding to full polarization, for which the ellipsoid of inertia is actually the polarization ellipse. In this case $\overline{\bf p}_1$ and $\overline{\bf p}_2$ are on the surface of the unit sphere and coincide with the Poincarana points. The Cartesian axes were omitted for clarity.}
\label{fig:threepoints}
\end{figure}

Polarization can then be fully described by the vector $\pmb{\mathsf s}$ and two directors (defined to within a global sign), ${\bf t}$ and $\pmb{\tau}$, all restricted to a unit ball. However, given relation~(\ref{eq:tpluss}), the vector $\pmb{\mathsf s}$ and the director ${\bf t}$ can be combined into two indistinguishable points with coordinates given by
\begin{align}
\overline{\bf p}_1=\pmb{\mathsf s}+{\bf t},\,\,\,\,\,\,\,\,\overline{\bf p}_2=\pmb{\mathsf s}-{\bf t},
\end{align}
which as shown earlier are also constrained to the unit ball. Therefore, we propose here a representation in terms of the two unambiguous and indistinguishable points $\overline{\bf p}_1$ and $\overline{\bf p}_2$, and the (sign-ambiguous) director $\pmb{\tau}$ (constrained to be normal to the line joining $\overline{\bf p}_1$ and $\overline{\bf p}_2$). This representation is simpler to visualize and has a more direct connection to the shape of the polarization path than the Stokes-Gell-Mann parameters, while containing the same information. Further, in the limit of full polarization $\overline{\bf p}_1$ and $\overline{\bf p}_2$ are both at the surface of the unit sphere and coincide with the Poincarana points, and as mentioned earlier $\pmb{\tau}$ becomes superfluous since it is required to point in the direction of the bisector of $\overline{\bf p}_1$ and $\overline{\bf p}_2$ and to have a magnitude equal to one minus half the separation between $\overline{\bf p}_1$ and $\overline{\bf p}_2$. This representation then reduces to the Poincarana two-point construction in the limit of full polarization, and generalizes it for partially polarized nonparaxial fields. Note, however, that $\overline{\bf p}_1$ and $\overline{\bf p}_2$ cannot be prescribed fully independently inside the sphere. For example, there are no physical states of polarization in which one of these points is at the surface and the other is not, independently of $\pmb{\tau}$. The three quantities are restricted by the inequalities in (\ref{eq:restrictions}). This construction is illustrated in Fig.~\ref{fig:threepoints} for a generic partially polarized case and for a case of full polarization.

Note that Eq.~(\ref{eq:Ppts}) can be now written as
\begin{align}
P_{\rm I}&=\sqrt{\frac14\left(\frac{|\overline{\bf p}_2-\overline{\bf p}_1|}2+|\pmb{\tau}|\right)^2+\frac34\frac{|\overline{\bf p}_1|^2+|\overline{\bf p}_2|^2}2},
\label{eq:Ppts2}
\end{align}
The second term inside the square root is reminiscent of the interpretation of degree of polarization in the paraxial case, which is given by the distance from the origin of the point(s) defining polarization. However, in this case there is also a contribution involving the separation of the points.

\subsection{Simple examples}
Let us illustrate this construction with some simple special cases.
\subsubsection{Regular polarization matrix}
Consider the type of polarization referred to by Gil and collaborators as polarimetrically regular \cite{SanJose2011,Gil2017,GilPRA2018,GilEPJ2018,GilOL2018,GilNJP2021}. These are the matrices for which the eigenvectors ${\bf e}_1$ and ${\bf e}_2$ correspond to coplanar ellipses, so that ${\bf e}_3$ must be linear and orthogonal to the plane containing these ellipses. Let us, for convenience, choose the reference frame in which ${\bf e}_3$ is aligned with the $z$ direction, while the $x$ direction is aligned with the major axis of the polarization ellipse for ${\bf e}_1$. The polarization matrix can then be written as
\begin{align}
\mathbf{\Gamma}=\left(\begin{array}{ccc}(\Lambda_1-\Lambda_2)a^2+\Lambda_2&-\ui(\Lambda_1-\Lambda_2)ab&0\\
\ui(\Lambda_1-\Lambda_2)ab&(\Lambda_1-\Lambda_2)b^2+\Lambda_2&0\\
0&0&\Lambda_3\end{array}\right),
\end{align}
where $a$ and $b$ are respectively the major and minor axis lengths of the polarization ellipses for ${\bf e}_1$ and ${\bf e}_2$, namely ${\bf e}_1=(a,\ui b,0)^{\rm T}$ and ${\bf e}_2=(\ui b,a,0)^{\rm T}$. Because the real part of all non-diagonal elements vanishes, it is easy to see that the normalized real-part eigenvalues $\bar{\lambda}_i$ correspond simply to normalized versions of the three diagonal elements. With this we can easily calculate
\begin{align}
{\bf t}=(\lambda_1-\lambda_2)(a^2-b^2)\,\hat{\bf x},\,\,\,\,\pmb{\tau}=2[(\lambda_1-\lambda_2)b^2+\lambda_2-\lambda_3]\,\hat{\bf z},\,\,\,\,\pmb{\mathsf s}=2(\lambda_1-\lambda_2)ab\,\hat{\bf z}.
\end{align}
A first thing to note from these expressions is that $\pmb{\tau}$ is parallel to $\pmb{\mathsf s}$. That is, for a polarimetrically regular matrix, $\pmb{\tau}$ is parallel to the spin density, and therefore any difference in the directions of $\pmb{\tau}$ and the spin density is a sign of polarimetric nonregularity. From these expressions we can also find the coordinates of the points $\overline{\bf p}_{1,2}$, which are given by
\begin{align}
\overline{\bf p}_{1,2}=P_{\rm II}\,[2ab\,\hat{\bf z}\pm(a^2-b^2)\,\hat{\bf x}],
\end{align}
where $P_{\rm II}=\lambda_1-\lambda_2$ was defined in Eq.~(\ref{DoPE}). Given the normalization condition $a^2+b^2=1$, it is easy to see that $|\overline{\bf p}_{1,2}|=P_{\rm II}$. 

\subsubsection{Incoherent mixture of a fully polarized and a fully unpolarized fields}
As mentioned in Section~\ref{sec:DOP}, a polarization matrix can only be separated into two parts that are respectively fully polarized and fully unpolarized if $\lambda_2=\lambda_3$, and in this case $P_{\rm I}=P_{\rm II}=Q_{\rm II}=\gamma$. This case is explicitly regular, and therefore is a special case of the previous example. By choosing the same coordinate axes as in the previous case, we obtain
\begin{align}
\overline{\bf p}_{1,2}=P_{\rm I}\,[2ab\,\hat{\bf z}\pm(a^2-b^2)\,\hat{\bf x}],\,\,\,\,\pmb{\tau}=P_{\rm I}\,2b^2\,\hat{\bf z}.
\end{align}
That is, all vectors are proportional to this degree of polarization. 

\subsubsection{Full polarization}
The case of a fully polarized field corresponds then simply to $P_{\rm I}=1$. In this case the points $\overline{\bf p}_{1,2}$ are over the surface of the unit sphere and correspond to the two Poincarana points, and $\pmb{\tau}$ bisects them and has a length equal to twice the square of the length of the minor semi-axis of the normalized polarization ellipse ($2b^2$). The mid-point between the Poincarana points corresponds to the normalized spin density, whose length is twice the product of the lengths of the semi-axes of the normalized polarization ellipse ($2ab$). Therefore, for a fully polarized field, $\pmb{\tau}$ is equal to $\pmb{\mathsf s}$ only for circular or linear polarizations, and it is shorter otherwise.  

\section{Concluding remarks and outlook}

This tutorial aimed to summarize different theoretical descriptions of the local polarization of general nonparaxial fields. The emphasis was on highlighting geometrical aspects. As has been noted already in many of the references cited here, while the extension from $2$ to $3$ relevant field components still allows using several of the same concepts, many qualitatively new aspects arise that are more than a simple increase in degrees of freedom. For example, the standard degree of polarization in the paraxial regime can be understood in at least two ways: as the fraction of light that is fully polarized, or as the magnitude of the normalized Stokes vector. These lead to two different measures of degree of polarization in the nonparaxial regime. Similarly, for nonparaxial light the region occupied by the Stokes parameters becomes not only of considerably higher dimensionality but also acquires a more complex shape, which is not fully enclosed by the subset of points corresponding to full polarization. It should be mentioned that other representations of polarization of nonparaxial light have been given that were not discussed here, such as the use of multipolar decompositions \cite{delaHoz2015}. The author apologizes in advance to those authors whose work was not described or might not have been properly cited. 

It is the author's hope that some of the ideas presented here can help motivate both theoretical and experimental efforts. Some of these are listed in what follows:

\subsection{Topological field distributions}
This article focused on the local description of polarization, and ventured into nonlocal aspects only very briefly in the discussions about geometric phases and M\"obius strips. However, as mentioned in the Introduction, the spatial distribution of polarization and the topological properties it can present given the physical restrictions imposed by Maxwell's equations is a very active and interesting field of research. This extended topology relies on the local topology of the space that polarization (or a specific aspect of it) inhabits. Perhaps the simplest nontrivial example of extended topological features are the M\"obius strips \cite{Bauer2015,FreundOC2010,FreundOL2010,Dennis2011}, since one needs only consider the polarization of points along a given closed curve, and not all aspects of this polarization but only the direction of the major axis of the ellipse. If this direction reverses sign while going around a closed loop in space, one has a polarization M\"obius strip. 

By adding one more variable and looking at the polarization distribution over a plane in space, one can see topological features such as the so-called {\it baby Skyrmions}, which are states for which a given parameter defined over a sphere is fully spanned. In polarization optics, these correspond to field distributions for which points over a plane in space are mapped onto a complete sphere that represents some aspect of polarization, either in the paraxial or nonparaxial regimes. In the paraxial case the spherical space in question can be simply the Poincar\'e sphere \cite{Beckley2010,Donati2016}, while in the nonparaxial case it can be the direction of the spin density for evanescent \cite{Du2019} or traveling \cite{Gutierrez-Cuevas2021} fields, the direction of the polarization ellipse's major axis \cite{Gutierrez-Cuevas2021}, or even the direction of the electric field at an instant \cite{Tsesses2018}. 
  
Field distributions over volumes have also been considered that span a complete compact parameter volume linked to polarization. One recent example is that of a full Skyrmionic field that spans a 3-sphere of polarization and phase, following a Hopf fibration structure \cite{Sugic2021}. Another  is a paraxial partially polarized field that covers completely the surface and interior of the Poincar\'e sphere \cite{Beckley2012} according to a simple 2-to-1 mapping, although this case is not Skyrmionic because the spanned polarization space is a 2-ball, which is not a closed (i.e. periodic) space but one with boundaries. 

The extended topological features just described rely on the local geometry/topology of the space of polarization. Understanding the full geometry of nonparaxial polarization can reveal interesting (open or closed) subspaces that can be spanned by a field distribution over physical spaces of different dimensions. By including in the analysis other variables, such as time or frequency, the dimensionality of the space can grow, allowing a field to span more of polarization's degrees of freedom. For example, current work regards a simple spatiotemporal light distribution where the complete 4D space of nonparaxial full polarization is covered \cite{Marco2022}.


\subsection{Higher dimensions}
\label{sec:CR}
As discussed earlier, the description of the polarization of a paraxial field, where only two components are important, requires three parameters in general, but only two in the limit of full polarization. For nonparaxial fields, where three components are involved, the number of parameters is eight in the general case and four in the limit of full polarization. It is interesting to think of the generalization of these ideas for vector fields with $N$ components. Such fields would require in general the specification of $N^2-1$ normalized Stokes parameters, where the subtraction of one results from normalization. In the case of full polarization (pure states) only $2N-2$ parameters are needed, where the subtraction of two results from normalization and extraction of a global phase. The normalized Stokes parameter vector ${\vec s}$ (where the normalization is proportional to the scalar $S_0$, which is the local intensity) can still be subdivided into three parts: 
\begin{itemize}
\item An $(N-1)$-sub-vector ${\bf s}_1$ that characterizes differences between the diagonal elements of the matrix. This sub-vector is constrained to a regular $(N-1)$-simplex whose $N$ corners are all at a unit distance from the origin and at equal distances $\sqrt{2N/(N-1)}$ from each other.
\item An $[N(N-1)/2]$-sub-vector ${\bf s}_2$ composed of the real parts of the off-diagonal elements, restricted to some curved convex region.
\item An $[N(N-1)/2]$-sub-vector ${\bf s}_3$ composed of the imaginary parts of the off-diagonal elements, and that is related to the generalization of the concept of spin density, and is restricted to a hyper-ellipsoid.
\end{itemize}
The shape of the inhabitable regions for these components results from ensuring that all eigenvalues of the polarization matrix are non-negative. For fully polarized (factorizable) matrices, the state can be represented by $N-1$ points over the surface of a unit sphere, e.g. following Majorana's prescription \cite{Majorana1932}. Generalizations of the Poincarana prescription might also exist, in which the distribution of the points is changed to facilitate the connection to geometric phases. Similarly, generalizations of the many-point/vector representation in the physical space might also exist.

\begin{figure}
\centering
\includegraphics[scale=0.35]{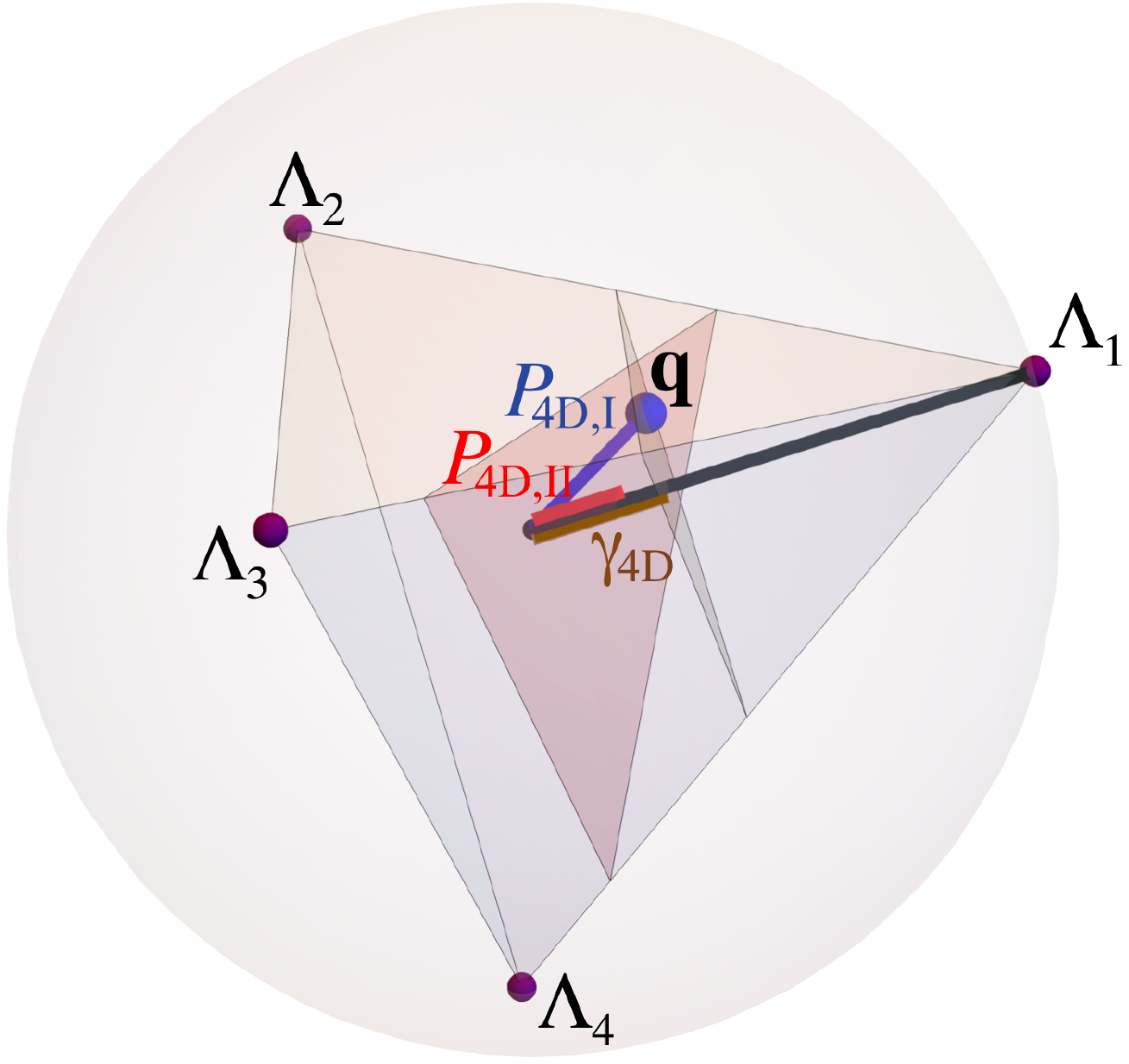}
\caption{Geometric interpretation of the measures of polarization $P_{4\rm D,I}$, $P_{4\rm D,II}$, and $\gamma_{4\rm D}$ in terms of the center of mass ${\bf q}$ (blue dot) of four point masses (purple dots) whose magnitudes are $\Lambda_i$, the eigenvalues of $\mathbf{\Gamma}_{4\rm D}$ and whose positions are the corners of a regular tetrahedron inscribed in a unit sphere: $P_{4\rm D,I}$ is the distance of ${\bf q}$ to the origin; $\gamma_{4\rm D}$ is the distance from the origin to the intersection of the line from the origin to the mass $\Lambda_1$ with a plane containing ${\bf q}$ and normal to this line; $P_{4\rm D,II}$ is the distance from the origin to the intersection of the line from the origin to the mass $\Lambda_1$ with a plane containing ${\bf q}$ and normal to the line segment joining the masses $\Lambda_1$ and $\Lambda_2$.}
\label{fig:4D}
\end{figure}

The barycentric interpretation for the measures of degree of polarization is also easily generalized by using $N$ point masses whose magnitudes are the eigenvalues $\Lambda_i$, and which are located at a unit distance from the origin and each equidistant to all other masses. This distribution must be placed in a space of dimension $N-2$ \cite{CoM}. The generalizations of $P_{\rm I}$, $P_{\rm II}$ \cite{SanJose2011}, and $\gamma$ are given by 
\begin{align}
P_{N{\rm D,I}}&=\sqrt{\frac{N}{N-1}\frac{{\rm Tr}(\mathbf{\Gamma}_{N{\rm D}}^2)}{({\rm Tr}\mathbf{\Gamma}_{N{\rm D}})^2}-\frac1{N-1}},\\
P_{N{\rm D,II}}&=\lambda_2-\lambda_1,\\
\gamma_{N\rm D}&=\lambda_1-\frac1N\sum_{i=2}^N\lambda_i.
\end{align}
Note that the interpretations of these parameters for $N=3$ still hold: $P_{N\rm D,I}$ corresponds to the radial coordinate of the center of mass in this multidimensional space, $P_{N\rm D,II}$ is a scaled version of the Cartesian coordinate of ${\bf q}$ in the direction joining the masses $\Lambda_1$ and $\Lambda_2$, and $\gamma_{N\rm D}$ is the Cartesian coordinate in the direction joining the origin and the mass $\Lambda_1$. These three measures coincide if all eigenvalues but the largest one are equal to each other (a condition automatically satisfied for $N=2$, that is, for paraxial light). Figure~\ref{fig:4D} shows this barycentric construction for $N=4$, for which the embedding space is three-dimensional and the four masses are the corners of a regular tetrahedron.

From the point of view of optical polarization this discussion of higher dimensionality might seem only of academic interest. However, there are situations in classical optics where a treatment that is mathematically analogous to that polarization for $N\ge4$ components is relevant. Perhaps the simplest example is the use of modal decompositions to express optical fields, in which $N$ corresponds to the number of linearly-independent modes used. Like polarization components, modes provide different channels that can be exploited by the degrees of freedom of light, and the analogue of full or partial polarization would be the coherence between these modes. In particular, paraxial optical beams can be expressed in terms of, say, Laguerre-Gaussian modes, which constitute a complete and orthonormal basis. This analogy between modal decompositions and polarization has led to the use of the Poincar\'e sphere construction to describe beam shape, first for fields involving only two modes \cite{Padgett1999} (which are then analogous to paraxial polarization), and then for more complicated fields \cite{Abramochkin2004,Habraken2010,Milione2011,Dennis2017,Alonso2017,Dennis2019,Gutierrez2019}. In this analogy, the axis $s_3$ corresponds not to spin but to orbital angular momentum. 
Due to the curvature of the parameter space, geometric phases can be observed under cyclic transformations of the beams, and they can be measured through several methods \cite{Galvez1999,Galvez2003,Milione2012,Malhotra2018}.  Further, beams composed of coherent superpositions of $N$ modes can be fully described in terms of $N-1$ points over a unit sphere by using the Majorana construction \cite{Gutierrez2020} since they are analogous to a fully polarized state.  

Mode superpositions that are partially coherent and their analogy with partial polarization in $N$ dimensions have also been studied within the context of waveguides supporting $N$ propagating modes. In particular, analogues of Stokes parameters can be used to characterize these modes \cite{Antonelli2012,Wood2015,Ji2019}, allowing generalizations of the Poincar\'e sphere \cite{Fernandes2017} and the use of the definition $P_{N\rm D,I}$ to characterize the overall level of coherence amongst the modes \cite{Mecozzi2014}.

As pointed out in this article and elsewhere \cite{Dennis2004}, the role of unitary transformations is not as important for the description of nonparaxial polarization as it is for paraxial light, since these transformations (other than rotations and inversions) do not correspond to the actions of optical elements. Note, however, that this is not necessarily the case for the higher-dimensional generalizations based on spatial modes mentioned in the previous two paragraphs. In theory, general unitary transformations may be physically realizable, e.g. through appropriate sequences of spatial light modulators,  so incorporating some sort of invariance to the relevant quantities is therefore desirable. In particular, for superpositions of Laguerre-Gaussian beams that can be represented as a collection of points on a sphere following Majorana's construction \cite{Gutierrez2020}, a specific subset of all possible unitary transformations, corresponding to passage through specific combinations of cylindrical lenses, amounts to a rigid rotation of all points over the sphere. That is, the properties of the representation in terms of what quantities are invariant must be dictated by the physical context. 

The examples of $N>3$ just discussed correspond to generalizations that use optical degrees of freedom other than polarization. However, there are at least three situations that deal exclusively with optical polarization where the effective number of field components is also larger than three: 
\begin{itemize}
\item One is the case of entangled photons. For example, a mixed state corresponding to a pair of paraxial photons ($N=4$), the density matrix involves $4^2-1=15$ degrees of freedom, and this number goes up to $6^2-1=35$ if we look at the nonparaxial situation ($N=6$). 

\item A second case is that of bichromatic (or multichromatic) fields \cite{Freund2003a,Freund2003b,Haitao2017,Pisanty2019} resulting, for example, from nonlinear harmonic generation. In their fully polarized form, these fields are composed of two frequencies, one being twice (or some other rational multiple of) the other. The electric field no longer traces an ellipse but a closed Lissajous figure over one temporal cycle. The polarization at a point (excluding a global phase and intensity) involves 10 degrees of freedom, and it would be interesting to investigate which representation of such state best captures the geometry and properties of the polarization curves. Clearly, the situation becomes more complex when more frequencies are involved.

\item Finally, another case corresponding to $N=6$ that deserves a closer treatment is that of nonparaxial electromagnetic polarization, where both the electric and magnetic fields are considered. As mentioned in the introduction, the polarization properties of each of these fields are essentially redundant in the paraxial regime, but not in the nonparaxial one where the path traced by one of these fields does not restrict the path traced by the other at the local level. (The restrictions imposed by Maxwell's equations arrive at the nonlocal level, given that they involve spatial derivatives.) The two fields can be made to have the same units through multiplication by the appropriate universal constants. Both fields might then have different average amplitudes, and the relative phase between their oscillations is physically relevant, so it makes sense to consider the joint polarization as a six-component vector, rather than two separate 3-component fields with their own normalizations and global phases. An appropriate representation must present desirable properties under the relevant transformations. 
\end{itemize}

\subsection{Novel measurement techniques} 

Many techniques have been proposed for imaging polarimetry, which allow the efficient measurement of 2D polarization over large arrays of points. These techniques are based on a variety of strategies such as the use of rotating waveplates, micro-polarizers arrays, or division of amplitude elements \cite{Tyo2006}. Other techniques proposed more recently are based on the use of spatially-varying birefringent elements \cite{Rubin2021} such as materials with stress-induced birefringence or metasurfaces, either for point-spread function engineering (for sparse objects) \cite{Ramkhalawon2013} or for polarization component separation \cite{Rubin2022}.  The generalization of these techniques for measuring simultaneously the 3D polarization at many points is a challenging problem for several reasons. First, the amount of information to be gathered at each point is more than twice as large. Second, capturing the polarization components that are aligned with the direction between the point of observation and the detector requires the use of large numerical aperture systems. Finally, the spatial scale of variation of polarization is typically much smaller, on the order of the wavelength, so the requirements for appropriate resolution and sampling density are much more stringent. As was mentioned in the Introduction, current techniques rely on scanning a Rayleigh scatterer and capturing and characterizing the scattered field over a large range of directions \cite{Lindfors2005,Lindfors2007,Deutsch2010,Bauer2015,BliokhPR2015,Aiello2015,BliokhNP2015,Lodahl2017,Eismann2020}. This allows measuring polarization of only one point at a time, and in order to obtain a polarization map one must scan the scatterer. Parallelizing this type of technique or developing new ones that permit measuring 3D polarization at many points simultaneously is an important challenge. 

One specific possible technological application of the geometric descriptions proposed here is in fluorescence microscopy. As discussed in Section~\ref{microscopy}, fluorophores typically have a dipolar emission pattern, and their orientation provides useful morphological information about the tissue or structure that is hosting them. Further, they often wobble at timescales that are much smaller than the detection time. This wobble is often assumed to be isotropic around a central direction, but this is not necessarily the case. The full characterization of this wobble provides also useful morphologic information. Techniques that allow characterizing simultaneously the position, orientation and wobble of multiple fluorophores are therefore a current thrust of research in microscopy \cite{Foreman2008,Aguet2009,Backlund2012,Backer2015,Zhang2018,Zhang2019,CHIDO,Backlund2018,Hulleman2021,Ding2021,Wu2022}. The performance of a given technique is often evaluated in terms of the lower bounds for the accuracy in the estimation of these parameters in the presence of Poisson noise, according to what is known as a Cramer-Rao bound \cite{Refregier2003,Vella2020}. As mentioned in this article, the wobbling properties of a fluorophore translate into the polarization properties of the light it emits, and therefore a meaningful measure of estimation must take into account the intrinsic geometric/topologic properties of the polarization matrix. The geometric representations for partial polarization might then provide self-consistent criteria for the definition of functions of merit for the estimation accuracy of the 3D orientation and wobble of a fluorophore. Further, many of the microscopy techniques just mentioned rely on using masks placed at the pupil of a microscope that encode the desired information into the shape of the point-spread function of the fluorophores \cite{Backlund2012,Backer2015,CHIDO,Hulleman2021,Ding2021}. The definition of meaningful functions of merit can then open the path for the design of masks that are optimal for this purpose. 
\appendix
\section{Appendix A: Pauli matrices}
\label{appA}
Here we give a brief summary of the Hermitian form of the Pauli matrices, following a numbering and sign convention consistent with the treatment of paraxial polarization presented here. These three matrices are given by
\begin{align}
\sigma_1=\left(\begin{array}{cc}1&0\\0&-1\end{array}\right),\,\,\,\sigma_2=\left(\begin{array}{cc}0&1\\1&0\end{array}\right),\,\,\,\sigma_3=\left(\begin{array}{cc}0&-\ui\\ \ui&0\end{array}\right).
\end{align}
They constitute a complete basis for $2\times2$ Hermitian matrices when complemented with the identity
\begin{align}
\sigma_0=\left(\begin{array}{cc}1&0\\0&1\end{array}\right).
\end{align}
This basis is orthogonal under trace, since
\begin{align}
\label{eq:orthogonalPauli}
{\rm Tr}(\sigma_m\sigma_n)=2\delta_{mn},
\end{align}
where $\delta_{mn}$ is the Kronecker delta. This means that any Hermitian matrix can be decomposed as a linear superposition of these matrices as
\begin{align}
\mathbf{\Gamma}_{\rm 2D}=\frac12\sum_{m=0}^3S_m\sigma_m,
\end{align}
where the coefficients of the expansion, namely the Stokes parameters, are given by
\begin{align}
S_m={\rm Tr}(\sigma_m\mathbf{\Gamma}_{\rm 2D}).
\end{align}

Other important properties of the Pauli matrices (for $m=1,2,3$) are the fact that they are traceless,
\begin{align}
{\rm Tr}(\sigma_m)=0,
\end{align}
that their square is the identity,
\begin{align}
\sigma_m^2=\sigma_0,
\end{align}
and that they satisfy the cyclic relations
\begin{subequations}
\begin{align}
\sigma_1\sigma_2&=-\sigma_2\sigma_1=-\ui\sigma_3,\\
\sigma_2\sigma_3&=-\sigma_3\sigma_2=-\ui\sigma_1,\\
\sigma_3\sigma_1&=-\sigma_1\sigma_3=-\ui\sigma_2.
\end{align}
\end{subequations}
In fact, these three properties yield the orthogonality relation in Eq.~(\ref{eq:orthogonalPauli}).

\section{Appendix B: Polarization statistics for a random superposition of paraxial plane waves}
\label{appB}
Consider the superposition of many plane waves traveling (approximately) in the $z$ direction, whose relative phases and polarizations are uncorrelated. (Whether they all have the same amplitude or not in the end is irrelevant, as subsets of them can be considered as individual elements.) Given their statistical independence, the probability density for the real or imaginary part of a given Cartesian component is the convolution of the probability densities for each, such that the central limit theorem can be applied and in the end we find a Gaussian probability density:
\begin{align}
P(A_x,B_x,A_y,B_y)=\exp[-\pi(A_x^2+B_x^2+A_y^2+B_y^2)],
\end{align}
where we used the notation $(E_x,E_y)=(A_x+\ui B_x,A_y+\ui B_y)$, with the field being normalized in dimensionless units so that the average intensity is $2/\pi$:
\begin{align}
\langle I\rangle=\int \exp[-\pi(A_x^2+B_x^2+A_y^2+B_y^2)](A_x^2+B_x^2+A_y^2+B_y^2)\,\ud^4E=\frac2\pi,
\end{align}
with the shorthand $\ud^4E=\ud A_x\ud B_x\ud A_y\ud B_y$.

Probability densities of different measurable quantities can be evaluated by taking the appropriate marginal projections:
\begin{align}
P_q(q)=\int P(A_x,B_x,A_y,B_y)\,\delta[q-{\cal Q}(A_x,B_x,A_y,B_y)]\,\ud^4E,
\end{align}
where $\delta$ is the Dirac distribution and ${\cal Q}$ is the function of the electric field associated with the quantity in question. For example, for the intensity we have
\begin{align}
P_I(I)&=\int \exp[-\pi(A_x^2+B_x^2+A_y^2+B_y^2)]\,\delta[I-(A_x^2+B_x^2+A_y^2+B_y^2)]\,\ud^4E\nonumber\\
&=\pi^2 I \exp(-\pi I),
\end{align}
Similarly, for the normalized Stokes parameter $s_1$ we get
\begin{align}
P_{s_1}(s_1)&=\int \exp[-\pi(A_x^2+B_x^2+A_y^2+B_y^2)]\,\delta\!\left(s_1-\frac{A_x^2+B_x^2-A_y^2-B_y^2}{A_x^2+B_x^2+A_y^2+B_y^2}\right)\,\ud^4E\nonumber\\
&=\pi^2\int \exp[-\pi(I_x^2+I_y^2)]\,\delta\!\left(s_1-\frac{I_x^2-I_y^2}{I_x^2+I_y^2}\right)\,\ud I_x\ud I_y,
\end{align}
where $I_i=A_i^2+B_i^2$. By now using the change of variables $I'=I_x+I_y,\Delta=I_x-I_y$ with Jacobian $1/2$ this can be written as
\begin{align}
P_{s_1}(s_1)&=\frac{\pi^2}{2}\int_0^{\infty}\int_{-I'}^{I'} \exp(-\pi I')\,\delta\!\left(s_1-\frac{\Delta}{I'}\right)\,\ud \Delta\ud I'\nonumber\\
&=\frac{\pi^2}{2}\int_0^{\infty} I'\,\exp(-\pi I')\,\ud I'=\frac12.
\end{align}
That is, the probability of finding any value for $s_1\in[-1,1]$ is the same. It is easy to see that the same probabilities apply to $s_2$ and $s_3$ since they can be calculated by a change of basis of the electric field components that leaves the form of the underlying probability density (which depends only on intensity) invariant. This result is then just a way to express that the probability distribution of polarizations over the surface of the Poincar\'e sphere is constant. Please note that, while we are using probability densities, the quantities are being calculated assuming that the field is fully polarized. 

Finally, let us also derive a result that is perhaps unsurprising: there is no correlation between intensity and polarization, as can be seen from the joint probability
\begin{align}
P_{I,s_1}(I,s_1)&=\int \exp[-\pi(A_x^2+B_x^2+A_y^2+B_y^2)]\,\delta\!\left(s_1-\frac{A_x^2+B_x^2-A_y^2-B_y^2}{A_x^2+B_x^2+A_y^2+B_y^2}\right)\nonumber\\
&\times\delta[I-(A_x^2+B_x^2+A_y^2+B_y^2)]\,\ud^4E\nonumber\\
&=\frac{\pi^2}{2}\int_0^{\infty}\int_{-I'}^{I'} \exp(-\pi I')\,\delta(I=I')\delta\!\left(s_1-\frac{\Delta}{I'}\right)\,\delta\!\left(s_1-\frac{\Delta}{I'}\right)\,\ud \Delta\ud I'\nonumber\\
&=\frac{\pi^2}2 I \exp(-\pi I)=P_I(I)P_{s_1}(s_1).
\end{align}
The same is true for the other two normalized Stokes parameters.

\section{Appendix C: Polarization statistics for a random superposition of paraxial plane waves}
\label{appC}
This appendix is a generalization of the results in Appendix~\ref{appB} to the nonparaxial regime, and hence uses similar notation. After writing this derivation, the author found out that Dennis \cite{DennisPersonal} had also derived it in unpublished work.

Consider superpositions of large numbers of plane waves propagating in directions that are uniformly distributed over the sphere of directions, and with polarization distributions that are uncorrelated to their propagation direction (except for the transversality condition). In the end, for each real or imaginary part of each of the three Cartesian components we end up with a superposition of statistically independent contributions so that the central limit theorem can be used, leading to a probability density
\begin{align}
P(A_x,B_x,A_y,B_y,A_z,B_z)=\exp[-\pi(A_x^2+B_x^2+A_y^2+B_y^2+A_z^2+B_z^2)],
\end{align}
where, as Appendix~\ref{appB}, the Cartesian components of the electric field are separated in real and imaginary parts as $E_i=A_i+\ui B_i$, in this case for $i=x,y,z$. The probability density for the intensity is now found to be
\begin{align}
P_I(I)&=\frac{\pi^3}2 I^2 \exp(-\pi I),
\end{align}
with average intensity $\langle I\rangle=3/\pi$. 
The marginals for the different parameters can be harder to calculate. In particular, we are interested in the marginal for the magnitude of the normalized spin density:
\begin{align}
|\vec s|=\frac{2\sqrt{(A_xB_y-A_yB_x)^2+(A_yB_z-A_zB_y)^2+(A_zB_x-A_xB_z)^2}}{A_x^2+B_x^2+A_y^2+B_y^2+A_z^2+B_z^2}.
\end{align}
To facilitate the computation, we note that, for any realization, there exists always a unit vector $\hat{s}$ such that $\hat{s}\cdot\vec E=\hat{s}\cdot(A_x+\ui B_x,A_y+\ui B_y,A_z+\ui B_z)=0$. That is, for any realization the field oscillates tracing an ellipse that is contained within a plane, and $\hat{s}$ is the normal to that plane. The superposition over all possible field realizations can then be separated into the sum over all $\hat{s}$ of the sum of all the cases where the field is normal to $\hat{s}$. (There are, of course, degenerate cases of linear polarization that are normal to multiple $\hat{s}$, but this is a subset of size zero.) For each subset corresponding to a given $\hat{s}$, the distribution of the field components is still an isotropic Gaussian on the remaining four free parameters, corresponding to the real and imaginary parts of the Cartesian components in a reference frame for which $\hat{s}$ is the direction of one axis. For this subset, the normalized spin density points in the direction of $\hat s$ and is mathematically analogous to the magnitude of the normalized Stokes parameter $s_3$ for paraxial light. As was shown in Appendix~\ref{appB}, this parameter has constant probability. Since the same is true for all reference frames, we arrive at the conclusion that 
\begin{align}
P_{|\vec s|}(|{\vec s}|)=1.
\end{align}
By using similar arguments, we can also show that there is no correlation between polarization and intensity.

\section{Appendix D: Gell-Mann matrices}
\label{appD}
Similarly to what was done in Appendix~\ref{appA} for the Pauli matrices, a very short summary of the definition and properties of the Gell-Mann matrices is given here. We use a sign and numbering convention consistent with the article. The eight $3\times3$ Gell-Mann matrices are
\begin{align}
&\Theta_{11}=\left(\begin{array}{ccc}1&0&0\\0&-1&0\\0&0&0\end{array}\right),\,\,\,\Theta_{11}=\left(\begin{array}{ccc}1/\sqrt{3}&0&0\\0&1/\sqrt{3}&0\\0&0&-2/\sqrt{3}\end{array}\right),\nonumber\\
&\Theta_{21}=\left(\begin{array}{ccc}0&0&0\\0&0&1\\0&1&0\end{array}\right),\,\,\,\Theta_{22}=\left(\begin{array}{ccc}0&0&1\\0&0&0\\1&0&0\end{array}\right),\,\,\,\Theta_{23}=\left(\begin{array}{ccc}0&1&0\\1&0&0\\0&0&0\end{array}\right),\nonumber\\
&\Theta_{31}=\left(\begin{array}{ccc}0&0&0\\0&0&-\ui\\0&\ui&0\end{array}\right),\,\,\,\Theta_{32}=\left(\begin{array}{ccc}0&0&\ui\\0&0&0\\-\ui&0&0\end{array}\right),\,\,\,\Theta_{33}=\left(\begin{array}{ccc}0&-\ui&0\\ \ui&0&0\\0&0&0\end{array}\right).
\end{align}
These matrices share many of the desirable properties of the Pauli matrices, but not others, the limitations not being a consequence of their choice but inherent to the space of $3\times3$ Hermitian matrices. 

Like the Pauli matrices, the Gell-Mann matrices are traceless:
\begin{align}
{\rm Tr}(\Theta_{mn})=0,
\end{align}
but unlike the Pauli matrices, the square of the Gell-Mann matrices do not give a matrix proportional to the identity but a non-negative diagonal matrix with trace equal to two:
\begin{align}
\Theta_{11}^2&=\Theta_{23}^2=\Theta_{33}^2=\left(\begin{array}{ccc}1&0&0\\0&1&0\\0&0&0\end{array}\right),\\
\Theta_{21}^2&=\Theta_{31}^2=\left(\begin{array}{ccc}0&0&0\\0&1&0\\0&0&0\end{array}\right),\\
\Theta_{22}^2&=\Theta_{32}^2=\left(\begin{array}{ccc}1&0&0\\0&0&0\\0&0&1\end{array}\right),\\
\Theta_{11}^2&=\left(\begin{array}{ccc}1/3&0&0\\0&1/3&0\\0&0&4/3\end{array}\right).
\end{align}
Unfortunately, there is also no simple generic form for the product of two different Gell-Mann matrices, but it is easy (if tedious) that all these products give matrices whose trace vanishes:
\begin{align}
{\rm Tr}(\Theta_{mn}\Theta_{m'n'})=2\delta_{mm'}\delta_{nn'}.
\end{align}
(It turns out that the commutator of two Gell-Mann matrices does give a result proportional to another Gell-Mann matrix, but this property is not needed for the purposes of this tutorial.) 
The Gell-Mann matrices then constitute a complete orthogonal basis for $3\times3$ Hermitian matrices when supplemented with a $3\times3$ identity matrix,
\begin{align}
\Theta_{0}=\left(\begin{array}{ccc}1&0&0\\0&1&0\\0&0&1\end{array}\right).
\end{align}
This matrix has trace 3, leading to a slight asymmetry in the expansion of an arbitrary Hermitian matrix $\mathbf{\Gamma}$:
\begin{align}
\mathbf{\Gamma}=\frac13S_0\Theta_{0}+\frac12\sum_{m,n}S_{mn}\Theta_{mn},
\end{align}
where the coefficients of the expansion, namely the Stokes-Gell-Mann parameters, are given by
\begin{align}
S_0={\rm Tr}(\Theta_0\mathbf{\Gamma})={\rm Tr}(\mathbf{\Gamma}),\,\,\,S_{mn}={\rm Tr}(\Theta_{mn}\mathbf{\Gamma}).
\end{align}

\section{Appendix E: An inequality restricting the phases of the correlations of three functions}
\label{appE}
Let us consider the relation between the correlations of three functions $f,g,h$. In particular, we seek to determine the range of complex values that the following normalized product of correlations can take:
\begin{align}
\Xi=\frac{\langle f^*g\rangle\langle g^*h\rangle\langle h^*f\rangle}{\langle|f|^2\rangle\langle|g|^2\rangle\langle|h|^2\rangle}.
\end{align}
To simplify the problem, consider that these functions are expanded in terms of a discrete orthonormal basis set. Without loss of generality, we can choose the basis so that only three basis elements are required, for example by using a Gram-Schmidt procedure such that the first element is a normalized version of $f$, the second element is a normalized version of the part of $g$ that is orthonormal to the first element, and the last element is a normalized version of the part of $h$ that is orthogonal to the first two elements. The expansion coefficients for the normalized versions of $f,g,h$ can then be written as $(1,0,0)$, $(\cos\theta_1,\sin\theta_1,0)\exp(\ui\gamma_1)$, and $[\cos\theta_2,\sin\theta_2\cos\phi\exp(\ui\eta),\sin\theta_2\sin\phi\exp(\ui\xi)]\exp(\ui\gamma_2)$, respectively. Notice that, without loss of generality, the phases of the elements were chosen so that the first coefficient for $f$ is real and the two coefficients for $g$ are in phase. The expression for the product of correlations then reduces to
\begin{align}
\Xi=\cos\theta_1\cos\theta_2[\cos\theta_1\cos\theta_2+\sin\theta_1\sin\theta_2\cos\phi\exp(\ui\eta)].
\end{align}

By substituting $\theta_{1,2}=\theta\pm\alpha$ and using trigonometric identities, $\Xi$ can be simplified to
\begin{align}
\Xi&=\frac14(\cos2\theta+\cos2\alpha)[(\cos2\theta+\cos2\alpha)+(\cos2\alpha-\cos2\theta)\cos\phi\exp(\ui\eta)]\nonumber\\
&=\frac14(c_1+c_2)[(c_1+c_2)+(c_2-c_1)c_3\exp(\ui\eta)],
\end{align}
where $c_1=\cos2\theta$, $c_2=\cos2\alpha$, and $c_3=\cos\phi$ are all within the range $[-1,1]$. As the three parameters $c_n$ and the phase $\phi$ vary, $\Xi$ spans a region over the complex plane; the goal of this appendix is to determine the edge of this region. It is straightforward to see that the edge is reached for extremal values of $c_3$ given the linear dependence of $\Xi$ on this parameter. Without loss of generality we then chose $c_3=1$ given that the two factors multiplying it can account for sign changes. The real and imaginary parts of $\Xi$ can then be written as
\begin{align}
\Xi_{\rm R}&=\frac14(c_1+c_2)[(c_1+c_2)+(c_2-c_1)\cos(\eta)],\\
\Xi_{\rm I}&=\frac14(c_2^2-c_1^2)\sin(\eta).
\end{align}
We now eliminate $\eta$ by solving for it in terms of $\Xi_{\rm R}$ and substituting the solution into the expression for $\Xi_{\rm I}$, which gives after some simplification
\begin{align}
\Xi_{\rm I}&=\pm\frac12\sqrt{[2\Xi_{\rm R}-c_1(c_1+c_2)][c_2(c_1+c_2)-2\Xi_{\rm R}]}.
\end{align}
The final step is to find the values of $c_1$ and $c_2$ that maximize the value of $\Xi_{\rm I}$ for fixed $\Xi_{\rm R}$. By taking derivatives with respect to both $c_1$ and $c_2$ and setting them to zero, we see that the only solutions correspond to $c_1=-c_2$, which is indeed stationary but is clearly not the bound we are looking for as it makes $\Xi=0$, and $c_1=c_2=\sqrt{\Xi_{\rm R}}$, which also does not correspond to the solution we are seeking since it makes $\Xi_{\rm I}=0$. The solutions must then correspond to points along the edge of the square region occupied by $(c_1,c_2)$. We therefore set $c_2=1$ and find the value of $c_1$ that maximizes $\Xi_{\rm I}^2$, with corresponds to $c_1=(4\Xi_{\rm R}-1)/3$. The resulting expression for the region inhabitable by $\Xi$ is that given in Eq.~(\ref{phase}). 
Note that the same boundary would have been found by setting instead $c_2=-1$, or by setting $c_1=\pm1$ and maximizing for $c_2$.

\section{Appendix F: Proof that relation~(\ref{phase2}) does not cause restrictions for $|\vec s|\le1/2$}
\label{appF}
We now show that the the inequality in (\ref{phase2}) only causes restrictions for $|\vec s|>1/2$. The key for this proof is to note that the inequality causes restrictions in the phase of $\Xi$ only for $|\Xi|\ge1/8$. 
Recall from Eq.~(\ref{Xi1}) that
\begin{align}
|\Xi|=\frac{3\sqrt{3}h_1h_2h_3}{1-3s_1^2+2s_1^3\sin3\alpha}.
\label{XiAbs}
\end{align}
It is easy to see also that $P_{\rm I}^2=|\vec s|^2=s_1^2+h_1^2+h_2^2+h_3^2$. We now seek for the values of these parameters that maximize $|\Xi|$ for fixed $P_{\rm I}=1/2$ to show that for these values $|\Xi|$ remains at or below $1/8$. Note that $h_1h_2h_3$ is maximized for fixed $h_1^2+h_2^2+h_3^2$ when all $h_m$ are equal. Let us then choose $h_m=h$, and set the numerator of the right-hand side of Eq.~(\ref{XiAbs}) to $3\sqrt{3}h^3$. We can then set $P_{\rm S}^2=s_1^2+3h^2=1/4$, solve for $h$ and substitute the result in Eq.~(\ref{XiAbs}), which is then a function of only $s_1$ and $\alpha$. The resulting expression for $|\Xi|$ is maximized for $s_1=0$, for which it gives precisely $|\Xi|=1/8$. Therefore, for any state for which $|\vec s|\le1/2$, $|\Xi|\le1/8$.


\section*{Funding}
Excellence Initiative of Aix-Marseille University - A$^*$MIDEX, a French ``Investissements d'Avenir" programme.
French National Research Agency (ANR) through award ANR-21-CE24-0014-01.

\section*{Acknowledgements}
I would like to thank Mark R. Dennis, Konstantin Bliokh, Luis A. Alem\'an-Casta\~neda, Rodrigo Guti\'errez-Cuevas, Anthony Vella, Isael Herrera, David Marco, Thomas G. Brown, Sophie Brasselet, Colin J.R. Sheppard, Fran\c cois Amblart, and Philippe R\'efr\'egier for useful conversations.

\section*{Disclosures}
The author declares that there are no conflicts of interest related to this article.

\end{document}